**Air entrainment by turbulent plunging jets: effect of jet roughness revisited.**

**Short title**: **Air flow rate entrained by plunging jets**

Redor Ivan[a], Guyot Gregory[a,b], Obligado Martin[a], Matas Jean-Philippe[c], Cartellier Alain[a]

[a]Univ. Grenoble Alpes, CNRS, Grenoble-INP, LEGI, Grenoble, 38000, France
[b]CEA, LITEN, Campus INES, 73375 Le Bourget du Lac, France
[c]Universite Claude Bernard Lyon 1, LMFA, UMR5509, CNRS, Ecole Centrale de Lyon, INSA Lyon, 69622 Villeurbanne, France

Alain.Cartellier@univ-grenoble-alpes.fr

**Highlights**:
- The roughness scenario prevails for heights of fall above 20 jet diameters.
- The entrained air flow rate grows as jet impact velocity times jet roughness.
- For non-aerated jets, the effective roughness is the maximum jet deformation.
- For aerated jets, the effective jet roughness is halved.
- A maximum air entrainment capacity is evaluated in both cases.

**Graphical abstract**

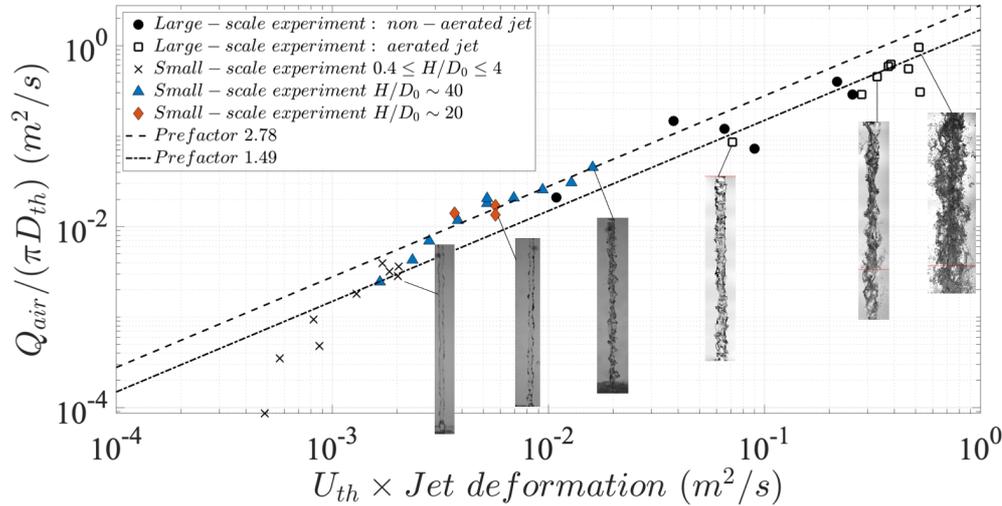

The air flow rate entrained below a free surface by coherent plunging jets has been measured in a small-scale and in a large-scale facility. For large enough fall height (here ≥ 20 jet diameter), the entrained air flow rate happens to be proportional to the jet perimeter at impact times the jet velocity at impact times the jet deformation evaluated as the standard deviation of the lateral position of one side of the jet. The prefactor for aerated jets (≈ 1.5) is about half the prefactor identified for non-aerated jets (≈ 2.8). Jets images are not to scale.

**Abstract**

The amount of air entrained by vertical water jets impacting a large pool is revisited. To test available phenomenological models, new data on the jet deformation at impact and on the entrained air flow rate were collected both on a small-scale (height of fall H about 1 m, jet diameter $D_0 = 7.6$ mm) and a large-scale (H up to 9 m, $D_0$ up to 213 mm) facilities. Conditions for which jet break-up occurred were not considered. For short heights of fall (H less than a few $D_0$), the jet deformation remains smaller than 0.1 jet diameter, and the entrained air flow rate happens to grow as $U_i^{3/2}$, where $U_i$ is the jet velocity at impact. This scaling agrees with the air film model proposed by Sene, 1988. At larger fall heights, even though conditions leading to jet break-up were avoided, the jets nevertheless exhibited complex topologies, including strong deformations and/or interface stripping and/or jet aeration. Further, the roughness model initiated by Henderson, McCarthy and Molloy, 1970 which stipulates that the entrained air flow rate corresponds to the air trapped within jet corrugations, was found valid for these conditions. More precisely, for corrugated jets, the effective roughness amounts to the maximum jet deformation, as measured from the 90% detection probability on the diameter pdf. Equivalently, the effective roughness can be defined as twice the total deformation of one side of the jet, experimentally evaluated as the standard deviation of the position of one jet edge. However, for jets experiencing strong stripping or aeration (the latter being identified by a threshold on the growth of the jet diameter with the falling distance), the effective roughness amounts to about 0.8 times the maximum jet deformation or equivalently 1.1 times the total deformation of one side of the jet. Compared with corrugated jets, the effective roughness is thus diminished by half. A possible origin of this drastic behavior change could be due to the impact of air friction on distorted liquid structures, such as waves or ligaments, which may lead to a decrease in the jet velocity at the jet periphery compared with the impact velocity expected from free fall and/or to corrugations with a different allure that are less efficient for trapping air. A criterion for the transition between the air film and the jet roughness scenarios has also been proposed.

For coherent (i.e. non atomized) plunging jets, we have identified a plausible upper limit of the entrained air flow rate with respect to the injected liquid flow rate. The latter equals 3 for corrugated jets, and it is about 2 for aerated jets. Local void fractions in the bubble cloud formed below the free surface are related to the gas flow rate fraction, leading to void fraction estimates useful for engineering purposes. Finally, we show for our data and for data from the literature that the entrained flow rate scaled by the injected liquid flow rate is comprised between two simple limits, expressed in terms of a Froude number based on velocity and diameter at impact. Globally, the air flow rate is shown to increase as the impact velocity $U_{th}^n$, with n between 1.7 and 2.1.

**Introduction**

Air entrainment occurs in a large variety of systems. It ensures the aeration of oceans by means of breaking waves (Kiger & Duncan, 2012; Tang et al., 2020), as well as that of torrents and diverse hydraulic structures through natural or artificial chutes (e.g., Chanson, 1997; Ervine, 1998; Chanson et al., 2021; Hoque and Paul, 2022). Air entrainment is exploited in industry notably, as contactors for chemical or biochemical engineering (according to Burgess et al., 1972, the use of a plunging jet as a reactor was patented in 1938). Aside from the variety of practical situations, air entrainment is related to several scientific issues. In particular, air entrainment mechanisms and conditions for the onset of air entrainment are known to be sensitive to interface deformation and break-up conditions. Besides, quantifying the amount of air entrained in and transferred to the liquid phase requires the determination of the characteristics of the bubble cloud that forms below the free surface. The latter include, for example, the depth of penetration, the bubble size distribution, or the spatial distribution of the gas flux and of the void fraction. Active research has been conducted on these topics since the 1950s, and notable progress has been made on the prototypal case of a single plunging jet impacting a large, deep pool (Bin, 1993; Kiger & Duncan, 2012; Müller and Chanson, 2020).

Recently, air entrainment by plunging jets has received renewed attention about the efficiency and safety of hydropower generation as bed erosion downstream overflowing dams becomes more critical in the context of climate change. To cope with such risks, an improved understanding of air entrainment processes is needed, but a clear consensus on the modeling of these systems is still missing, even when considering the elementary situation consisting of a single plunging jet. Such lack of consensus is clearly illustrated by the variety of correlations currently proposed for quantities such as the entrained air flow rate (see the discussion by Kiger & Duncan, 2012), the penetration depth, or the rate of oxygen transfer (see the discussion by Kumar et al., 2021).

In the following, we consider air-water systems in ambient pressure conditions involving a single vertical jet impinging a pool large enough to avoid any influence of side and bottom walls. We consider steady boundary conditions. We focus on coherent turbulent jets, coherent implying in this work that the liquid core of the jet remains continuously connected with the nozzle. In other words, we do not consider broken-up jets. For these flow conditions, we recently proposed a phenomenological model for the depth of penetration that proves reliable for small (about hundreds of microns) up to medium-scale (about tens of centimeters) diameter jets (Guyot et al., 2020; Dev et al., 2024). The present contribution focuses on measurements of the gas flow rate $Q_{air}$ entrained below the free surface and on the comparison of these measurements to existing models.

## 1. State of the art

### 1.1 Entrainment regimes and mechanisms:

Before examining the amount of gas entrained below the free surface by a single plunging jet, it is worth discussing air entrainment mechanisms. A variety of processes are evoked in the literature, but the contribution of many of these remains speculative as direct evidence is lacking. The difficulty can be caused by the absence of adapted measurements, as is the case for the aeration by entrainment of a foamy layer formed at the pool surface (e.g., Sene, 1988). It can also be caused by the absence of undisputable validation of proposed models, as is the case for the contribution of an air boundary layer modeled by an idealized smooth interface with a steep density gradient (e.g., Van de Sande and Smith, 1973). Our starting point here, which will be subjected to a posteriori verification, is that two main mechanisms can be identified for coherent jets:

- Air entrainment directly related to the roughness of the jet, as illustrated in Fig.1, where a fraction of the air enclosed within jet corrugations is expected to be dragged inside the pool. Such a scenario was initially proposed by McCarthy, Henderson and Molloy in the 1970s (quoted in Burgess, Molloy and McCarthy, 1972) and it has been further considered notably by Sene, 1988; Evans et al., 1996; Bagatur and Sekerdag, 2003. A companion mechanism corresponds to "isolated" bulges impacting the pool and creating an air cavity that drags air below the free surface, as investigated and modeled by Oguz, 1995, Oguz et al., 1998 and Zhu et al., 2000. A contribution from the air boundary layer that develops along the jet is sometimes added to this scenario (Szekely, 1969 quoted in Van de Donk, 1981, Van de Sande and Smith, 1973).

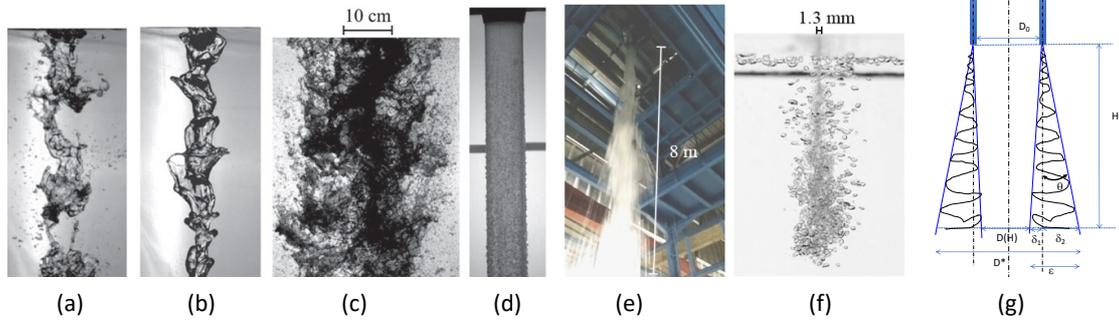

Fig.1: Examples of corrugated turbulent jets (a) $D_0$=37 mm and $U_0$=11 m/s, (b) $D_0$=83 mm and $U_0$=3 m/s, (c) $D_0$=83 mm and $U_0$=5 m/s, (d) Same conditions as (c) but at the injection. (e) View of the large-scale experimental facility for $D_0$ = 213 mm, $U_0$ = 2 m/s. (f) View of a bubble cloud formed below the free surface in the small-scale experiment for $D_0$ = 1.3 mm and $U_0$ = 8 m/s. (a), (b), (c) are at the same scale. (g) Sketch of the jet deformation according to Burgess et al., 1972.

- Air entrainment by a thin continuous air film that forms a sheath around the liquid jet below the free surface, as illustrated in Fig.2. Such film elongates as the jet velocity increases, and above some critical velocity, it breaks into bubbles that are entrained deeper into the receiving pool. This film mode has been identified in high-viscosity liquids (Lin and Donnelly, 1966; Cartellier and Lasheras, 2003; Lorenceau et al., 2004). It also exists for low-viscosity fluids such as in air-water systems, as exemplified in Fig.2 (see also Bonetto and Lahey, 1993; Chanson and Cummings, 1994; Kiger and Duncan, 2012), when the jet velocity becomes large enough.

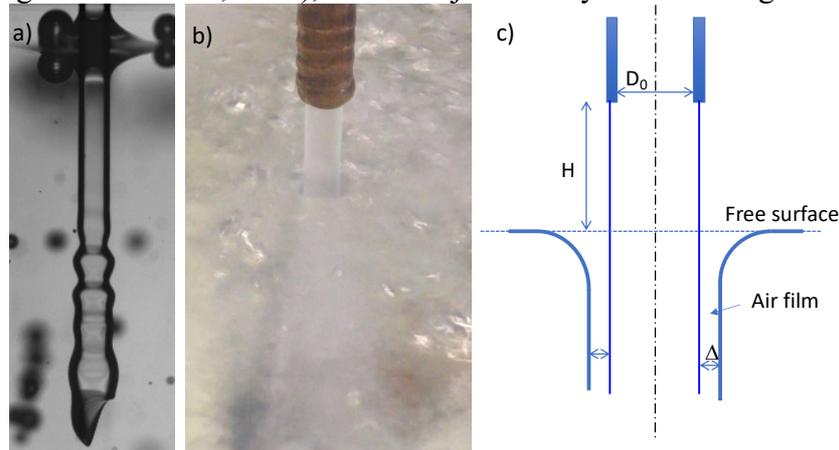

Fig.2: Gas film formation beneath the free surface for a very smooth jet: a) with a viscous fluid (canola oil, Cartellier and Lasheras, 2003), b) with air and water in the small-scale experiment described in Section 2 (note the "trumpet" formed at the interface and also the conical trace of the air film boundary visible below the free surface), c) sketch of the interface deformation and the gas film in the vicinity of the free surface.

The roughness of the jet is expected to play a central role in controlling which of these two entrainment mechanisms is preponderant. The jet roughness is a complex function of the jet diameter, velocity, and height of fall but also of the design of the water circuitry and the nozzle: this design may affect turbulence and/or secondary motions at injection and/or instabilities such as the internal boundary layer stability (e.g., Ramirez de la Torre et al., 2020). Jet roughness may also be controlled by perturbations, such as vibrations, either external or imposed (Davoust et al., 2002). Both entrainment scenarios can therefore be observed in the same facility depending on the selected flow selected, and boundary conditions. In particular, a transition in the air entrainment mode is often reported for water jets when the jet velocity exceeds about 4.5 to 5 m/s (e.g., van de Sande and Smith, 1972, 1973). That transition is attributed to a shift between an entrainment mode related to jet roughness and an air film mode. When the injection velocity increases, all other flow parameters being kept the same, the transit time of disturbances between the nozzle and the free surface decreases, and the jet roughness at impact diminishes and could cause air entrainment to be controlled by the "smooth jet" mechanism. Such a transition in the air entrainment mode is expected to strongly impact the dependence of the air flow rate upon the jet velocity. We will come back to the competition between these scenarios in Section 1.3. We first discuss in the following subsection the conditions corresponding to the onset of air entrainment.

### 1.2 Onset of air entrainment by a plunging jet:

It is commonly observed that air entrainment occurs when the jet velocity at impact $U_i$ exceeds some critical value $U_c$. Yet, for turbulent jets, unambiguous measurements of $U_c$ are difficult because, close to such critical velocity, the air entrainment process is usually intermittent, while it becomes more continuous as the impact velocity increases significantly above $U_c$. The intermittency is attributed to localized disturbances traveling from time to time along the jet and impacting the pool. Moreover, different definitions of air entrainment inception have been considered in the literature. For example, McKeogh and Ervine, 1981 defined $U_c$ as the onset of continuous air entrainment, while Cummings and Chanson, 1999 measured $U_c$ as "a 'primary' entrainment event which occurs during an interval of 5 minutes". For turbulent water jets in air under normal pressure and temperature conditions, and depending on flow conditions (nozzle design, height of fall, turbulence in the jet etc.), $U_c$ was found to vary between about 0.8 m/s – a value observed for "rough" jets – and about 3.7 m/s – a value observed on "very smooth" jets (Bin, 1993; Cummings and Chanson, 1999; Chanson, 2009; Harby et al., 2014). So far, no general model is providing $U_c$ for turbulent jets. In an attempt to rationalize existing data, Cummings and Chanson, 1999 empirically correlated the critical capillary number $Ca_c = \mu_L U_c/\sigma$, where $\mu_L$ is the dynamic viscosity of the liquid (water in their case), and $\sigma$ denotes surface tension, with the turbulent intensity Tu in the liquid (Tu is defined as the root mean square of velocity fluctuation relative to the mean velocity). They found:

$$Ca_c = 0.0109\ [1 + c_1 \exp(- c_2\ Tu)] \qquad\qquad (1)$$

Where coefficient $c_1$ is comprised between 3.5 and 3.75 and coefficient $c_2$ is comprised between 70 and 80 (Cummings and Chanson, 1990; Chanson, 2009; Bertola et al., 2018a). The transition between the two asymptotic behaviors, namely from $Ca_c \approx 0.051$ (that is $U_c \approx 3.7$ m/s) in the "smooth jets" limit to $Ca_c \approx 0.011$ (that is $U_c \approx 0.8$ m/s) in the "rough jets" limit, arises when the velocity fluctuation Tu increases from $\approx 10^{-3}$ to about $\approx 10^{-1}$.

Concerning the limit of applicability of eq.(1), it is known that the critical capillary number strongly increases as the jet diameter decreases below the capillary length $a_c = \{\sigma/(\rho\ g)\}^{1/2}$ (see Annex A). In particular, for a water jet in the air with a diameter $D_0 \approx$ $a_c/10$, $Ca_c$ reaches 0.1, that is, the critical velocity $U_c$ becomes as large as about 7.5m/s (Cartellier and Lasheras, 2003). Since eq.(1) has been built from jets with a diameter ranging from 1 to 9 times the capillary length scale $a_c$ ($a_c$ = 2.7 mm for air-water in standard conditions), it is therefore valid for jets with a diameter exceeding the capillary length scale. Besides, eq.(1) has been built from fall heights ranging from 0 to 60 times the jet diameter. The smooth jet limit has been obtained either with turbulence-damping devices in the nozzle and/or for very short heights of fall so that interface disturbances have very little time to develop before impact. Overall, eq.(1) applies to coherent (i.e., non-atomized, but possibly with some droplets stripped off the jet by the relative wind), non-aerated (i.e., without air imprisoned inside the jet) turbulent water jets whose diameter is comprised between $a_c$ and 10 $a_c$.

The other piece of information available on the critical entrainment velocity concerns viscous smooth jets. In that case, the onset of air entrainment is abrupt and is thus easier to detect in experiments (Cartellier & Lasheras, 2003). A model has been proposed for the onset. Indeed, for viscous, smooth jets, an air film forms below the free surface (Fig.2). The latter ends with a cusp whose equilibrium is driven by the viscous stress at the tip counterbalanced by capillarity. The cusp penetrates deeper below the free surface as the jet velocity increases. In addition, the cusp becomes sharper as the jet velocity exceeds $\sigma/\mu_L$ (where $\mu_L$ is the dynamic viscosity of the liquid). As shown by Eggers (2001), air entrainment occurs when that equilibrium no longer holds so that a crack forms in the liquid, leading to a criterion for the onset of entrainment that writes:

$$Ca_c = \mu_L\, U_c\, /\, \sigma = c_3 \ln(\mu_L/\mu_G) \qquad\qquad (2)$$

Where $\mu_G$ is the dynamic viscosity of the gas (or of the entrained phase in a liquid-liquid system). That scaling has been corroborated by experiments performed in liquid-liquid and air-liquid systems (Lorenceau et al., 2003; Cartellier & Lasheras, 2003) with a prefactor $c_3$ in eq.(1) that slightly varies with flow conditions. For smooth plunging jets with a diameter comprised between 0.2 and 4 times the capillary length scale, eq.(2) was found valid for a viscosity ratio $\mu_G/\mu_L$ between $10^{-6}$ and approximately $10^{-3}$. Over that range, the critical capillary number predicted by eq.(2) decreases from 7-8 down to about 0.5-1. The above-mentioned experiments have, therefore, proved that eq.(2) holds at least for $Ca_c$ above about unity. For smooth viscous jets (Lin and Donnely, 1966; Cartellier & Lasheras, 2003), the limit $Ca_c \approx 1$ coincides with jet Reynold numbers less than about 1000.

In air-water jets, the critical capillary number for air entrainment evolves between $10^{-2}$ and $5\ 10^{-2}$: it is thus much smaller than unity. These onset conditions also correspond to jet Reynold numbers larger than 1000. Hence, eq.(2) is not applicable to air-water systems, possibly because the derivation of eq.(2) does not account for inertia nor for the apparition of an asymmetrical gas film or trumpet (Cartellier & Lasheras, 2003).

Concerning the prediction of the critical entrainment velocity, there is thus a gap to be filled between the empirical formula eq.(1) established for jet Reynolds numbers above 2200 and the prediction from eq.(2) valid for jet Reynolds numbers below $\approx$ 1000.

### 1.3 Phenomenological models for the air flow rate entrained by a plunging jet

The air flow rate $Q_{air}$ entrained by a plunging jet has been measured in diverse flow conditions using a variety of techniques. As reviewed, notably by Bin, 1993; Chanson, 1997; Kiger and Duncan, 2012 and by Ervine, 1998 for various hydraulic structures, several empirical correlations have been proposed from collected data, and some of those have a good predictive capability within their range of validity. However, these correlations exhibit quite different dependencies of $Q_{air}$ versus the jet velocity and/or the jet diameter, meaning that no definite agreement has been reached on the scaling laws driving the entrained air flow rate. In addition, the identification of relevant scaling laws from experiments could be delicate. For example, for water jets with a diameter between 3 and 10 mm, Van de Sande and Smith (1972, 1973, 1976) identified three regions corresponding to $Q_{air} \propto U_0^3$ at large fall heights or to $Q_{air} \propto U_0^2$ at low fall heights for jet velocities below 4 to 5 m/s, to $Q_{air} \propto U_0^{1/2}$ for jet velocities between 4-5 m/s and 10-15 m/s, and to $Q_{air} \propto U_0^2$ for velocities above 10-12 m/s. Bin (1993) re-analyzed these data, and he showed that a $Q_{air} \propto U_i^{1.56}$ behavior is equally valid over the three regions.

In an attempt to identify relevant scaling laws, we first revisit in the following section available phenomenological models before presenting a dedicated experimental campaign (Section 2) with adapted measuring techniques (Section 3) to confront predictions and experiments (Section 4). The discussion of available models and their predictive capability is organized along the two main mechanisms introduced above, namely the air film scenario and the jet roughness scenario. Unless otherwise stated, the experiments considered involve vertical water jets in the air under ambient thermodynamic conditions.

### 1.3.1 Air film scenario:

Two models have been proposed that involve the presence of a continuous air film below the free surface. On the one hand, Sene, 1988 proposed an equilibrium between viscous forces that maintain the thin air film open and an assumed hydrostatic pressure distribution in the receiving pool. Such an equilibrium leads to a Couette-Poiseuille flow inside the gas film. Sene, 1988 solved the resulting equation for the velocity profile in the film by assuming that the gas cannot sustain a reverse flow. The film thickness $\Delta_{Sene}$ derived by Sene, 1988 writes:

$$\Delta_{Sene}^2 = 2 \ \mu_{air} \ U_i \ / \ (\rho_L \ g) \tag{3}$$

Where $U_i$ is the jet velocity at impact. The gas flow rate per unit length $Q_{air}/L$ where L is the length of the contact line between the jet and the free surface (for a cylindrical jet $L=\pi D_i$) is deduced from the gas velocity profile. Sene, 1988 found:

$$Q_{air} \ / \ L = (1/3) \ U_i \ \Delta_{Sene} = (2^{1/2}/3) \ [\mu_{air} \ / \ (\rho_L \ g)]^{1/2} \ U_i^{3/2} \tag{4}$$

In that model, the entrained gas flow rate increases as $U_i^{3/2}$. Note that the above reasoning does not include any capillary effect.

On the other hand, Lorenceau et al., 2004 considered viscous fluids for which the air film becomes quite long. From an equilibrium between viscous tension in the thin film and surface tension associated with the presence of the meniscus (the latter remains close to a hydrostatic meniscus), they derived the air film thickness $\Delta_{cap}$, namely:

$$\Delta_{cap} = 0.5 \ a_c \ Ca_G^{2/3} \tag{5}$$

Where $a_c$ is the capillary length, and the air capillary number is defined as $Ca_G = \mu_G U_i/\sigma$. where $\mu_G$ is the dynamic viscosity of the entrained phase. The prefactor 0.5 in eq.(5) has been identified from experiments in various systems and using diverse couples of fluids (including liquid-liquid systems) that cover capillary numbers $Ca_G$ based on the viscosity of the upper phase ranging from $10^{-4}$ up to a few units. Owing to the Couette flow taking place in the gas film, the gas flow rate injected in the pool per unit length $Q_{air}/L$ equals $U_i \Delta_{cap} / 2$. Therefore:

$$Q_{air} / L = U_i \, \Delta_{cap} / 2 \; \approx 0.25 \; a_c \, [\mu_{air}/\sigma]^{2/3} \, U_i^{5/3} \qquad (6)$$

Lorenceau et al., 2004 predict $Q_{air}$ to increase as $U_i^{5/3}$, which is close to the $U_i^{3/2}$ dependency proposed by Sene. Despite this similitude, the two proposals exhibit significant differences. First, and contrary to Sene, 1988's proposal, the entrained gas flow rate in the Lorenceau et al., 2004 's model depends on surface tension. Second, the quantitative predictions from these two models are quite different. For water jets in the air under normal pressure and temperature conditions, the film thickness predicted by the viscous-capillary equilibrium grows from 5 μm to 100 μm when the jet velocity increases from 1 to 100 m/s. This is 5 to 10 times smaller than the thickness predicted by Sene, 1988 as the latter is comprised between 50 and 550 μm. Accordingly, the entrained gas flow rate per unit perimeter predicted by the viscous-capillary model of Lorenceau et al., 2004 is between 5 and 7 times smaller than the one predicted by the viscous-hydrostatic approach of Sene, 1988. Let us also mention that none of the above models account for liquid inertia.

Concerning the range of validity of the above proposals, the air film thickness predicted by Lorenceau et al., 2004 has been shown to hold for a 1.5 mm diameter glycerol jet in air at velocities $U_0$ from $\approx 0.5$ to 3 m/s, that is, for $10^{-4} \leq Ca_G \leq 6 \; 10^{-4}$ and jet Reynolds numbers from 750 to 4500. In addition, and thanks to experiments in liquid-liquid systems, eq.(5) was found valid for capillary numbers based on the dynamic viscosity of the upper fluid $Ca_G$ comprised between $10^{-4}$ up to a few units. The associated air flow rate was not measured but, owing to Lorenceau et al.'s model, eq.(6) is expected to hold whenever the film thickness follows eq.(5).

Meanwhile, and to our knowledge, Sene,1988's film thickness model eq.(3) has never been directly tested as no air thickness measurements are available for air-water systems. The few crude estimates of the gas thickness $\Delta$ by Chanson and Cummings, 1994 indicate that for $U_0$ between 2 and 7 m/s, $\Delta$ lies in the range 0.3-3 mm to be compared with the 0.08-0.15 mm interval predicted from eq.(3). Concerning the entrained air flow rate, Sene, 1988 considered that the data of Van de Sande and Smith, 1973 for jet velocities above 10 m/s (flow conditions: air-water jets at a 30° inclination, $D_0$ from 3 to 10 mm, $H/D_0 <$ 0.64) support the scaling $U_i^{3/2}$ predicted by eq.(4). By analyzing Van de Sande and Smith, 1973 data for jet velocities above 10 m/s, the entrained air flow rate per unit perimeter happens to grow as $U_i^n$ with exponents n between 1.73 to 1.8 for each data series. These exponents are compatible with Sene, 1988's proposal as well as with Lorenceau et al., 2004 's model, and that compatibility is observed over the following range of parameters: $6.8 \; 10^4 \leq Re_{impact} = U_i \, D_i/\nu_{water} \leq 1.1 \; 10^5$; $2 \; 10^{-3} \leq Ca_G = \mu_G U_i/\sigma \leq 3.2 \; 10^{-3}$; $0.14 \leq Ca_L = \mu_L U_i/\sigma \leq 0.23$. However, over that range, the gas flow rate predicted by eq.(4) only amounts for 4% of the measured value. Similarly, the prediction from eq.(6) amounts to less than 1% of the measured gas flow rate. Hence, even though the scalings of $Q_{air}/L$ with the jet velocity proposed by both models correspond to those observed in Van de Sande and Smith, 1973 experiments, the quantitative predictions are off by significant factors that can hardly be attributed to prefactor(s) adjustments in the proposed phenomenological models. In conclusion, the Lorenceau et al., 2004 film model has been found valid over a narrow range of flow parameters (in particular, $10^{-4} \leq Ca_G \leq 6 \ 10^{-4}$) while the gas film proposal from Sene, 1988 has, so far, not received any undisputable experimental confirmation.

### 1.3.2 Jet roughness scenario

In the 1970s, Henderson, McCarthy and Molloy argued that all the air trapped within the corrugations of the jet is entrained below the free surface of the receiving pool. Hence, the entrained gas flow rate $Q_{air}$ is predicted to be equal to the area occupied by the corrugations times the jet velocity (see Burgess, Molloy and McCarthy, 1972 and references therein), namely:

$$Q_{air} = (\pi/4) \ (D^{*2} - D_0^2) \ U_0 \tag{7}$$

Where the outer envelope of the jet is assimilated to a circle of diameter $D^*$ as illustrated in Fig.1. Eq.(7) can be rewritten:

$$Q_{air} / Q_w = (D^*/ D_0)^2 - 1 \tag{8}$$

Where $Q_w$ denotes the liquid flow rate. In Henderson et al.'s proposal, the diameter of the pure liquid core at impact is assumed to be the same as the jet diameter at injection $D_0$, and the jet velocity at impact is considered equal to the jet velocity at nozzle $U_0$. Henderson et al. performed experiments for a fixed nozzle diameter ($D_0$=2.54 mm), for jet velocities between 10 and 18 m/s, and for fall heights between $\approx$5 and 75 $D_0$. They did not provide data directly supporting their proposal. Instead, they evaluated the global interfacial area from the global reaction rate in their system, and found it proportional to $U_0 D^*/D_0$, thus indirectly validating eq.(7) in their experimental conditions.

Van de Sande and Smith, 1973 tested eq.(8) on water jets in air for nozzle diameters between 2.85 and 6.8mm, injection velocities between 2.5 and 20 m/s, and heights of fall H from 10 to 51 $D_0$, but smaller than the break-up limit. All datasets except for two points correspond to H < 0.6 $L_B$, where $L_B$ is provided in Van de Sande and Smith, 1976. They selected long nozzles (length > 50 diameters) and jet Reynolds numbers above $5.10^4$ to ensure fully developed turbulent pipe flows at the nozzle exit. The diameter of the corrugated jet outer envelope $D^*$ was measured from long-exposure photographs with backlighting. Their results correlate as follows:

$$D^*/ D_0 = 0.085 \ (We_{air} \ Re_{length})^{1/6} \tag{9}$$

where $We_{air} = \rho_{air} U^2 D_0/\sigma$ and $Re_{length} = U H/\nu_{air}$. Here, U is the jet velocity, $\rho_{air}$ denotes the air density, and $\nu_{air}$ its kinematic viscosity, $\sigma$ is the surface tension of water with air. Since only air-water systems in ambient conditions were considered, the dependencies on physical properties in eq.(9) were not checked. Van de Sande in his PhD (1974) did not distinguish between the jet velocity at impact $U_i$ and at nozzle $U_0$ because he claimed that friction equilibrates gravity so that $U_i$ equals $U_0$. Eq.(9) was found to be valid for $We_{air}$

Re$_{length}$ larger than 7 10$^6$, a criterion that corresponds to large jet velocities U$_0$ (the latter must be above 6 m/s for the D$_0$ = 10mm injector and above 8 m/s for all others injectors). At lower values of We$_{air}$ Re$_{length}$, the predicted gas flow rate approaches zero and can even become negative because D* becomes less than D$_0$. The ratio Q$_{air}$/Q$_w$ can be deduced by injecting eq.(9) into eq.(8). In Fig.3, these predictions (using the parameters U$_0$ and D$_0$ at injection in eq.(9)) are compared with Van de Sande and Smith (1973) experimental data collected for a jet inclination of 30° with the vertical, for D$_0$ = 3, 3.8, 4.9, 6.8 and 10 mm and H=0.1 m. Let us underline that the 30° jet inclination is not a key issue for that comparison as its impact on Q$_{air}$ is at most ±15% when compared with a vertical jet (see the discussion in Bin, 1993 and Bin, 2019). Fig.3 shows that the measured values of Q$_{air}$/Q$_w$ happen to be comprised between 0.9 and 2.5 times the prediction from eq.(8). To reach a better agreement with experiments, Van de Sande and Smith, 1973 proposed to add a contribution from the air boundary layer that develops along the jet, assuming that the latter is entrained below the free surface. However, this proposal is contradicted by the variations of the ratio Q$_{air}$/Q$_w$ with flow conditions. Indeed, their model states that the boundary layer contribution increases with the jet velocity (see Fig.8 in Van de Sande and Smith, 1973) while, as shown in Fig.3, the difference between the measured and the predicted ratio Q$_{air}$/Q$_w$ *decreases* when the jet velocity increases. Letting aside this uncertain air boundary layer contribution, the experiments from Van de Sande and Smith (1973) indicate that the model based on eq.(8) and eq.(9) tends to be valid only at large jet Reynolds numbers (say Re $\gg\approx$10$^5$), and at large air Weber numbers (presumably beyond 15-20). The origin of the failure of the model based on eq.(8) and eq.(9) at lower Reynolds and Weber numbers remains unclear.

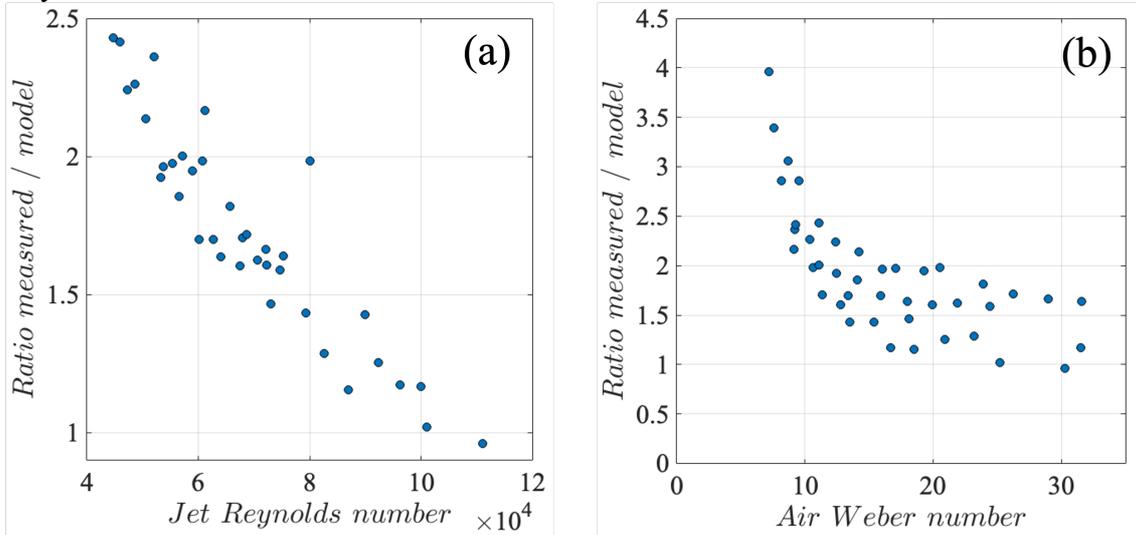

Fig.3: Ratio of the measured air flow rate to the air flow rate predicted from eq.(8) and eq.(9) using U$_0$ and D$_0$ the jet Reynolds number at injection and b) the air Weber number at injection. Experimental data from Van de Sande and Smith (1973) series for D$_0$ = 3, 3.8, 4.9, 6.8 and 10 mm, H = 0.1 m, 30° inclination angle and for flow conditions such that We$_{air}$ Re$_{length}$ > 7 10$^6$.

Cumming I.W. (1975) modified eq.(8) by considering the jet parameters at impact. For a vertical cylindrical jet that does not experience any distortion of its shape, the jet velocity taking into account free fall (without friction) under gravity is U$_{th}$(H) = (U$_0^2$ + 2gH)$^{1/2}$, and the jet diameter deduced from continuity writes D$_{th}$ = 2 R$_{th}$ = D$_0$ (U$_0$/U$_{th}$)$^{1/2}$. Cumming I.W. proposed to write Q$_{air}$ = (π/4) (D*$^2$ - D$_{th}^2$) U$_{th}$, or equivalently:

$$Q_{air} / Q_w = (U_{th}/U_0) (D*/ D_0)^2 - 1 = (D*/ D_{th})^2 - 1 \qquad (10)$$

Cumming I.W., 1975 tested eq.(10) on his experiments performed for $D_0$ from 4.6 to 8.9 mm, for $U_0$ from 1.8 to 7.8 m/s and for $H/D_0$ from 11 to 64. As shown in Fig.4, his data are somewhat scattered (measurement uncertainties are not provided). Cumming, 1975 states that "The jets used in this study were very rough, and droplets were continually flung from the jet at higher speeds". The maximum roughness measured by Cumming, 1975 is about 0.35 $D_0$, indicating that his flow conditions correspond indeed to coherent turbulent jets. A linear fit forced through the origin of Cumming's data roughly validates eq.(10) with a correlation coefficient of 0.9 (dotted line in Figure 4). Interestingly, the slope equals 0.91, indicating that 91% of the air trapped within the jet boundaries is entrained below the free surface. The dispersion around the fit is significant, lying between -73% and +93%.

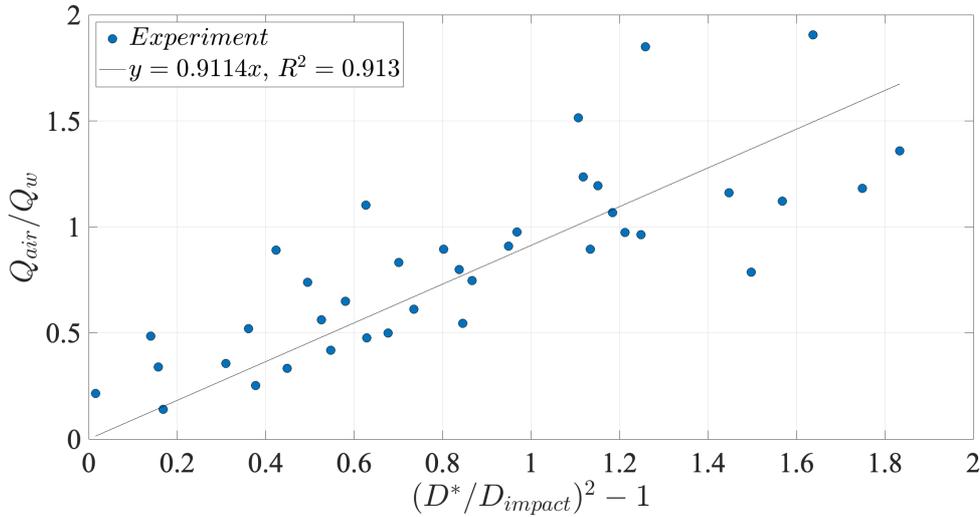

Fig.4: Test of eq.(10) using the experimental data of Cumming, 1975 (Tables 22, 23 and 24 from his PhD, 1975). These data series correspond to $4.6 \leq D_0 \leq 8.9$ mm, $1.8 \leq U_0 \leq 7.8$ m/s, $11 \leq H/D_0 \leq 64$.

Further tests were provided by Ervine, McKeogh and Elsawy (1980) and McKeogh and Ervine (1981), who performed experiments on water jets in the air for $D_0$ from 6 to 25 mm, $U_0$ up to 10 m/s, fall heights up to 4 m, and turbulent intensities between 0.3 and 8%. They measured the entrained air flow rate, using slightly inclined jets (the inclination is not provided). To exploit their data, we assumed that the jet inclination they used was always small enough so that it had no significant effect on the entrained air flow rate. These authors quantified the jet deformation using photographs of the jet. Their procedure is not precisely explained; they just indicate that measurements become tricky when the deformation exceeds 0.6-0.7 times the jet radius. According to the sketches they provide (see Fig.7 and 10 in Ervine et al., 1980), the deformation $\varepsilon_{max}$ they measured corresponds to the maximum deformation of one side of the jet with respect to the *median position* of the interface, that is $D^* = D + 2\ \varepsilon_{max}$, where D=2R is mean jet diameter at the considered position. Let us mention that, in these experiments and over the entire velocity range considered, the growth rate of the jet deformation with the fall height H decreases with the turbulent intensity: $\varepsilon_{max}$ increases as $(H/D_0)$ for smooth jets (Tu < 1%), as $(H/D_0)^{0.73}$ for jets with 1% < Tu < 5%, and as $(H/D_0)^{1/2}$ for rough jets (Tu > 5%). Ervine, McKeogh and Elsawy, 1980 rewrote eq.(8) observing that $(D^*/D)^2 - 1 = 2\ (\varepsilon_{max}/R) + (\varepsilon_{max}/R)^2$. Their experiments cover the range $0.05 \leq \varepsilon_{max}/R \leq 1$, where the largest deformation $\varepsilon_{max} \approx R$ corresponds to the break-up limit. Let us underline that it is unclear whether or not the authors have measured the jet radius versus the distance to the injector. Instead, they seem to have considered D=2R as equal to the diameter at the nozzle $D_0$ in their analysis. Ervine et al. (1980) argue that eq.(8) slightly underestimates $Q_{air}/Q_w$ measurements for $\varepsilon_{max}/R \leq 0.79$ (they attributed the difference to an extra contribution due to an air boundary layer that develops along the falling jet). Moreover, they also observe that eq.(8) overestimates measurements for $\varepsilon_{max}/R \geq 0.79$: they indicate that sinuous deformations become then significant and that "the crests of the undulations break away from the main body of the jet", meaning that their jet experienced some atomization by stripping. However, when closely examining their results (see Fig.15 in McKeogh and Elsawy, 1980), the measured ratio $Q_{air}/Q_w$ happens to be equal to $(D^*/D)^2 -1$ with a dispersion comprised between -16% and +53%. Meanwhile, for the alternate fit proposed by Ervine et al. (1980), namely $Q_{air}/Q_w = (1/4) [(D^*/D)^2 - 1 - 0.1]^{0.6}$, the dispersion is ±30%. It is therefore difficult to conclude that the latter proposal is more relevant than a linear fit, and we can reasonably consider that the experiments of Ervine, McKeogh and Elsawy, 1980 and of McKeogh and Ervine, 1981 bring some support to eq.(8) over the range of parameters investigated.

Ervine and Falvey, 1987 measured the lateral spread of a horizontal jet produced with a nozzle diameter $D_0 = 50$ or 100 mm for ejection velocities $U_0$ from 3 to 29.6 m/s. They found:

$$\delta_2 / x = 0.358 \ U'/ U_0 \qquad and \qquad \delta_1 \approx 1/5 \ to \ 1/7 \ \delta_2 \qquad (11)$$

Where $\delta_1$ and $\delta_2$ are defined in Fig.1-g, and where x is the distance from the nozzle. The divergence angle θ of the jet (Fig.1), corresponding to eq.(11), happens to be quite small: it is about 0.2° for Tu=1%, and 2° for Tu=10%. Ervine and Falvey, 1987 did not provide the exact flow conditions corresponding to the measurements that lead to eq.(11). In particular, they mention that, above some velocity, the jet experiences "free surface aeration". According to their Fig.7, air penetrates inside the jet up to 37% of the jet radius. However, they do not quantify the non-aerated / aerated transition from their experiments, and therefore, we do not know if eq.(11) holds for non-aerated jets, for aerated jets, or both. Ervine and Falvey, 1987 seem to consider that for $D_0$=100 mm (see their Fig. 4), the case $U_0$=5 m/s is non aerated, while the case $U_0$=25 m/s is aerated. They indicate that the transition depends on the turbulent intensity Tu.

Sene, 1984 & 1988 proposed a model for rough jets based on two arguments. First, and in line with previous proposals, Sene argues that the entrained air corresponds to the air trapped within the corrugations of the jet interface. Sene introduced a characteristic height ε of the jet corrugations at impact, where ε is counted in a direction perpendicular to the mean jet velocity. Considering the length L of the contact line between the jet at impact and the pool, and the jet velocity at impact $U_i$, Sene suggests that $Q_{air}/L$ is proportional to $U_i$ ε. His suggestion leads to:

$$Q_{air} / L = K_{12} \ U_i \ \varepsilon \qquad (12)$$

Where $K_{12}$ is a prefactor to be determined. Initially, Sene wrote eq.(12) for a plane jet, and ε represents the deformation of the unique jet boundary involved. When applied to a cylindrical jet, and considering that L = π $D_{th}$ at impact, eq.(12) leads to :

$$Q_{air} / Q_w = 4 K_{12} (U_i/U_{th}) (\varepsilon/D_{th}) = 2 K_{12} (\varepsilon/R_{th}) \qquad (13)$$

Where for the second equality, it has been assumed that the jet velocity at impact equals $U_{th}$, the velocity reached by the free fall jet free gravity without friction ($U_{th}$ is defined before eq.(10)). In eq.(13), $\varepsilon$ represents the interface deformation on *one side of the jet*. Eq.(13) happens to be a linearized version of Henderson et al.'s model or of its Cumming's 1975 derivative eq.(10) for small deformations. Indeed, if one writes $D* = D_{th} + 2 \varepsilon = 2 R_{th} + 2 \varepsilon$, then eq.(10) transforms into :

$$Q_{air} / Q_w = 2 \varepsilon/R_{th} + (\varepsilon/R_{th})^2 \qquad (14)$$

Therefore, eq.(13) happens to be equivalent to eq.(14) in the limit $\varepsilon \ll R_{th}$. Eq.(13) departs from eq.(14) by more than 10% when the deformation $\varepsilon/R_{th}$ exceeds 0.2.

Sene borrowed his second argument from sub-surface turbulence in free surface flows, as he proposed to estimate $\varepsilon$ as $u'^2/g$ where u' is the root mean square of velocity fluctuations for the velocity component normal to the jet interface. Since, for a jet of fixed turbulent intensity $Tu = u'/U_i$, u' remains proportional to $U_i$, Sene wrote:

$$\varepsilon \propto U_i^2 /g \qquad (15)$$

And the gas flow rate per unit contact length predicted by Sene becomes:

$$q_{air} = Q_{air} / L \propto U_i^3/g \qquad (16)$$

Ervine, 1998 rewrote eq.(16) under the form $q_{air} \propto Tu^2 U_i^3/g$ to underline the role of the turbulent intensity in Sene's model. Sene, 1984, 1988 measured the entrained air flow rate in experiments involving supported planar jets at various inclinations. He also measured an interface roughness $\varepsilon$. Sene did not exploit the position of the interface, but he used instead liquid concentration profiles in a direction perpendicular to the jet, deduced from a resistivity probe. The mean interface position was set as the position for which the liquid concentration equals 0.5, and he used the standard deviation of the liquid fraction around that mean to evaluate the interface roughness. If one assumes that the interfacial waves have a fixed shape, the two quantities are indeed unambiguously related. Unfortunately, Sene did not provide data for $\varepsilon$, so that a direct test of his model using his experiment dataset under the form of eq.(12) is not possible. Yet, Sene tested the dependency of $Q_{air}/L$ versus the jet velocity, and he found the $U^3$ behavior predicted by his proposal eq.(16) to be globally valid for the experimental conditions he considered, namely for $U_0$ from 1.4 to 3.15 m/s, for planar jets thicknesses from 12 mm to 47 mm and plunge angles from 25° to 75°.

Evans, Jameson and Rielly, 1996 investigated confined jets for diameter $D_0$ ranging from 2.38 mm to 7.12 mm and velocities $U_0$ from 7.8 to 15 m/s. They pursued the approach initiated by Henderson et al. by measuring the jet roughness on photographs of the jet for heights of fall from 3 to 30 $D_0$. Most of their experiments used water with a small quantity of surfactants to avoid coalescence. For an unconfined jet, they correlated the jet roughness as:

$$2\,\delta_{Evans}/D_0 = D^*/D_0 - 1 = 0.0085\ Oh^{0.83}\ Re^{0.63} \qquad (17)$$

where the Ohnesorge number is defined as $Oh = \mu_{water} / (\rho_{water}\,D_0\,\sigma)^{1/2}$ and $Re = U_0\,H /\nu_{water}$. Their measurements deviate at most of $\pm20\%$ from that correlation. In their experiments, the roughness $\delta$ varies from nearly 0 up to 0.6 jet radius. They collected data on $Q_{air}$ in confined conditions, and as proposed by Van de Sande and Smith, 1973, they compare their measurements with the sum of a contribution arising from eq.(8) (where they account for the jet diameter at impact as measured in unconfined conditions in place of $D_0$) and a contribution from the air boundary layer that develops along the jet. They claim that their model agrees within $\pm20\%$ with experiments.

Oguz, 1998 investigated a microbubble generator, a device that has strong similarities with a confined plunging jet. His system allowed reliable measurements of the entrained air flow rate. Besides, the axial evolution of the corrugation of the thin water jets ($D_0=1.6$, 2 and 2.4 mm) was quantified with an imaging technique. From these data, Oguz, 1998 found eq.(12) valid in his system, for a jet roughness evolving between $0.002\,D_0$ and $0.02\,D_0$, and for jet velocities equal to 6.1 m/s and 8.3 m/s. A marked discrepancy was found for $U_0=4.7$ m/s, whose origin is unclear.

Oguz, Prosperetti and Kolaini, 1995 and Zhu, Oguz and Prosperetti, 2000 demonstrated that a bump on a smooth jet could induce the formation of an air cavity. They show that the volume of entrained air is proportional to $D_0^3\ Fr^{1/3}$, where $Fr = U_i^2/(gD_0)$, and where the prefactor depends on the size of the bump with respect to $D_0$. The pinch-off of the cavity formed below the free surface occurs at time $t_c = 6\ (D_0/2g)^{1/2}\ Fr^{-1/6}$. Considering a corrugated jet as a succession of bumps on a smooth jet, Kiger and Duncan (2012) transformed the results of Zhu et al., 2000 into the following proposal:

$$Q_{air} / Q_w \propto Fr^{1/3} \qquad (18)$$

Davoust et al. (2002) designed a measuring technique providing v'/U where v' is the root mean square of the fluctuation of the velocity component normal to the interface: a quantity they named the "dynamical roughness" $\varepsilon_{dyn}$. They show that, for long wavelengths, $\varepsilon_{dyn} = v'/U$ is proportional to the wave amplitude divided by the axial wavelength of the interfacial deformation. For the measuring system they designed, this quantity should exceed $4\pi$ times the wave amplitude to be properly detected. Measurements of $\varepsilon_{dyn}$ were performed close to the nozzle for a vertical water jet in the air with $D_0=14$ mm and $U_0=2.04$ m/s. Davoust et al. (2002) obtained quasi-linear relationships between the entrained air flow rate and the dynamical roughness $\varepsilon_{dyn}$, respectively for the laminar case and for the turbulent case. The increase is especially steep for a turbulent jet since a 3% increase in $\varepsilon_{dyn}$ induces a 35% increase in the entrained air flow rate (see Fig.8b in Davoust et al., 2002). These results demonstrate the direct connection existing between the entrained air flow rate and the jet roughness. Unfortunately, their results concern a very narrow range of $\varepsilon_{dyn}$ as the latter evolve from $3.0\ 10^{-5}$ to $3.2\ 10^{-5}$. Moreover, $\varepsilon_{dyn}$ was measured near the nozzle while the air flow rate was measured at $H/D_0=32.2$, and no information is provided on the evolution of $\varepsilon_{dyn}$ with the height of fall. In a companion publication, El Hammoumi et al. (2002) chose an alternate route where the jet roughness is no longer considered as an explicit parameter. Instead, these authors directly sought a relationship between the entrained air flow rate and the input parameters characterizing the plunging jet. Based on dimensional analysis, they established empirical correlations (one for the laminar case and one for the turbulent case) between the Weber number $\rho_{air} V_{air}^2 D_0/\sigma$, where the velocity $V_{air} = Q_{air}/[\pi D_0^2/4)]$ quantifies the air flow rate, and the following four parameters, the jet Weber number at injection (defined as $\rho_L U_0^2 D_0/\sigma$), the Ohnesorge number at injection ($Oh = \mu_L/(\rho_L D_0 \sigma)^{1/2}$), the Froude number at injection ($U_0^2/(gD_0)$) and $H/D_0$. Hence, the jet roughness no longer explicitly appears in these correlations, even though it is somehow related to the above-mentioned parameters.

Bagatur and Sekerdag (2003) tested Henderson et al.'s proposal on rectangular nozzles with rounded ends. They used two nozzles, one labeled $D_0$=4.7 mm with a thickness $a_0$=3.5 mm and a width $l_0$=5.5 mm, and another one labeled $D_0$=7.5 mm, with a thickness $a_0$=5.0 mm and a width $l_0$=9.8 mm. Jet velocities $U_0$ ranged between about 2 and 12 m/s, with fall heights H equal to 150 mm or 200 mm. The jet was at 45° from the vertical. They measured (with a millimeter scale) the lateral width l* at impact (the jet extent in the other direction is not provided): they observed a linear relationship between $Q_{air}/Q_w$ and l*/$l_0$ -1, and with a slope close to unity (the slope of the best fit is about 0.95). Although the nozzles were not much elongated (the elongation, defined as the nozzle width over the nozzle thickness, evolves from 1.57 to 1.96), they validated Henderson et al.'s model by considering the jet width only. Surprisingly, they measured a maximum jet width equal to 4.1 times the nozzle width, and, for these conditions, the ratio $Q_{air}/Q_w$ reaches 3.

Table 1 summarizes the contributions from literature that test the dependency of jet roughness with flow parameters, and the phenomenological models providing the entrained air flow rate that are relevant for the jet roughness scenario. Notable divergences appear between these contributions, putting in evidence that there is no definitive agreement on how the entrained air flow rate is related to jet roughness.

| Jet roughness scenario (air – water systems) | Geometry | D0 | U0 | H/D0 | H compared with break-up length LB | Ratio experiment/phenomenological model for the entrained gas flow rate | jet roughness: data and/or correlation | maximum measured jet roughness |
|---|---|---|---|---|---|---|---|---|
| Burgess, Molloy and McCarthy 1972 | vertical cylindrical jet | 2.54 mm | 10 to 18 m/s | 5 to 75 | LB not provided presumably H < LB | indirect validation of eq (8) | no jet roughness data | not available |
| Van de Sande and Smith, 1972, 1973, 1976 | cylindrical jets inclined at 30° (60° not considered) | 1.95 to 10 mm | 2.5 to 20 m/s | 10 to 51 | H / LB < 0.64 (except for 2 data at D0=1.95mm for which H / LB = 0.7 or 0.9) | Qair was measured for $1.95 \leq$ D0 $\leq 10$ mm. The ratio with eq (8) ranges from 0.9 to 2.5 for U0 $\geq$ 8m/s | D* measured for $2.85 \leq$ D0 $\leq 10$ mm. correlation : $D^*/D_0 =$ 0.085 $(Weair\,Re_{weph})^{1/6}$ valid for U0 $\geq$ 8m/s | Max(D*) = 2.4 D0 => Max(ε/D0) = 0.7 |
| Cummings I.W. 1975 | vertical cylindrical jet | 4.6 to 8.9 mm | 1.8 to 7.8 m/s | 11 to 64 | LB not provided presumably H < LB | ratio with eq (10) $\cong$ 0.91 with parameters at impact but poor correlation coefficient and significant dispersion -73% / +93% | no jet roughness data | $0.61 \leq D^*/D0 \leq 1.2$ => Max(ε/D0)= 0.1 $1 \leq D^*/Di \leq 1.7$ => Max(ε/Di)= 0.35 |
| Ervine, McKeogh and El Sawy 1980 ; McKeogh and Ervine 1981 | nearly vertical cylindrical jet (unknown angle) | 6 to 25 mm | 2 to 10 m/s | 20 to 391 | from 0.09 up to 1.76 | ratio with eq (10) $\cong$ 1 - parameters at impact but significant dispersion 30% +50% Qair underestimated for ε/R $\leq$ 0.79 and overestimated for ε/R $\geq$ 0.79 | Jet roughness raw data. Roughness linearly increases with the jet velocity. | Maximum roughness at U0=2m/s, D0=25mm: ε/D0 = 0.56 (ε/Di=0.81) for Tu=1% ; ε/D0 = 0.71 (ε/Di=1.18) for 1% < Tu < 5% ; ε/D0 ~ 0.84 (ε/Di=1.63) for Tu >5% |
| Ervine and Falvey 1987 | horizontal jet | 50 or 100 mm | 3 - 29.6 m/s | not relevant | not relevant | not relevant | Jet roughness δ2 / x =0.358 U/U0 for Tu between 1 and 9% | not available |
| Sene 1984 and 1988 | supported planar jet angle from 25 to 90° | 5 to 47 mm | 1.8 - 3.0 m/s | no free fall | not relevant | no test of models | Jet roughness measured from concentration profiles but no data provided | - |
| Evans, Jameson and Rielly 1996 | vertical cylindrical jet | 2.38 to 7.12 mm | 7.8 to 15 m/s | 3 to 30 | not provided | Qair measured in confined systems, and comparison with model achieved by including a boundary layer contribution | jet roughness data measured in unconfined conditions and correlation 2.8Evans/D0 = D*/D0 – 1 = 0.0085 Oh^0.83 Re^0.63 | roughness $\leq$ 0.6 R |
| El Hammoumi et al., 1994, 2002 | vertical cylindrical jet | 14mm | 2.04 m/s | 32.2 | In this system, the roughness grows with the distance from the nozzle until it saturates. | Support the connection between dynamic roughness and entrained gas flow rate but no test of models | data for v'/V from 3 10^-5 to 3.2 10^-5 | - |
| Oguz, 1998 | vertical cylindrical jet confined by a gas annulus (micro bubble generator) | 1.6 to 2.4mm | 4.7 - 8.3 m/s | - | | Validates Sene proposal eq (12) for 6.1 and 8.3 m/s jet velocities. Significant differences at 4.7 m/s. | jet roughness from 0.002 to 0.02 D0 | Max roughness=0.02 D0 |
| Bagatur and Sekerdag 2003 | rectangular jet at 45° | thickness 3.5 and 5 mm for widths 5.5 and 9.8mm respectively | 2 - 12 m/s | $\approx$ 40-60 | not provided, presumably | ratio with eq (8) $\cong$ 0.95 when using jet width | Jet roughness data | Max(width*) = 4.1 jet width |

Table 1: Contributions testing phenomenological models based on jet roughness.

In that context, Ma et al. (2010) re-analyzed several air-water experiments for jet diameters or sheet thickness exceeding 10 mm. They replot collected data as a dimensionless entrained gas flow rate $(Q_{air}/L) / (U_c^3/g)$ versus $U_i/U_c$, and they recovered the two behaviors predicted by Sene. They observed that most data follow the $Q_{air} \propto U_i^3$ dependency predicted by eq.(16) for the surface roughness scenario, while the $Q_{air} \propto U_i^{3/2}$ dependency predicted by eq.(4) holds when an air cavity forms. Ma et al. (2010) argued that "the transition from entrainment via surface roughness to entrainment via the formation of a cavity" corresponds to an impact velocity equal to 4.5 $U_c$, that is 4.5 m/s when $U_c$ is set to 1 m/s which is a reasonable value for the quite turbulent flow conditions they considered. Hence, the scaling laws governing the entrained air flow rate seem to have been identified, even though Ma et al. (2010) underlined that the prefactors of $Q_{air}(U_i)$ relationships vary significantly from one experiment to the other, as they are expected to depend on fluid properties, on the turbulence level, on the height of fall, etc.

However, and although Ma et al.'s results appear quite encouraging, the first step in Sene's proposal that relates the entrained gas flow rate with the jet roughness has never been independently tested. Notably, the interface deformation predicted by eq.(15) assumes that gravity is counterbalancing inertia. This argument is hardly relevant for vertical jets, and the vast majority of the data considered by Ma et al. (2010) do concern vertical jets. Also, in eq.(15), Sene proposed that the roughness evolves as $U_i^2$, but such a dependency is not universal. In particular, Van de Sande and Smith, 1973 observed that D* increases as $U^{1/2}$, McKeogh and Ervine, 1981 measured a deformation linearly increasing with the jet velocity. Moreover, Ervine and Falvey, 1987 reported a roughness only driven by the turbulent intensity, and Evans, Jameson and Rielly, 1996 observed a deformation D* growing as $U^{0.63}$. Furthermore, a closer examination of figure 2 from Ma et al., 2010 indicates that the transition criterion they proposed is only valid for short heights of fall, say for $H/D_0$ less than $\approx$10. For larger $H/D_0$ (say above 20-30), they no longer observe the $Q_{air} \propto U_i^{3/2}$ behavior that they associate with the air cavity scenario. This is fully consistent with the fact that the jet roughness increases with $H/D_0$, and in consequence, with all other flow conditions being fixed, an increase in the height of fall leads, at some point, to the surface roughness scenario. In other words, if one considers an impact velocity exceeding 4.5 m/s, both $Q_{air} \propto U_i^3$ and $Q_{air} \propto U_i^{3/2}$ behaviors are possible depending on the flow conditions. In this work, we suggest that the transition is controlled by the magnitude of the jet roughness compared with that of the gas film thickness. Indeed, as discussed in Section 4.3, when the former exceeds the latter, a stable gas film solution is no longer possible because the incoming jet corrugations periodically impact the inverted gas meniscus formed below the free surface (Fig.2) and perturb the air flow at the film entrance. If so, the system is expected to shift from the gas film scenario to the jet roughness scenario.

To clarify these issues, our goal is to test experimentally each of the two steps in Sene's rationale. Aside from a revisit of data available from the literature, our objective is to develop experiments in which one can independently measure all parameters involved in equations (12) to (15): the entrained air flow rate per unit length of the contact line, the diameter and the velocity of the jet at impact and the deformation of the jet boundaries. We remark that we do not investigate here the connection between the jet deformation and the flow characteristics such as nozzle design, height of fall, internal flow including turbulence, or possible secondary motion at the injector exit. Instead, our goal is to produce a wide range of jet topologies in order to identify the potential domain of validity of Sene's proposal and also to investigate the nature of the interface deformation relevant to air entrainment. To this end, both a small-scale facility and a large-scale facility were exploited to collect experimental data and test modeling proposals. Both facilities use air and tap water under ambient conditions and with jet diameters larger than the capillary length scale. Furthermore, both experiments involved the same experimental techniques, allowing to perform a direct comparison between both datasets.

## 2 Experimental facilities

### 2.1 Small-scale facility

The small-scale facility, built at LEGI, consisted of a circular jet produced from a 7.6mm diameter nozzle, plunging vertically in the middle of a large reservoir (lateral dimensions 0.966 m by 0.47 m, height 1.184 m) equipped with an overflow to keep the free surface at a fixed height. During the experiments, the water level remained between 0.76 m and 0.82 m above the bottom of the tank. The jet impacted the center of the reservoir, far enough from the side walls to avoid any influence of them. The jet was fed with tap water: the circuitry consisted of a filter (20 μm pore size), a 5-meter-long plastic tube followed by a 0.5-meter-long straight solid pipe with an internal diameter of 15 mm, equipped with internal roughness to promote turbulence. A sudden contraction connects this tube and the nozzle. The latter consisted of a 5 cm smooth straight pipe with a 7.6 mm internal diameter. The verticality of the injector was carefully checked: its inclination with the vertical axis (aligned with gravity) was less than 1°. The injector can be moved vertically, allowing the height of fall H to be varied from 4 mm ($\approx 0.5$ $D_0$) to 317 mm ($\approx$ 41 $D_0$). The nozzle can also be moved within a horizontal plane about 5 cm around its medium position by way of two translating stages with a 0.1 mm resolution. We exploited these features to map the bubble cloud formed under the free surface. Indeed, for a probe held at a fixed location, the jet was displaced in a horizontal plane around the probe position, therefore providing information on the radial structure of the bubble cloud. The process was repeated for a probe immersed at different depths below the free surface.

The liquid flow rate $Q_w$ was measured within a 5% uncertainty using rotameters. It ranged from 200 to 1600 liters per hour, leading to a mean liquid velocity at injection $U_0$, approximately ranging from 1 to 10 m/s. The jet Reynolds number at injection $Re_0 = U_0$ $D_0/\nu_{water}$ was in the interval $9\ 10^3$ to $7.5\ 10^4$, while the air-based Weber number, namely $We_{air} = \rho_{air}\ U_0^2\ D_0/\sigma$, ranged from 0.2 to 12. According to Richardson (1954), friction with air becomes significant when $We_{air}$ exceeds 10: the only condition for which that criterion is fulfilled in the small-scale facility corresponds to the largest injection velocity considered, namely $U_0$=9.8 m/s.

The experimental conditions considered for air entrainment measurements are provided in Fig.5, which represents the height of fall H scaled by $D_0$ versus the injection velocity $U_0$. The break-up length $L_B$, defined here as the shortest distance to injection for which the continuous connection through the liquid with the injector is lost, was measured from high-speed image analysis at various injection velocities (see Section 3). For all the conditions considered, the jets were never atomized at impact, and an $L_B$ shorter than the height of fall could never be observed. The jet topologies are illustrated in Fig.6. Coherent turbulent jets are observed in all cases except for one condition indicated by a circle in Fig.5. Indeed, when the injection velocity drops below about 1.2 m/s, a Rayleigh-Plateau break-up mode takes place, and the capillarity instability leads to varicose disturbances whose wavelengths are of the order of the jet diameter (see the first image to the left in Fig.6). Note also that isolated drops were sometimes extracted from the jet (as shown in the image collected for $U_0$=4.9 m/s in Fig.6) but such events were too rare to affect the air entrainment process. Thus, most conditions considered in the small-scale facility correspond to coherent turbulent jets without atomization.

The critical velocity of air entrainment $U_c$ has not been directly measured. Instead, it has been evaluated by extrapolating the measured entrained air flow rate to zero. We found $U_c \approx 1.14$ m/s for $H/D_0 = 40$, and $U_c \approx 1.2$ m/s for $H/D_0 = 4$. These values are close to each other and are consistent with data available in the literature: they correspond to $Ca_c \approx 0.016$. Note that the turbulence intensity Tu in the small-scale facility deduced from eq.(1) and from $Ca_c$ was about 3%.

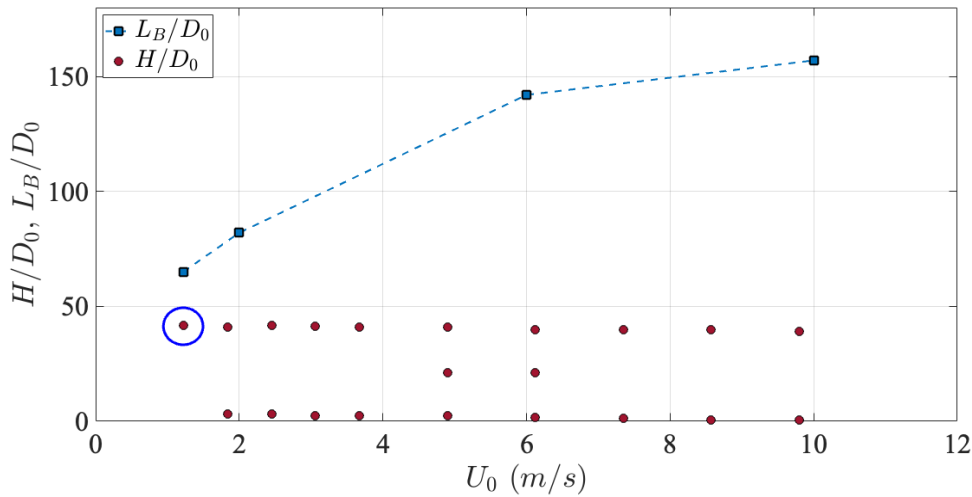

Fig.5: ● Experimental conditions investigated on the small-scale facility with a fixed nozzle diameter ($D_0$=7.6 mm). Dimensionless break-up length $L_B/D_0$ measurements are also reported (the measurement uncertainty is smaller than symbol size). The condition within the blue circle corresponds to a jet deformation due to a Rayleigh-Plateau instability (see illustrations in Fig.6).

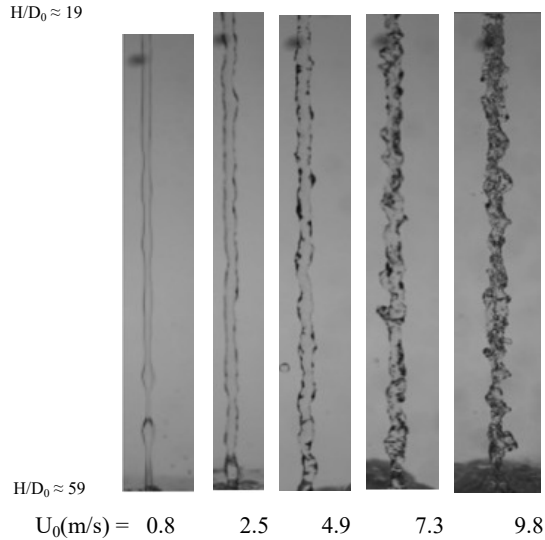

Fig.6: Jets observed in the small-scale facility for heights of fall comprised between 150 mm and 450 mm and for various injection velocities. The Rayleigh-Plateau mode is illustrated on the left-hand-side image (case $U_0 = 0.8$ m/s): it occurs for $U_0$ less than about 1.2 m/s. For the largest velocities investigated in the small-scale facility, one observes turbulent coherent jets without atomization.

## 2.2 Large-scale facility

The large-scale facility is sketched in Fig.7. This facility was built at CERG (CERG, Centre d'Etudes et de Recherche de Grenoble). Four different nozzles with internal exit diameters of 23, 82.9, 163.2, and 213 mm were used to generate vertical jets. The angle to the vertical was checked to be less than 0.6°: this angle was measured from the lateral deviation of the center of the jet over a 2 m height, as determined from the image analysis presented in Section 3. These nozzles were fed from a large chamber (406 mm in diameter, 2 meters long) equipped with a 50 cm long honeycomb to damp secondary flows, followed by smooth contractions. The length of the convergent was held fixed (equal to 1.38 m) in order to maintain the same fall height when changing the nozzle diameter. The injection velocity $U_0$ ranged up to 25 m/s for nozzle diameters below 100 mm. The velocity $U_0$ was up to about 10 m/s for $D_0$=163.2 mm and up to about 5 m/s for the largest $D_0$ = 213 mm nozzle. The turbulent intensity was measured 10 mm downstream of the nozzle on the jet axis with an FGP pressure sensor (model XPM5-S126). The longitudinal velocity fluctuation equals 0.14 m/s at 5 m/s, and it monotonically decreases with the jet velocity down to less than 0.01 m/s at 25 m/s. Accordingly, the turbulent intensity Tu drastically diminishes with the jet velocity: Tu is less than 3% for $U_0$=5 m/s, it becomes less than 1% for $U_0$ above 8 m/s, and less than 0.1% for $U_0$ above 12 m/s.

The fall height was varied between 2 m and 9.5 m. The receiving basin consisted of a pool 5 meters deep and 5 meters in diameter, equipped at its center of a 2 m diameter well whose maximum depth is 23 m. The water was recirculated from this basin to the injector by two pumps (their total power was 30 kW). At the steady state, the level of the free surface is essentially set by the initial water volume in the facility, and it varies due to the volume of air entrained beneath the free surface. However, and owing to the size of the receiving basin, the water level changed only by a few millimeters when varying flow conditions. To reduce fouling issues on optical probes (discussed in Section 3), the tap water was filtered (the mesh size of the filter was 30 micrometers). The flow rate was monitored with two Krohne Optiflux electromagnetic flow meters, one for the range 2-50 m³/h and one for the range 50-500 m³/h. The facility was previously exploited to investigate the depth of penetration of the bubble cloud (Guyot et al. 2020, 2022).

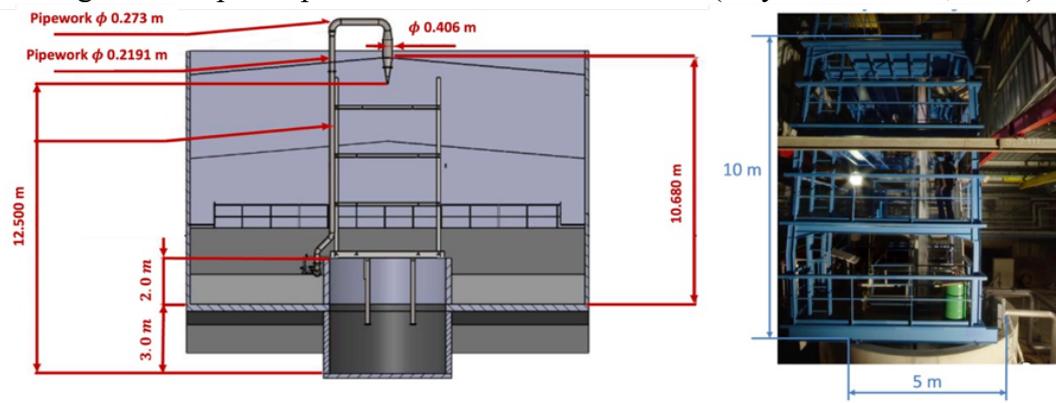

Fig.7: Sketch and image of the large-scale experiment.

The experimental conditions for which entrained gas flow rate measurements were achieved are given in a $D_0$ versus $U_0$ plane in Fig.8-a. For these conditions, the jet Reynolds number at injection $Re_0$ was in the interval $12\ 10^4$ to $1.6\ 10^6$, while the air Weber number ranged from 8.6 to 552. The fall height is indicated next to each ($D_0$, $U_0$) condition in Fig.8-a. For the condition ($D_0$ = 82.9 mm, $U_0$ = 2.5 m/s), three heights were considered, namely 3.25, 5 and 9 meters. The same information is presented in a $H/D_0$ versus $U_0$ plane in Fig.8-b. Raw data are provided in Table 4 and Annex (Table C-3).

As illustrated in Fig.9, all conditions correspond to coherent jets (possibly with complex interface deformations as discussed later) except for a single condition, namely $D_0$=23 mm, $U_0$=7.5 m/s, and H=9 m for which the jet was broken up: that condition is indicated by the circled data in Fig.8-a and it is illustrated in the bottom row of Fig.9. The break-up length $L_B$ measured for the $D_0$=23 mm nozzle is provided as a function of the injection velocity in the insert of Fig.8-b. $L_B$ ranges from 5.2 m to 6.5 m and happens to be almost independent of $U_0$.

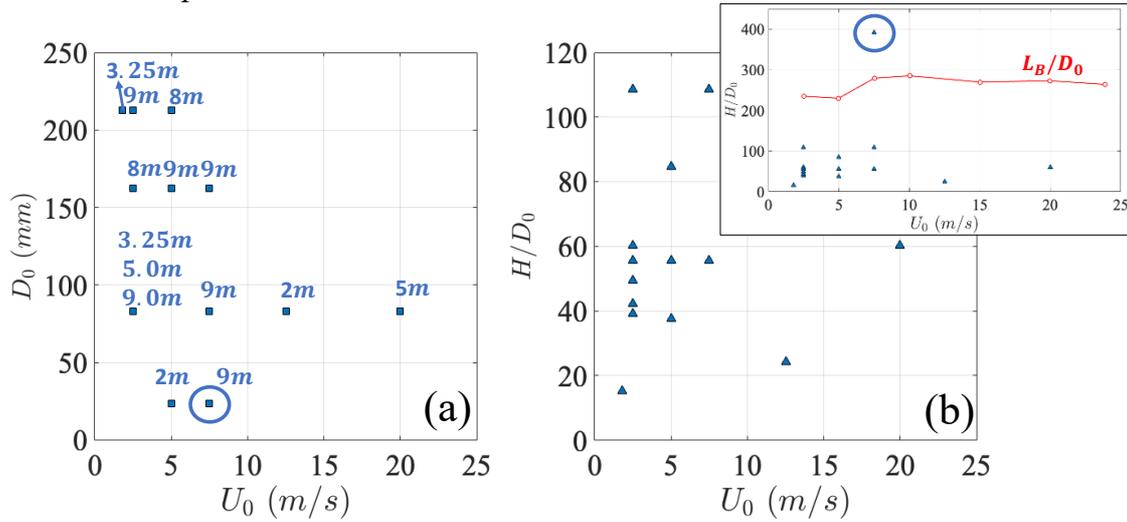

Fig.8: Experimental conditions for which entrained gas flow rates were measured in the large-scale facility. The conditions are presented in a $D_0$ versus $U_0$ plane with the heights of fall considered (a) and in a $H/D_0$ versus $U_0$ plane (b). The circled data correspond to a broken-up jet. The insert provides the break-up length (red dots and line) measured for the $D_0$=23.6 mm nozzle as a function of $U_0$: all experimental conditions are reported, including one condition beyond break-up (i.e., $D_0$=23.6 mm, $U_0$=7.5 m/s, H=9 m).

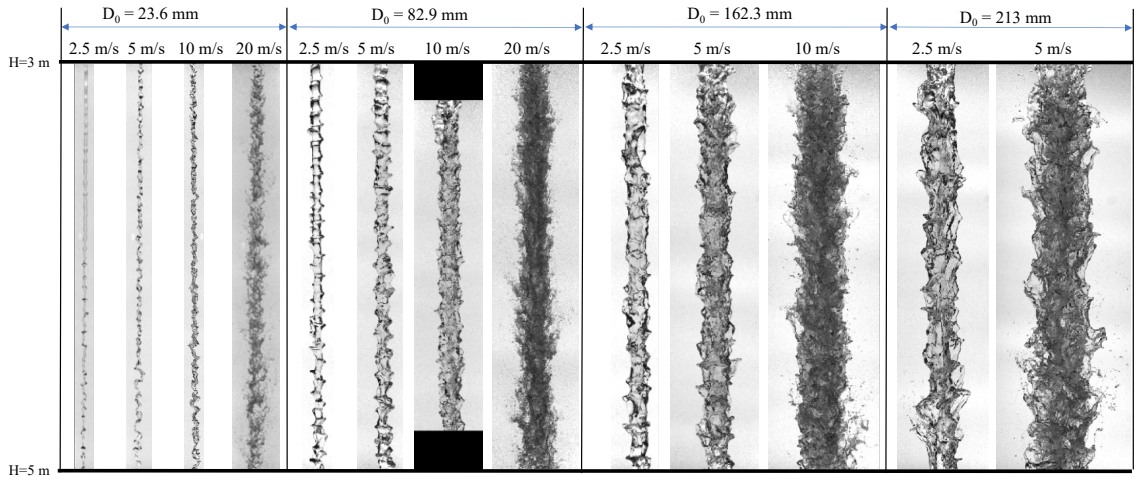

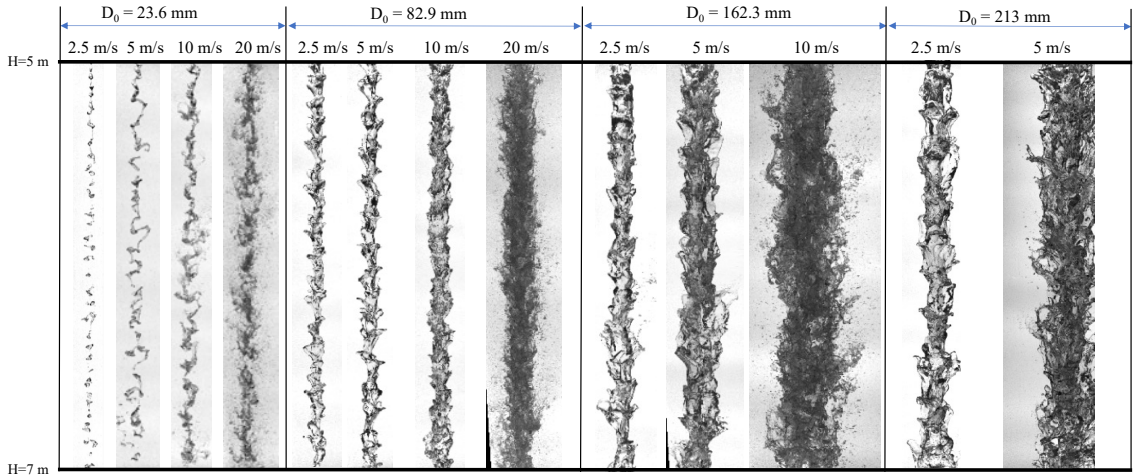

Fig.9: Illustration of jet topologies observed in the large-scale facility for various injection diameters and velocities and for heights of fall comprised between 3 m and 5 m (top row) and between 5 m and 7 m (bottom row).

The marked difference in the behaviors of the break-up length with the jet velocity between the experiments deserves some comments. Let us summarize the discussion provided in Annex B. At moderate air Weber number $We_{air} = \rho_{air} U_0^2 D_0/\sigma$, the characteristic break-up time originates from a capillary instability, and the break-up length evolves as $L_B/D_0 \propto We_L^{1/2}$ where $We_L = \rho_L U_0^2 D_0/\sigma$ is the liquid Weber number. When the air Weber number $We_{air}$ increases above some threshold, a shear instability becomes dominant compared with a capillary destabilization, and the resulting break-up length no longer depends on the jet velocity. Instead, it is set by the density ratio as $L_B/D_0 \propto [\rho_{liquid}/\rho_{air}]^{1/2}$ (Eggers & Villermaux, 2008). According to our experiments (see Annex B), the threshold air Weber number $We_{air}$ for this behavior is about 20, a value compatible with previous findings.

To further characterize the flow conditions, the fall height relative to the break-up length has been evaluated based on the above results. For all the flow conditions considered in the small-scale facility (Fig.5), the fall height is at most half the break-up length. In the large-scale facility, the fall height is at most 41% of the break-up length for all flow conditions except the one corresponding to jet break-up (Fig.8). Hence, in all cases but one, the jets are far from break-up conditions.

## 3   Measuring techniques and methods

In this section, we expose the techniques used for the jet characterization and for measuring the entrained air flow rate.

### 3.1 Characterization of the jet topology

As most of the jets considered here remain coherent, our objective was to characterize the jet diameter as well as the jet deformation. Globally, the fraction of the gas trapped within two crests that is entrained below the free surface should depend on the shape of the deformation. For a wave with a wavelength very large compared to its depth, only a small fraction of the trapped gas is expected to be entrained below the free surface. On the opposite, a short and deep wave should be able to entrain most of the gas trapped between two successive crests. In that perspective, it would be worthwhile to characterize both the depth and the wavelength of successive corrugations (e.g., Davoust et al., 2002 Ramirez de la Torre et al., 2020). As a first step, such measurements of joint variables were not undertaken here. Instead, we started with a simpler approach by quantifying the sinuous and the varicose components of the jet deformation. Such jet characteristics were determined from images gathered using backlighting in both facilities. As exemplified in Fig.10, the radial positions of the right and left jet contours were identified on images using either a canny filter or a threshold on the grey level. The strategy was adapted according to the image characteristics (a grey-level threshold happened to be better adapted in the presence of multiple droplets). In addition, the sensitivity to the threshold selection was analyzed to identify optimal criteria. Isolated droplets separated from the jet were then eliminated, and the contours of the main liquid lump were identified.

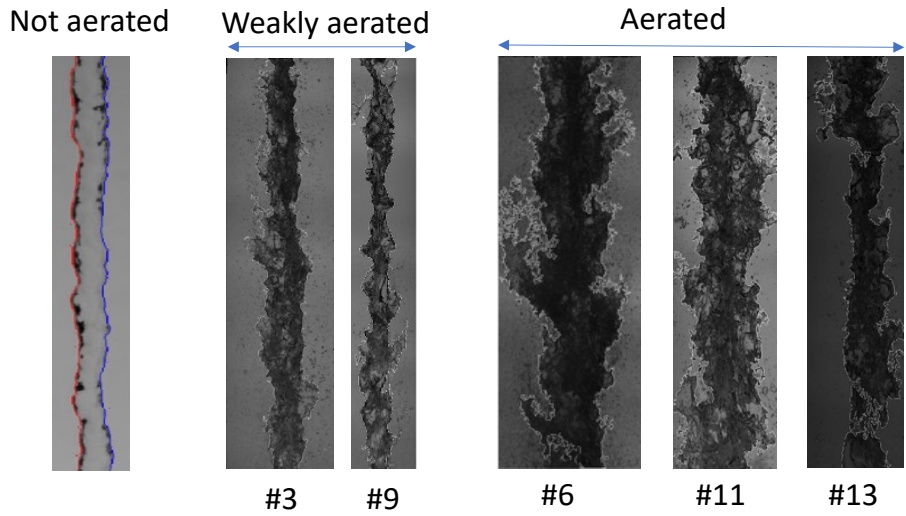

Fig.10: Examples of edge detection on different jets: non-aerated jet produced in the small-scale facility (image to the left), weakly aerated jets (runs #3 and #9), and aerated jets (runs # 6, #11, and #13) produced in the large-scale facility. Images have different scales. The run numbers refer to Table 4.

At a given distance H from the nozzle, their difference provides the local and instantaneous jet diameter D(H,t) while their half sum defines the local and instantaneous center of the jet C(H,t) relative to the mean jet axis (Fig.11). Considering a set of uncorrelated images, the mean jet diameter <D> was evaluated as the arithmetic mean of instantaneous jet diameters D(H,t). We also evaluated the standard deviation of the positions of the left-hand side, i.e., std(left edge), and of the right-hand side, i.e., std(right edge), jet boundaries. The sum of these standard deviations provides the twice total deformation, namely 2 $\varepsilon_{total}$ = std(left edge) + std(right edge), where $\varepsilon_{total}$ represents the deformation of one side of the jet. In addition, the sinuous deformation was evaluated as the standard deviation - noted std(.) hereafter - of the lateral position of the jet center, namely $\varepsilon_{sinuous}$ = std(C). Similarly, the varicose deformation was evaluated as the standard deviation of the jet diameter, that is $\varepsilon_{varicose}$ = std(D). From the definition of the total deformation, one should have 2 $\varepsilon_{total}$ = $\varepsilon_{varicose}$ + $\varepsilon_{sinuous}$. To test the reliability of the image processing routine, we performed independent measurements of $\varepsilon_{varicose}$, $\varepsilon_{sinuous}$, and $\varepsilon_{total}$. In the small-scale facility, the equality 2 $\varepsilon_{total}$ = $\varepsilon_{varicose}$ + $\varepsilon_{sinuous}$ happened to be fulfilled within 5% for all flow conditions except at very low heights H (namely for H less than 2D₀). Indeed, in the latter cases, the deviation can reach 30% because the interface deformations were very small (less than two hundred of micrometers), and they were, in fact, smaller than the spatial resolution of the imaging device. Similarly, in the large-scale facility, the above-mentioned equality was found equally valid for all flow conditions.

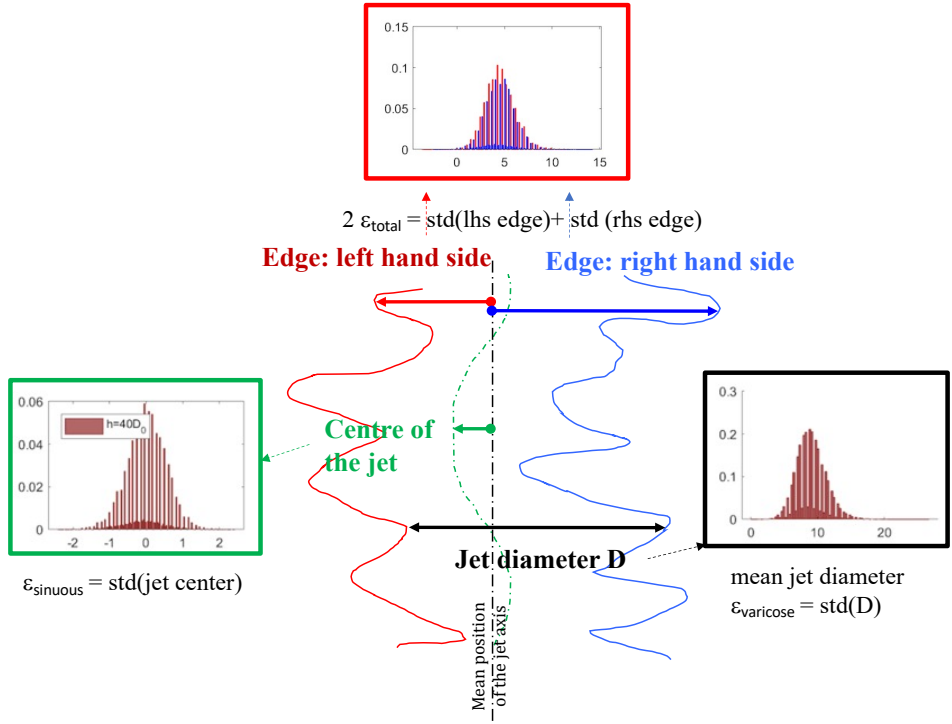

Fig. 11: Measurements of the jet diameter and of the sinuous and varicose deformations. The measured distributions of the jet diameter, of the position of the jet center and of jet edges have been obtained on the small-scale facility for H= 40 $D_0$, $Q_L$= 1600 l/h. As explained in the text, the total deformation deduced from edges is equal to the sum of sinuous and varicose deformations.

On the small-scale facility, the jet evolution during the free fall was investigated using a high-speed Phantom Miro camera (1280 x 256 pixels, exposure time 70 μs) with its optical axis perpendicular to the jet. Backlighting was ensured with LED (6000 lumens) illuminating a 120 cm by 60 cm area. Two sections of the jet were successively imaged: the first section was comprised between the injector exit at H=0 m and 60 cm downstream, and the second section was comprised between H=50 cm and H=110 cm. The spatial resolution was 0.47 mm per pixel (that is almost 2 pixels/mm). For each jet velocity, a first video was made of the first section of the jet at a low acquisition rate (typically 70 frames per second) in order to gather about 10000 uncorrelated images from which we deduced the jet diameter and the deformation of the edges. A second movie was acquired at a larger acquisition rate (typically 1000 frames per second) to quantify the jet velocity. The latter was obtained by correlating grey patterns of successive narrow horizontal bands on raw images, with a resolution of about ±5%. As the grey pattern is due to light reflections and/or transmissions by the deformed jet, the velocity that is measured is expected to be representative of a kind of mean jet velocity averaged over one diameter, at least for sufficiently smooth jet deformations. The nozzle was then moved up, and the same process was repeated for the lower portion of the jet. For each jet velocity, these four successive movies were collected without interrupting the water flow to ensure that conditions remained as stable as possible.

For the large-scale facility, videos of the jet were collected with a Phantom V2640 camera (resolution 2048 ×1952 pixels). The field of view was set to 2 m high (i.e., in the vertical direction) and 1.9 m wide. Backlighting was ensured by seven neon lights with a diffuser that homogenized the light flux across the whole field of view. Given the large-scale nature of the facility, ensuring an acceptable uniformity of lighting conditions over the whole field of view was challenging. In practice, the contrast was not always satisfactory along the image boundaries, and narrow bands along the image boundaries need to be discarded from the analysis. To characterize the jet along its fall, four sections were considered that correspond to distances from the nozzle from 1 to 3 m, from 3 to 5 m, from 5 to 7 m, and from 7.5 to 9.5 m. For the first three intervals, the line of sight of the camera was perpendicular to the jet axis. For the last interval, and because of technical constraints, the camera was positioned at a finite viewing angle with respect to the jet axis, as illustrated in Fig.7-a. For each position, a calibration grid was used to determine the magnification and to correct for optical distortion when needed. The typical resolution was of 1 mm/pixel. For each flow condition, namely $U_0$ and $D_0$, and for each section, a 10000 frames movie was collected at 100 frames per second. Such an acquisition rate ensures statistically independent images over the range of jet velocities considered here.

The jet velocity was also measured in the large-scale facility. In this case, the camera was run at between 1500 to 2000 frames per second. The field of view was limited to a 10 cm wide band centered on the jet axis. Taking advantage of jet distortions, including surface disturbances, ligaments…, the mean velocity was determined by correlating these 10 cm wide bands shifted by 8 to 16 pixels along the vertical coordinate. The resulting resolution on velocity is about ±10%. However, as we will see in Section 4, the jets often experienced more complex deformations in the large-scale facility than in the small-scale facility. Hence, the sources of grey-level non-uniformities in the images include surface disturbances but also gas inclusions or interfaces trapped inside the jet and/or droplets detached by stripping. Owing to this complexity, the interpretation of the velocity measured by image correlation as a mean jet velocity is possibly less convincing.

For the small-scale facility, and as shown in Fig.12-left, the jet characteristics, including its diameter, the position of its center relative to the symmetry axis, and the distributions of the position of jet edges are nearly Gaussian for all the flow conditions indicated in Fig.5 except for injection velocities below about 1.2 m/s. In the latter case, the jet experiences Rayleigh instabilities, and the distribution of diameters becomes non-axisymmetric. As Rayleigh conditions are discarded here, the standard deviations of the various distributions considered unambiguously quantify the jet deformation.

For the large-scale facility (Fig.12-right), the distributions of the diameter, of the jet center position, and of the position of each jet edge are also nearly Gaussian. Exceptions occur in presence of a strong sinuous deformation of the jet, deformations that are reminiscent of the buckling observed on fast thin jets (Stockman and Bejan, 1982; Rezayat et al., 2021). In these cases, and because the jet diameter is counted along a direction normal to gravity, the jet diameter distribution exhibits a tail toward large sizes that increases the mean diameter measured from images. Among the explored flow conditions (see Fig.8), a buckling-like behavior was only observed for the $D_0$=82.9 mm nozzle at low injection velocity (run #4 at $U_0$ = 2.5 m/s and run #5 at $U_0$ = 5 m/s): that behavior was possibly induced to the slight ovality of that injector.

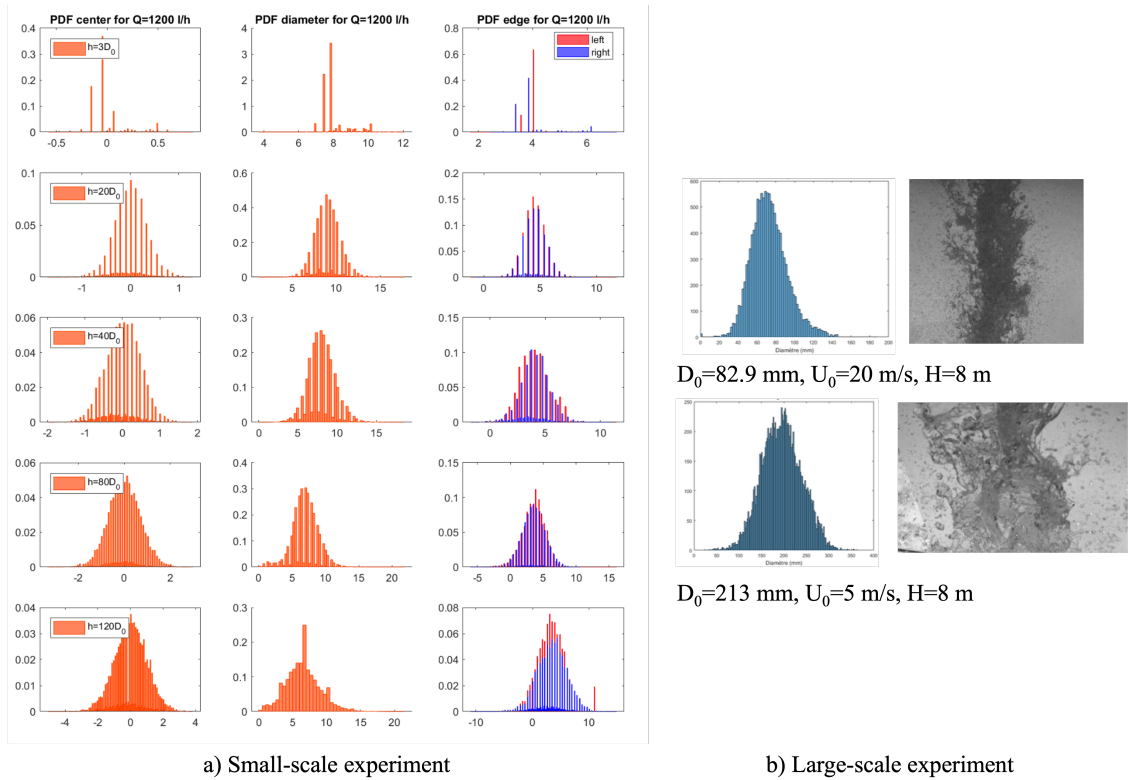

a) Small-scale experiment        b) Large-scale experiment

Fig.12: a): Typical measured distributions of the center position of the jet with respect to the symmetry axis (left column), of the jet diameter D (middle column), and of the positions of right and left edges (right column) recorded on the small-scale facility at $U_0$=7.35 m/s for heights of fall from 3 to 120 $D_0$. b): diameter distributions measured in the large-scale facility with snapshots of the corresponding jet.

Except for these few specific cases, the jet deformations are well captured, and this is so even in the presence of droplet stripping (as for the two bottom figures in Fig.12-right). The jet characteristics determined from the above procedure are discussed in Section 4.

### 3.2 Characterization of the bubble cloud

The bubble cloud formed under the free surface was characterized using optical probes that provide statistics on bubble arrival time, interface velocities and gas dwell times, from which one can deduce relevant quantities such as void fraction, bubble size distribution, bubble velocity distribution, interfacial area density and local gas flux (Cartellier, 1999). Such sensors are well adapted to plunging jets owing to the magnitude of the velocities involved. Indeed, the measured mean bubble velocities exceed about 0.9 m/s in all the flow conditions considered here, so that the uncertainty of de-wetting probe measurements remains less than 10% in air-water systems (Vejrazka et al., 2010). In addition, the quasi-unidirectional flow that takes place just below the free surface is favorable for the quantification of the global quantities by integrating transverse profiles. In practice, we used multimode conical probes that exploit the de-wetting times to evaluate velocities (Cartellier and Barrau, 1998). In the small-scale facility, the probe response was quite stable. In the large-scale facility, the conical probes happened to experience significant fouling from time to time, partly because of the larger velocities involved and also because of the difficulty in controlling the water cleanliness in an industrial environment. Fouling is known to affect the de-wetting dynamics and thus velocity measurements. Therefore, to ascertain reliable bubble velocity measurements in the large-scale facility, the calibration of conical probes was routinely checked and corrected when needed using a newly developed Doppler probe that exploits an "absolute" principle to provide the bubble velocity. The principle of operation of Doppler probes, including design, signal processing, and performances are detailed in Lefebvre et al. (2022). Examples of velocity and chord distributions (obtained by direct detection only) collected in the large-scale facility with a conical probe and with a Doppler probe are provided in Fig.13.

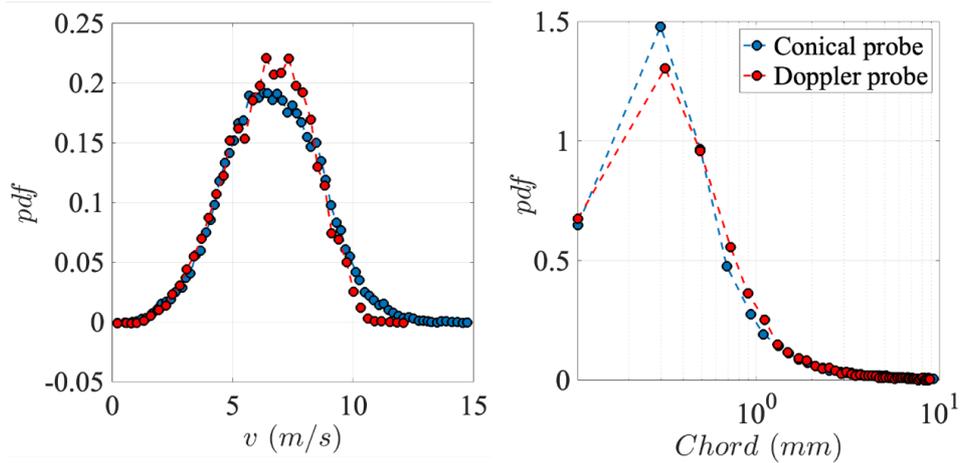

Fig.13: Examples of bubble velocity and chord distribution from direct detection collected at the same location with a conical probe and with a Doppler probe ($D_0$ = 82.9mm, $U_0$ = 2.5 m/s, H = 5 m, Probe position: on the jet axis and 330 mm below the free-surface. The mean velocity is 6.7m/s (respectively 6.5m/s) and the mean chord is 1.1 mm (respectively 0.89 mm) for the conical probe (respectively for the Doppler probe).

In the small-scale facility, the upward-facing probe was immersed at a distance comprised between 3 and 6 $D_0$ below the free surface, the most common depth being about 5 $D_0$. The probe was held fixed, and the profiles were obtained by moving the nozzle. The center of the jet was tracked by first moving the nozzle in a horizontal plane in order to detect extrema of phasic velocity, void fraction and/or gas flux. The transverse profiles were then collected using lateral displacements of 1 or 2 mm, perpendicular to the probe's direction until void fraction, velocity and gas flux approached zero. Typically, 15 to 20 data points were collected for each profile. To ensure converged statistics, the measuring duration was set to 120 seconds or 50000 bubbles detected, whatever event occurred first. Far from the jet axis and for the lowest liquid flow rates, the number of detected bubbles dropped sometimes, throughout the measurements, down to a few hundred. The success rate in terms of velocity detection (i.e., the fraction of bubble signatures providing direct velocity measurements) was quite high: it was typically up to 90% and even 98% in the center of the jet, while it dropped down to 70-80% on the far edges. Typical profiles gathered are provided in Fig.14 for H/$D_0$ ≤ 4 and in Fig.15 for H/$D_0$ ≈ 40.

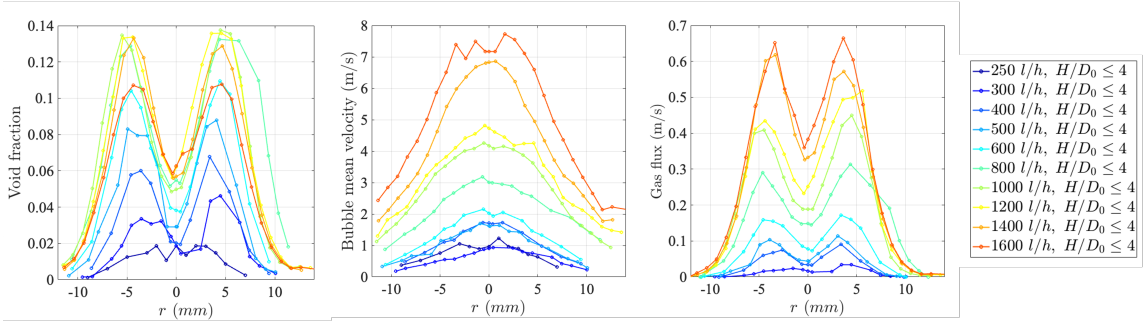

Fig. 14: Full transverse profiles of the void fraction, of the mean bubble velocity (accounting for direct measurements only) and of the gas flux for $H/D_0 \leq 4$ and different liquid flow rates. Small-scale experiment.

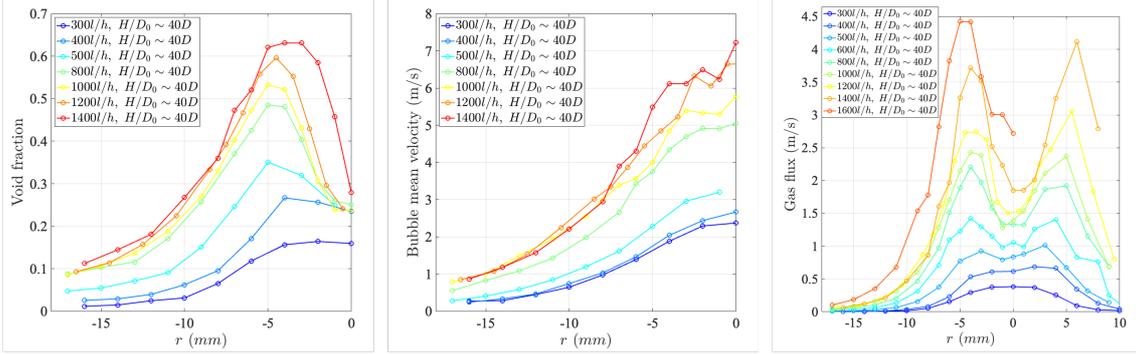

Fig. 15: Half transverse profiles of the void fraction and of the mean bubble velocity (accounting for direct measurements only) and full transverse profiles for the gas flux for $H/D_0 \approx 40$ and different liquid flow rates. Small-scale experiment.

At the chosen depths, and for all H, the mean bubble velocity is maximum at the center and smoothly decays with the distance to the axis as expected from the presence of the incoming liquid jet and from the lateral dispersion of bubbles. The void fraction exhibits two maxima that correspond to the bubble production zone, located at the periphery of the impacting jet. The lower void fraction value on the axis arises from the merging of the two mixing layers that develop from the jet boundaries at impact. The gas flux profiles also exhibit the trace of the bubble production zones. An estimate of the magnitude of the void fraction will be presented in Section 5.2.

In the large-scale facility, in order to limit the measurement duration, we used an array consisting of two de-wetting probes and one Doppler probe. All probes were directed upwards and were placed one cm apart. The array was mounted on a traverse allowing translations along three orthogonal axes. The measurements were achieved between 4 to 8 $D_0$ below the free surface. Once the flow conditions were stabilized (notably, some air trapped in the circuitry needed to be eliminated), the center of the jet was identified using quick scans performed along the two coordinates of the horizontal plane (perpendicular to the jet axis). Then, a complete profile was achieved using 5 mm (close to the jet center) to 20 mm (far from the jet center) displacement steps depending on flow conditions. For de-wetting probes, the measurement duration was set to 100000 bubbles detected or 180 seconds, whatever event occurred first. For the Doppler probe, 4 minutes (respectively 1 minute) long records were used to achieve velocity (respectively void fraction) measurements. For each run, the mean bubble velocity measured with the Doppler probe was used to check for the presence of some fouling and if so, to re-calibrate the de-wetting probe response. For the latter, the percentage of direct velocity detection happened to be quite high, with a success rate always above 90%. Each profile consists of 10 to 25 (depending on the complexity of its shape) measuring locations. To test the reliability of the whole procedure, the local measurements have been repeated for several flow conditions, and the results happened to be fairly reproducible. Typical results are provided in Fig.16.

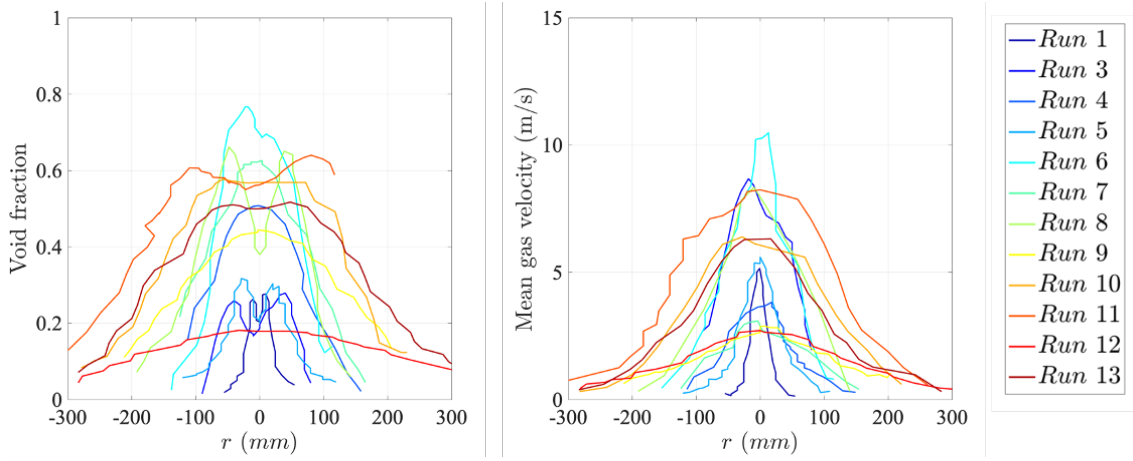

Fig.16 Void fraction and mean gas velocity profiles below the free surface for various injection velocities and for different $H/D_0$. Large-scale experiment.

### 3.3 Evaluation of the entrained gas flow rate

The gas flow rate was evaluated by integrating the local gas flux $\varphi(r)$:

$$Q_{air} = \int_0^{R_{max}} 2\pi r\ \varphi(r)\ dr$$
$$= \int_0^{R_{max}} 2\pi r\ \alpha(r)\ V_G(r)\,dr \qquad (19)$$

where r is the distance to the jet axis, and $V_G(r)$ is the local mean gas phase velocity defined as $\varphi(r)/\alpha(r)$. The local gas flux $\varphi(r)$ is itself obtained by the sum of gas chords detected per unit time by optical probes at position r. In practice, the center of each profile was accurately determined from the raw data. Then, two integrations were performed along a horizontal axis (axis y) assuming axisymmetric profiles, one from the axis origin (y=0) and for positive y and another one from the axis and for negative y. The comparison between these two quantities provides a test of the axial symmetry. The entrained air flow rate is evaluated as the average of these two integrals. The sensitivity of the global gas flux estimate to the outer limit of integration $R_{max}$ was thoroughly tested. For that, we examined the growth of the integral with the limit of integration $R_{max}$, and the integration process was stopped when the integral value reached a clear asymptote with residual contributions amounting at most for less than a few percents. Let us underline that, some distance away from the two-phase jet, bubbles are moving upward but, owing to the probe orientation, these bubbles do not provide valid velocity detections, and thus they are not contributing to the downward directed gas flux. The integration procedure was further tested by measuring the flux from two profiles collected along orthogonal directions in a horizontal plane. That test was achieved for one flow condition in the small-scale experiment, and the deviation between the two integrals was found to be within ±11%. Also, on the same facility, a 12% difference on the flux was observed when measurements were independently performed by two operators. The axial symmetry assumption happened to be valid within 10% for most flow conditions. In the small-scale facility, larger deviations were recorded mainly for jet velocities below about 2 m/s. In the large-scale facility, deviations were less than 25% except for two conditions: the asymmetry was indeed 28% for $D_0$=162 mm, $U_0$=7.5 m/s, H=9 m, and it reached 41% for $D_0$=83 mm, $U_0$=7.5 m/s, H=9 m. The origin of such large asymmetries is unclear (unfortunately, these measurements were not repeated owing to the complexity and cost of running the large-scale experiment). They could possibly be related to the influence of secondary flows that could appear over large time scales

We will analyze how the entrained gas flow rate evolves with flow parameters in Section 4. Before that, let us examine the conditions for gas detection by optical probes. Concerning the bubble size, Fig.17 provides the Sauter mean diameter of bubbles $D_{32}$ computed as 3/2 times the mean gas chord detected by conical probes assuming spherical inclusions (Liu and Clark, 1995), on the flow axis and at small depths below the free surface. Due to the high success rate, the Sauter mean diameter associated with direct velocity detection is always very close to the Sauter mean diameter including interpolated velocities (the interpolation issue, where missing velocity measurements on isolated events are estimated using information from the first neighbors of the event, is notably discussed in Cartellier, 1998 and in Lefebvre et al., 2022). Therefore, only direct measurements are presented in Fig.17. In the small-scale facility, the Sauter mean diameter is about 1 mm for small height of fall and about 3 mm at large H. In all cases, the Sauter mean diameter slightly decreases with the impact velocity, if one sets aside the two smallest velocities for H/$D_0 \approx 40$ for which the jet deformation corresponds to a Rayleigh mode. In the large-scale facility, and for the conditions H=8 m, $D_0$ = 82.9 mm and 162.3 mm at $U_0$ = 2.5m/s and $D_0$ = 213 mm at $U_0$ = 5 m/s, the Sauter mean diameter on the axis evolves between 3 and 3.7 mm. The latency length of the conical probes used here was in the range 30 to 50 µm: it was thus always much smaller than the mean size of bubbles present in both facilities.

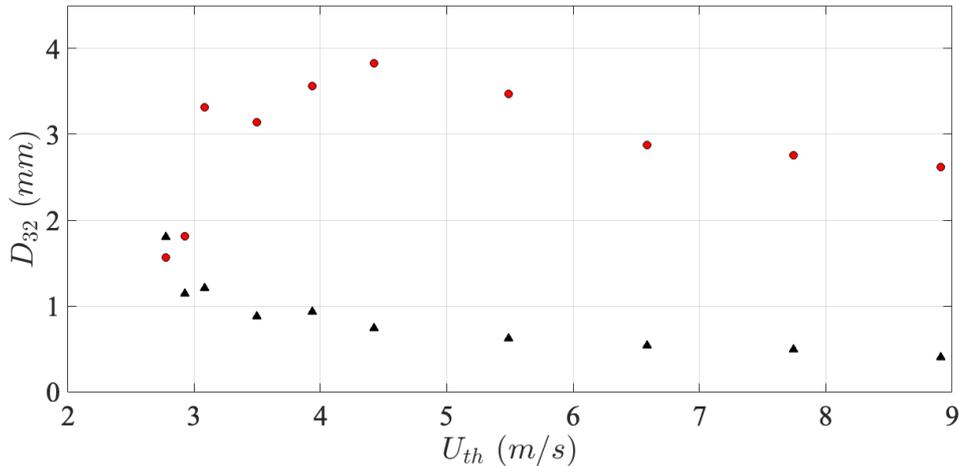

Fig. 17: Measured Sauter mean diameter of bubbles on the jet axis and at small depth below the free surface (from 3 to 6 $D_0$) versus the impact velocity for small (H/D ≤ 4) and large (H/D ≈ 40) fall heights. Small-scale facility.

Concerning velocities, Fig.18 provides the arithmetic mean bubble velocity (only direct values are presented as mean values with and without interpolation are very close), and the mean gas phase velocity measured on the axis versus the impact velocity. For the small measurement depths considered here (the latter is between 3 and 6 $D_0$), the mean bubble velocity as well as the gas phase velocity remain globally proportional to the impact velocity in both experiments. In the small-scale facility, the magnitude of the mean bubble velocity (respectively the gas phase velocity) is well approximated by 0.8 (respectively 0.7) times the impact velocity. The fact that the velocity measured below the free surface is less than the impact velocity, is expected because of the growth of the jet cross-section beneath the free surface due to lateral entrainment. In the large-scale facility, the data are somewhat dispersed, possibly because of the stronger variability of jet topologies. In addition, the measured velocities evolve between 0.8 $U_{th}$ down to 0.2 $U_{th}$: they are globally lower than in the small-scale experiment mainly because of jet aeration (the latter is discussed in the next Section).

Note that the absolute bubble velocity with respect to fixed probes was always above 0.8 m/s in the experimental conditions considered. This value is large enough to ensure a reliable probe response in air-water systems (Vejrazka et al., 2010). That conclusion applies to classical conical probes as well as to Doppler probes whose latency length ($\approx$ 6 µm) is smaller than that of conical probes.

Let us observe that, in both facilities and for all the flow conditions considered, the time of flight of the bubbles from the free surface to the measuring location, which is of order of the measurement depth divided by the jet velocity, is much smaller than the bubble response time evaluated as $D_{32}^2 / \nu_{liquid}$. That ratio typically ranges from 20 to 100. Hence, the measured velocity of the bubbles is not due to the entrainment by the liquid phase. Instead, it is due to the initial velocity imposed during bubble formation.

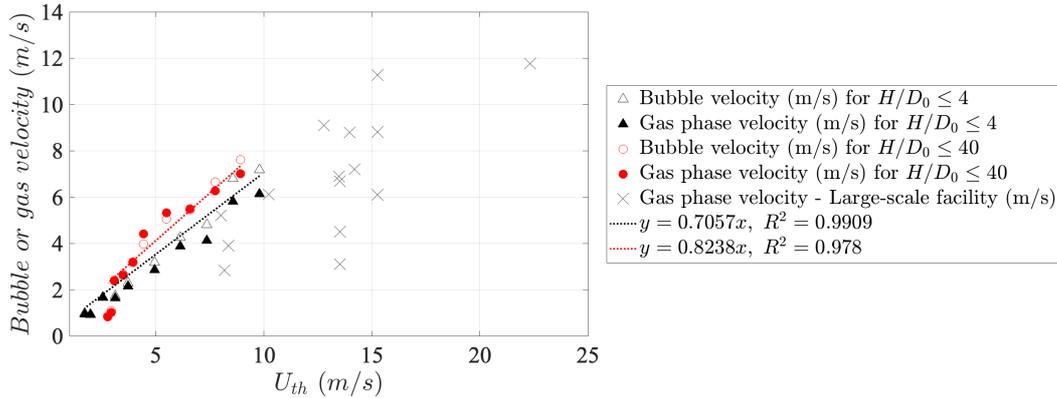

Fig. 18: Measured velocities (arithmetic mean bubble velocity and mean gas phase velocity) versus the jet velocity at impact for two heights of fall in the small-scale experiment and for various flow conditions in the large-scale experiment. The regressions are based on data collected in the small-scale experiment.

## 4. Analysis and Discussion

### 4.1 Impact conditions

Let us first investigate the impact conditions, as these are required in most models presented Section 1. Measurements in the small-scale experiment for $U_0$ from 1 to 10 m/s are compared with the theoretical free fall velocity $U_{th}$ and with the theoretical mean jet diameter $D_{th}$ in Fig.19. For $H/D_0 =$ 20, 40 and 60, the jet velocity at impact happens to equal the free fall velocity within ±10%. If one sets aside the smallest injection velocity that corresponds to a jet in the Rayleigh regime, nearly all data agree within ±5%. The mean value of the jet diameter was measured by considering a moving average over a 1 cm high window. For $H/D_0$ from 4 to 40, the measured diameter corresponds to the free fall diameter within 0/+15%, except one data at 22% at the largest liquid flow rate. Hence, the theoretical estimates of velocity and diameter at impact based on free-fall happen to be valid for all the flow conditions considered in Fig.5.

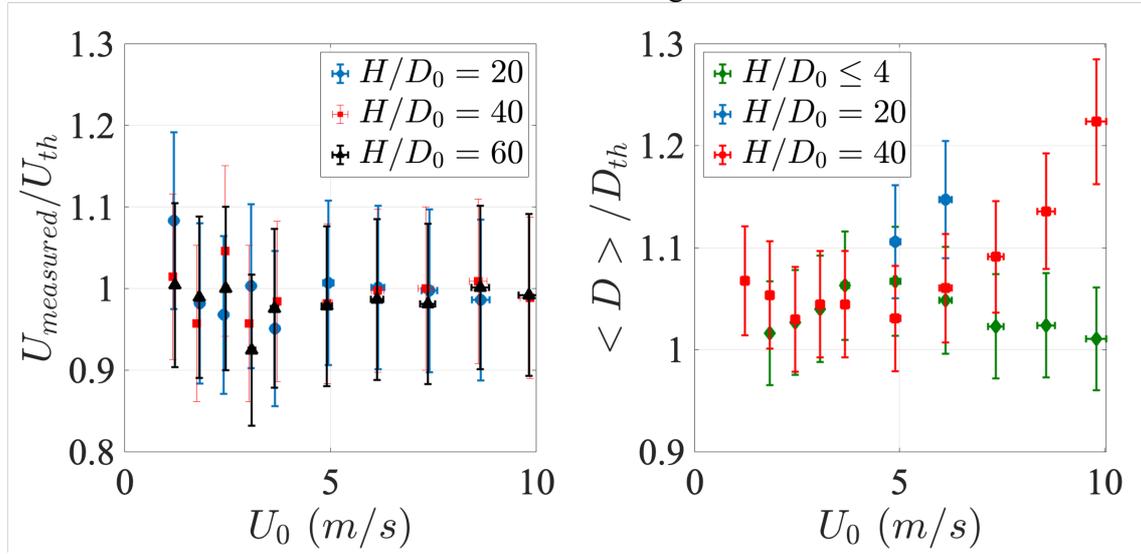

Fig.19: Left: Ratio of the jet velocity measured after a fall height H to the free-fall velocity $U_{th}$. Right: Ratio of the measured mean jet diameter <D> after a fall height H to the free-fall impact diameter $D_{th}$. Both ratios are close to one. Data collected in the small-scale experiment for the flow conditions of Fig.5.

In the large-scale facility, the velocity has been measured using image correlation at H=8 m and for flow conditions $U_0$, $D_0$ similar to those of Fig.8. The results are provided in Fig.20: they show that the velocity measured at impact corresponds to the free fall velocity within ±10%.

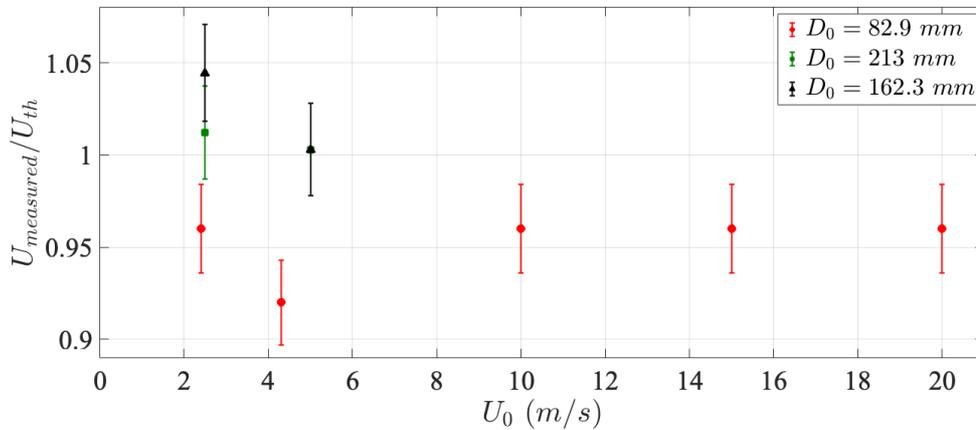

Fig.20: Ratio of the measured jet velocity at H=8 m to the free-fall velocity $U_{th}$. Data collected in the large-scale experiment.

The evolution of the jet diameter with the fall height has been measured for the flow conditions shown in Fig.8. Fig.21 presents typical raw data collected. Some defects arise at the boundaries between vertical sections, due to lighting non-uniformities. To avoid these shortcomings, narrow bands should be discarded from the analysis on the top and/or the bottom sides of each section.

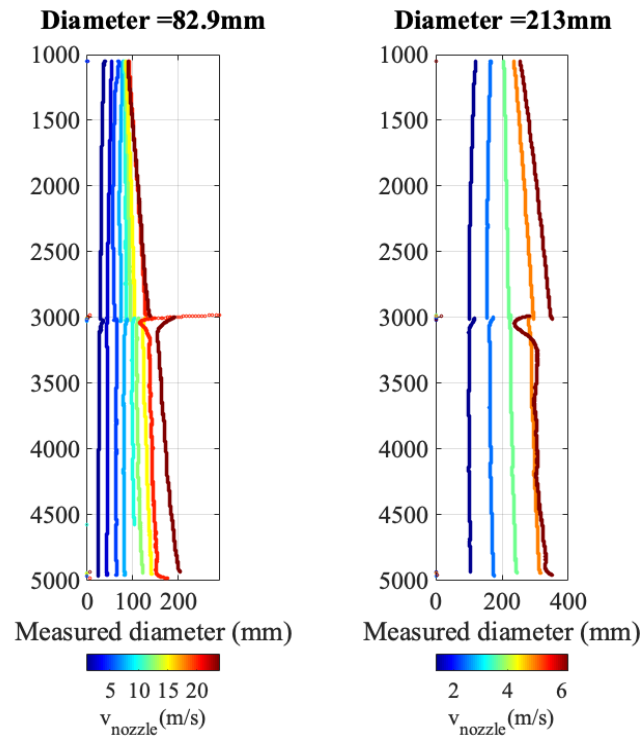

Fig.21: Examples of the evolution of the mean jet diameter <D> with the fall distance from the nozzle (the ordinate H is indicated in mm). The color coding refers to the jet velocity at nozzle. Note that the data corresponding to the image top and bottom boundaries have been kept for sake of completeness. These regions are subject to lighting inhomogeneities and should be discarded from the analysis. Large-scale experiment.

At small to moderate jet velocities (that velocity limit evolves with the jet diameter), the measured mean jet diameter <D> equals $D_{th}$ within 20% or less for all heights up to $H = 7$ m, meaning that these jets behave as "regular" coherent, turbulent jets, as observed in the small-scale facility. At larger velocities, the evolution of the jet diameter drastically changes as the jet diameter *monotonously increases* with the fall height. That increase is quite significant: for example, for $D_0 = 82.9$ mm, $U_0 = 20$ m/s, the measured diameter at $H = 5$ m is twice the expected free fall diameter. This is the mark of strongly aerated jets for which the air deeply penetrates inside the liquid, a situation illustrated by some images in Fig.9. Clearly, the large-scale experiment is able to generate quite different jet structures from those obtained in the small-scale experiment.

To identify the flow conditions for which the jet becomes aerated, we examined the growth of the jet diameter between 1 m and 5 m from the injector as the key criterion. Owing to the resolution of imaging techniques used here, *weak aeration* is said to occur whenever <D>/$D_{th}$ exceeds 1.2 to 1.3. Also, jets are said to be *aerated* when <D>/$D_{th}$ becomes larger than 1.5. The aerated conditions thus defined and encountered in the large-scale facility are provided in Fig.22 in a $D_0$, $U_0$ map of analyzed flow conditions. Globally, as shown by the tentative boundary drawn between aerated and non-aerated jets at $H = 5$ m, for a given nozzle diameter, jet aeration occurs above some critical injection velocity. That critical injection velocity decreases with the nozzle diameter nearly as $\approx 1/D_0$. That steep decrease means that large jets are quite prone to aeration. This observation is also accentuated by the fact that the critical injection velocity is as low as $\approx 2$ m/s for the largest jet considered here (namely $D_0 = 213$ mm). Therefore, a strong jet aeration is the situation expected for most hydraulic applications. Note that the tentative boundary drawn in Fig.22 roughly corresponds to a constant jet Reynolds number about $0.5 \cdot 10^5$, but the validity of such a tentative criterion deserves to be tested.

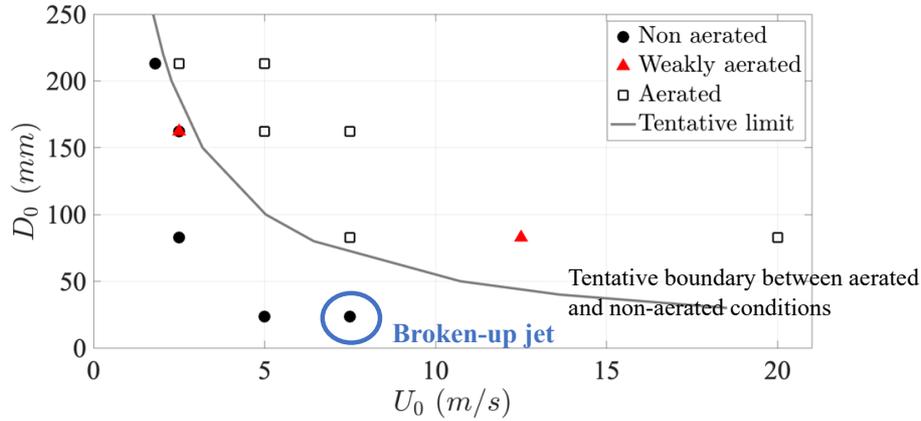

Fig. 22: Experimental conditions in the large-scale experiment represented in a ($D_0$, $U_0$) plan and tagged with the jet aeration criterion evaluated at H = 5 m.

### 4.2 Jet deformation

In the small-scale facility, the nozzle diameter was held fixed while both the jet velocity at injection and the fall height were varied. As shown in Fig.23, the total deformation $\varepsilon_{total}$ on one jet side compared with the radius $R_{th}$ evolves in the range 4.6% to 42%. At very low fall heights (H/$D_0 \leq 4$), the total deformation remains weak, from 4.6 to 8.6%, and it does not significantly evolve with the jet velocity. That situation is illustrated by the central image in Fig.2. At larger fall height, the sinuous, the varicose and the total deformation all monotonously increase with the jet velocity (Fig.23): the increase with $U_0$ as well as with the impact velocity $U_{th}$ is close to be linear. At these large H/$D_0$, the total deformation on one jet side evolves from 18% to 42% of $R_{th}$. Let us underline that, for all the flow conditions considered, the sinuous deformation is nearly proportional to the varicose deformation: in average, one has $\varepsilon_{sinuous} \approx 0.57 \, \varepsilon_{varicose}$.

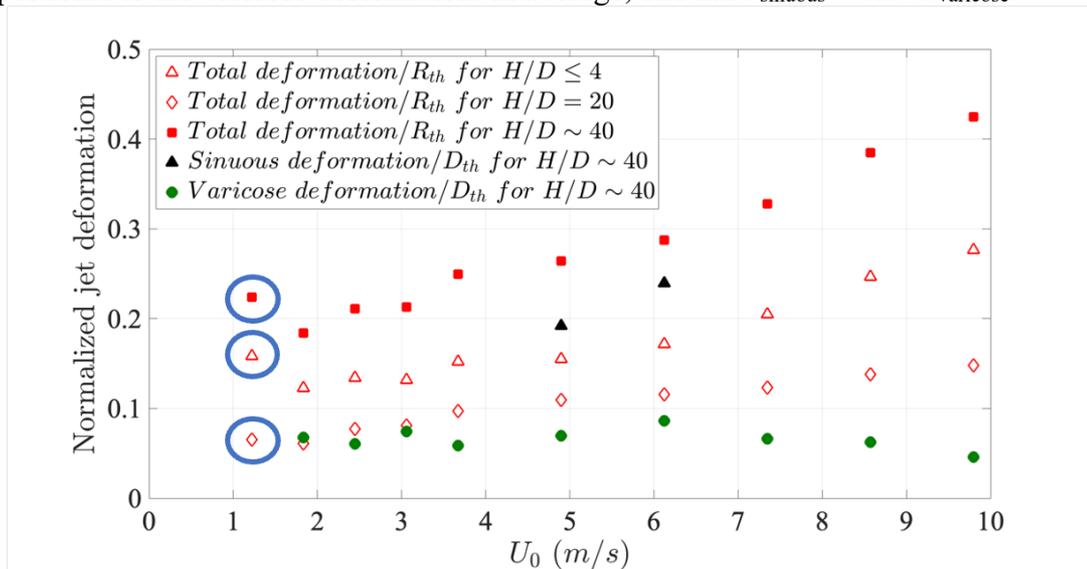

Fig. 23: Evolution of the sinuous deformation and of the varicose deformation scaled by $D_{th}$ and evolution of the total deformation (counted on one jet side) scaled by $R_{th}$ with the jet velocity at nozzle. The circled data correspond to a jet deformation due to a Rayleigh instability. Data collected in the small-scale experiment ($D_0$=7.6 mm) for the flow conditions of Fig.5.

In the large-scale facility, the nozzle diameter has also been varied in addition to H and $U_0$. Besides, the jet topology experienced huge modifications as discussed above (see also Fig.9). As shown in Fig.24, the total deformation $\varepsilon_{total}$ relative to one side of the jet ranges from 10% to 90% of the free fall radius $R_{th}$. Moreover, and except for the buckling behavior of the run#7, aerated jets correspond to the largest deformations. In the large-scale facility, the sinuous and varicose components remain globally correlated (see Table C-2) but there is no longer a nearly univocal connection as observed in the small-scale experiment. Instead, the ratio of the amplitude of the sinuous versus the varicose deformation fluctuates from about 0.3 up to 2.8.

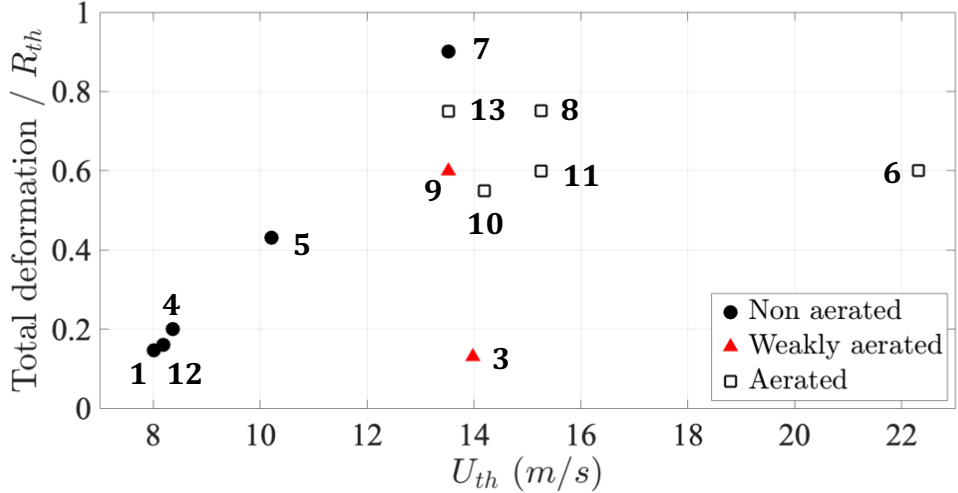

Fig. 24: Evolution of the total deformation scaled by the impact radius $R_{th}$ versus the impact velocity. The run numbers refer to Table 4. The jet aeration is indicated. Data collected in the large-scale experiment.

As an alternate way to characterize the jet deformation, we also quantified from the diameter distributions, the diameter $D_{90}$ that corresponds to a 90% detection probability. The $D_{90}$ could be seen as an estimate of the maximum lateral extent of the jet that Henderson and coworkers introduced in eq.(7). Let us define the deformation $\varepsilon_{90}$ on one side of the jet such that $D_{90} = D_{th} + 2 \varepsilon_{90}$. In Fig.25, the deformation $\varepsilon_{90}$ is plotted versus $\varepsilon_{total}$ in the two facilities. In both cases, there is a connection between $\varepsilon_{90}$ and $\varepsilon_{total}$, but the two quantities are not strictly equivalent. In the small-scale facility and at large $H/D_0$, one has $\varepsilon_{90} = 1.93 \ \varepsilon_{total}$ with an average deviation of about 3% and maximum deviations in the range -15% to +19%. In the large-scale facility, one has $\varepsilon_{90} \approx 1.47 \ \varepsilon_{total}$ with an average deviation of about 26%, but with much stronger maximum deviations as the latter range from +96% to -48%. Note that the run #6 appears isolated in Fig.25-right. Let us underline that the run #6 corresponds to a large injection velocity (20 m/s) that leads to a very efficient stripping of the jet (see Fig.9 and Fig.30): ligaments and droplets are produced in number at the jet boundary leading to very difficult conditions for image analysis. It is thus probable that a fraction of detached drops, that overlap in dense regions, get included within the jet boundary for this run.

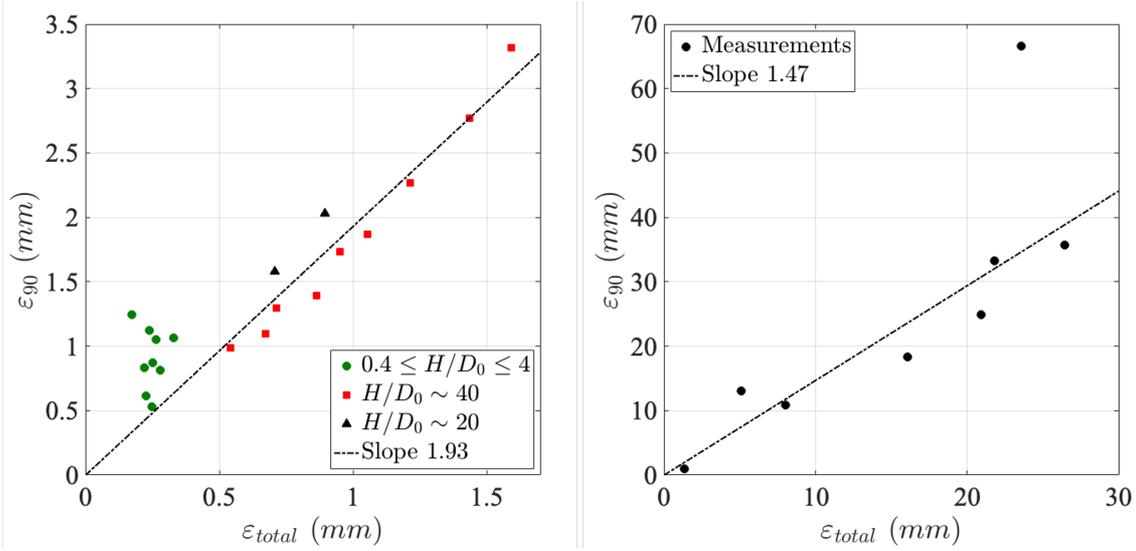

Fig.25: Deformation $\varepsilon_{90}$ versus the total deformation $\varepsilon_{total}$ in the small-scale facility (left) and in the large-scale facility except run #2 (right). The isolated data in Fig.25-right corresponds to run #6.

### 4.3 Entrained gas flow rate

Owing to the different behaviors of the jet evolution with the fall height observed in the two facilities, the scenarios presented in Section 1.3 will be tested in the following using the jet perimeter at impact evaluated as $\pi D_{th}$ instead of considering the measured diameter. For the same reason, the velocity $U_{th}$ will be considered as the reference velocity at impact.

#### 4.3.1 Small-scale experiments

Let us start the discussion for large heights of fall, namely $H/D_0 \approx 20$ and 40. As a test of the roughness scenario on the small-scale experiment, the measured air flow rate per unit perimeter $Q_{air}/(\pi D_{th})$ is plotted in Fig.26 versus $U_{th}$ times the total deformation of one side of the jet $\varepsilon_{total}$. The data for large $H/D_0$ (i.e. $\approx 20$ and 40) happen to be fairly well aligned along the line of equation:

$$Q_{air} \: / \: (\pi \: D_{th}) = 2.78 \: U_{th} \: \varepsilon_{total} \qquad (20)$$

The correlation coefficient equals 0.98. This result supports the first argument of Sene in the roughness scenario that leads to eq.(12). According to eq.(20), the effective roughness $\varepsilon_{eff}$ relative to one jet side amounts to 2.78 $\varepsilon_{total}$ that is $\approx 2.8$ times the deformation of one side of the jet as quantified by the standard deviation of the jet boundary position.[1]

---

[1] When $Q_{air}/(\pi D_{th})$ is plotted versus $U_{th} \: \varepsilon_{varicose}$ (where $\varepsilon_{varicose}$ is relative to the two sides of the jet), a linear behavior is also observed (the correlation coefficient is 0.96) with a prefactor equals to 1.87. That prefactor is consistent with eq.(20) and with the linear connection between sinuous and varicose deformations that holds for the small-scale facility. Hence, the effective roughness for the small-scale experiment at large $H/D_0$ also corresponds to 1.87 $\varepsilon_{varicose}$.

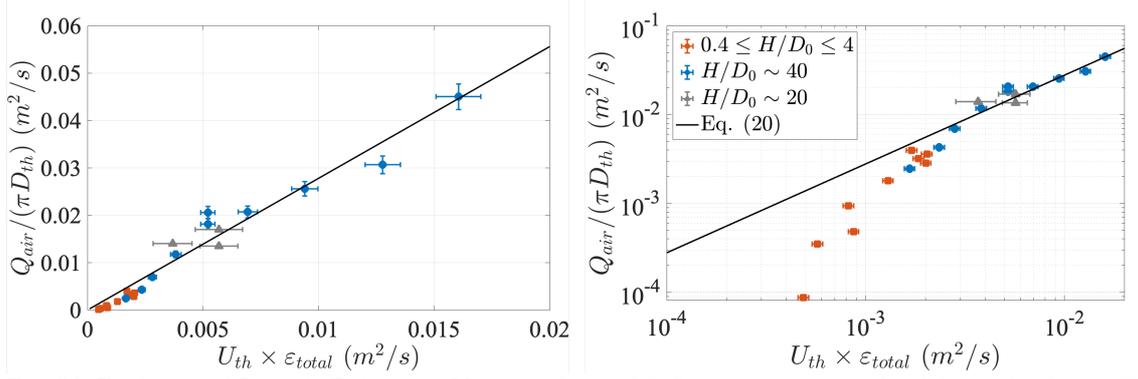

Fig. 26: Evolution of $Q_{air} / (\pi D_{th})$ versus $U_{th}$ times the total deformation on one side of the jet for the small-scale experiment (left). Same data in a log-log plot (right).

Alternately, we considered eq.(14) that accounts for the exact area occupied by the jet corrugation instead of its linearized version eq.(13). Here, we seek to determine the effective deformation $\varepsilon_{eff}$ relevant for air entrainment that would lead to $Q_{air}/Q_{water} = \{2\ \varepsilon_{eff}/R_{th} + (\varepsilon_{eff}/R_{th})^2\}$ as requested by eq.(14). The data of $Q_{air}/Q_{water}$ at large $H/D_0$ (i.e. $\approx 20$ and $\approx 40$) provide the following behavior:

$$Q_{air} / Q_{water} = C_0\ \{2\ K_{21}\ \varepsilon_{total}/R_{th} + (K_{21}\ \varepsilon_{total}/R_{th})^2\} \qquad (21)$$

with coefficient $C_0$ equals 1.0005 when the prefactor $K_{21}$ equals 2.11. The corresponding correlation coefficient is 0.956. The identification of the effective deformation on one jet side, $\varepsilon_{eff}$, that enters in eq.(14) leads to $\varepsilon_{eff} = K_{21}\ \varepsilon_{total}$. Hence, using the exact formula for the corrugated area, the proposal of Henderson et al. is recovered for large fall heights providing that the effective deformation $\varepsilon_{eff}$ relative to one side of the jet amounts to 2.11 $\varepsilon_{total}$, that is about two times the deformation of one side of the jet as quantified by the standard deviation of the jet boundary position.

We also evaluated an effective deformation relative to air entrainment with the data on $\varepsilon_{90}$ (see Section 4.2). Again, both the linearized version eq.(10) and the full formulae under the form eq.(21) were exploited. Table 2 summarizes the estimates of the effective deformation $\varepsilon_{eff}$ with respect to air entrainment for the small-scale facility at large $H/D_0$. In the last row of Table 2, the effective deformation has been expressed as a function of $\varepsilon_{total}$ using the relationship between $\varepsilon_{90}$ and $\varepsilon_{total}$ identified in Section 4.2. The estimates made exploiting either $\varepsilon_{total}$ or $\varepsilon_{90}$ are very close when using the linearized formula. This is also true with the full formula. The slight difference between the effective deformation evaluated from the linearized formula (2.7 or 2.8 $\varepsilon_{total}$) and that evaluated with the full formula (2.0 or 2.1 $\varepsilon_{total}$) is due to the behavior of eq.(13) and eq.(14). Indeed, these functions significantly depart from one another when the deformation $\varepsilon/R_{th}$ increases above 20%, so that adjusting a linear law on large deformation data can easily lead to a 25-30% difference on the slope estimation. As the deformations involved here are significant (see Fig.23), the use of the full formula is to be preferred.

Globally, the results of Table 2 validate both Sene's and Henderson et al.'s proposals despite some dispersion. The averaged dispersion around the mean trend is less than 25%, which is reasonable. However, the maximum dispersion recorded is rather large as it reaches 88% for the linearized formula and 63% for the full formula. Overall, the effective deformation in the small-scale facility and at large fall heights, happens to be about 2 times the deformation of one side of the jet as quantified by the standard deviation of the jet boundary position.

| Effective deformation with respect to air entrainment in the small-scale experiment at large H/D0 | | | |
|---|---|---|---|
| Measured deformation exploited | from the linearized formula eq.(12) | from the full formula (see eq.(21)) | Range of deformation covered by experiments |
| $\varepsilon_{total}$ | 2.78 $\varepsilon_{total}$ 1 side | 2.11 $\varepsilon_{total}$ 1 side | 0.18 ≤ $\varepsilon_{total}$ 1side /Rth ≤ 0.42 |
| $\varepsilon_{90}$ | 1.38 $\varepsilon_{90}$ 1 side ≈ 2.66 $\varepsilon_{total}$ 1 side | 1.06 $\varepsilon_{90}$ 1 side ≈ 2.06 $\varepsilon_{total}$ 1 side | 0.35 ≤ $\varepsilon_{90}$ 1 side / Rth ≤ 0.65 |

Table 2. Effective jet deformation relative to air entrainment determined from experiments at large $H/D_0$ in the small-scale facility.

We will come back to the roughness scenario when the results on the large-scale experiment will be presented. For the time being, let us focus on short fall heights (namely $H/D_0 \leq 4$).

### 4.3.2 Small-scale experiments at small fall heights

As shown in the insert of Fig.26, the data for short fall heights ($H/D_0 \leq 4$) deviate from eq.(20) notably at low impact velocities (more precisely, for impact velocities less than about 3 times the critical velocity). For these flow conditions, the jet roughness is small ($\varepsilon_{total}$ is less than 0.1 $D_{th}$, see Fig.23), and an air film scenario may be more appropriate. Hence, measured entrained air flow rate are compared with the predictions from Sene (eq.4) and from Lorenceau et al. (eq.6). As shown in Fig.27, the viscous-capillary model is not adapted to the air-water experiments considered here as it strongly underestimates the entrained flow rate over the whole velocity range. Measured air flow rates are closer to the viscous-hydrostatic proposal of Sene for velocities below about 2 m/s. However, the divergence with Sene's model increases at larger velocities. At larger velocities, say above about 4 – 5 m/s, the entrained gas flow rate increases as $U_{th}^{3/2}$ as in Sene's proposal. An air film scenario may possibly be valid in that range of flow conditions, provided that one accounts for an entrainment efficiency larger than that of a pure Couette flow. A more in-depth investigation of air entrainment by smooth jets deserves to be undertaken to clarify actual mechanisms in that regime.

Although the range of flow conditions considered here is too limited to identify the domain of existence of an air film scenario, a reasonable criterion can be tentatively proposed. Indeed, let us consider a smooth enough jet so that an air film mode is present, and let us progressively increase the jet roughness, all other parameters being fixed. At some point, one would expect that the air film can no longer survive the perturbations that incoming jet deformations (such as waves, liquid bulges...) would induce on the air flow inside the film and/or on the air film entrance conditions. In order to determine if such a transition happens in our data gathered in the small-scale facility, let us compare the measured total jet deformation on side of the side $\varepsilon_{total}$ with the gas film thickness $\Delta_{Sene}$ predicted by eq.(3). Although Sene model may be valid only at very low jet velocities, let us use $\Delta_{Sene}$ as a plausible reference for the air film thickness. Fig.28 provides the variations of the ratio $\varepsilon_{total}/\Delta_{Sene}$ as a function of the impact velocity, for several $H/D_0$ ratios. It happens that there is a clear separation between low fall height conditions ($H/D_0 \leq 4$) for which:

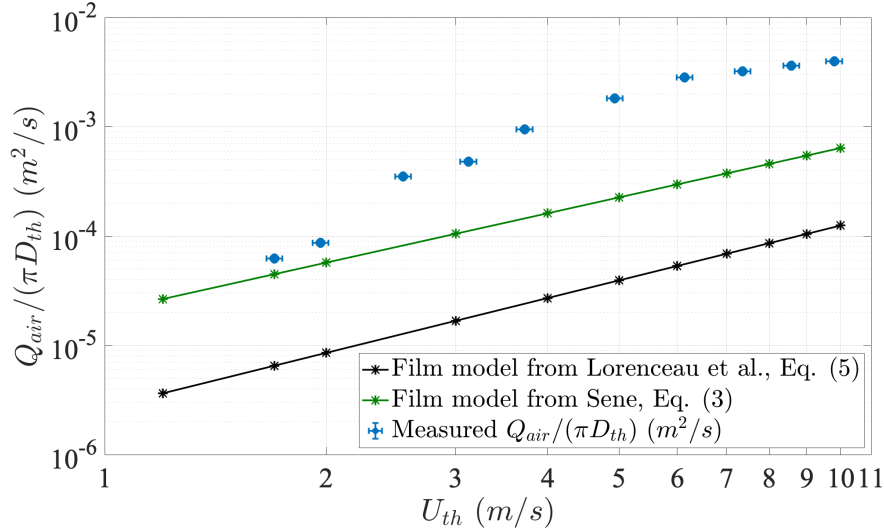

Fig.27: Comparison of measured air flow rate per unit perimeter $Q_{air}$ / ($\pi\ D_{th}$) with the predictions from eq.(4) and from eq.(6). Data from small-scale experiments with $0.4 \leq H/D_0 \leq 4$.

    i)       the jet deformation $\varepsilon_{total}$ is always less than $\approx 3$ times $\Delta_{Sene}$,

    ii)      the jet deformation $\varepsilon_{total}$ decreases with the impact velocity down to $\approx \Delta_{Sene}$,
and conditions with a larger fall height ($H/D_0 \approx 20$ or $40$) for which:

    i)       $\varepsilon_{total}$ is always above $5\ \Delta_{Sene}$,

    ii)      the jet deformation $\varepsilon_{total}$ significantly increases with the impact velocity (it
             reaches $8\ \Delta_{Sene}$ in the flow conditions considered).

Hence, the comparison between the gas film thickness that would exist for a smooth jet and the actual jet deformation seems to be a relevant criterion to delimitate the conditions leading to an air film scenario from those corresponding to a jet roughness scenario. Complementary experiments are however required to fully validate that proposal.

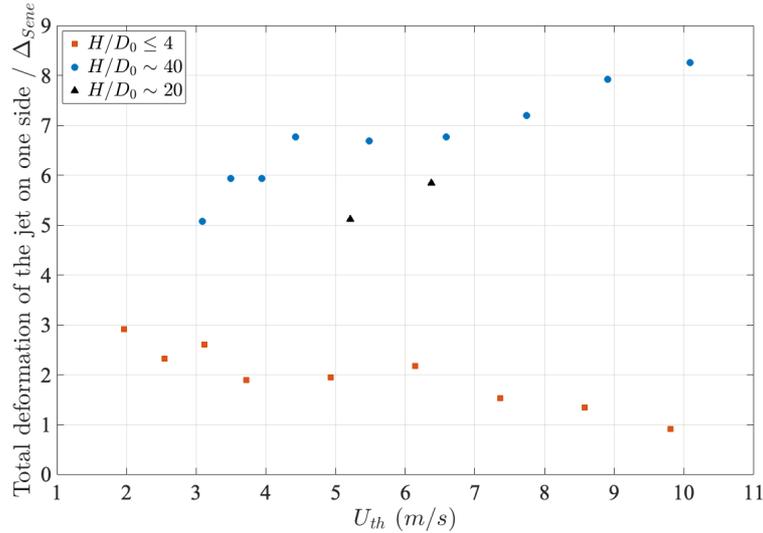

Fig.28: Evolution of the ratio of the total jet deformation on one side of the jet with the air film thickness predicted by Sene's model (eq.(3) for different heights of fall and impact velocities. Data from the small-scale experiment.

### 4.3.3 Large-scale experiment

To test the roughness scenario on the large-scale experiments, the measured air flow rate $Q_{air}$ divided by the jet perimeter at impact evaluated as $\pi\,D_{th}$ has been plotted in Fig.29 versus the impact velocity $U_{th}$ times the total deformation on one side of the jet. Despite some scatter (see later comments), the data collected in the large-scale facility remain globally aligned along a linear behavior. A fit of these large-scale experiments, provides:

$$Q_{air}\,/\,(\pi\,D_{th}) = 1.49\ U_{th}\ \varepsilon_{total} \tag{22}$$

with a correlation coefficient 0.96. Note that the run #6 has not been accounted for to establish eq.(6): the specific behavior of that run, already identified in Fig.25 regarding deformation, will be commented later. Eq.(22) is parallel to the trend observed in the small-scale experiment, but with a smaller prefactor: the apparent deformation amounts to $1.49\ \varepsilon_{total}$ for large-scale experiments, to be compared with $2.78\ \varepsilon_{total}$ in the small-scale facility.[2] Besides, the data collected in the large-scale facility (still disregarding the run #6) correspond to the full version of Henderson et al.'s model under the form of eq.(21), with a coefficient $C_0$ equals to 1.0005 when the prefactor $K_{21}$ equals 1.075. The correlation coefficient is 0.92. Thus, the proposal of Henderson et al. is recovered for the large-scale experiment provided that the effective deformation on one side of the jet is evaluated as $1.08\ \varepsilon_{total}$.

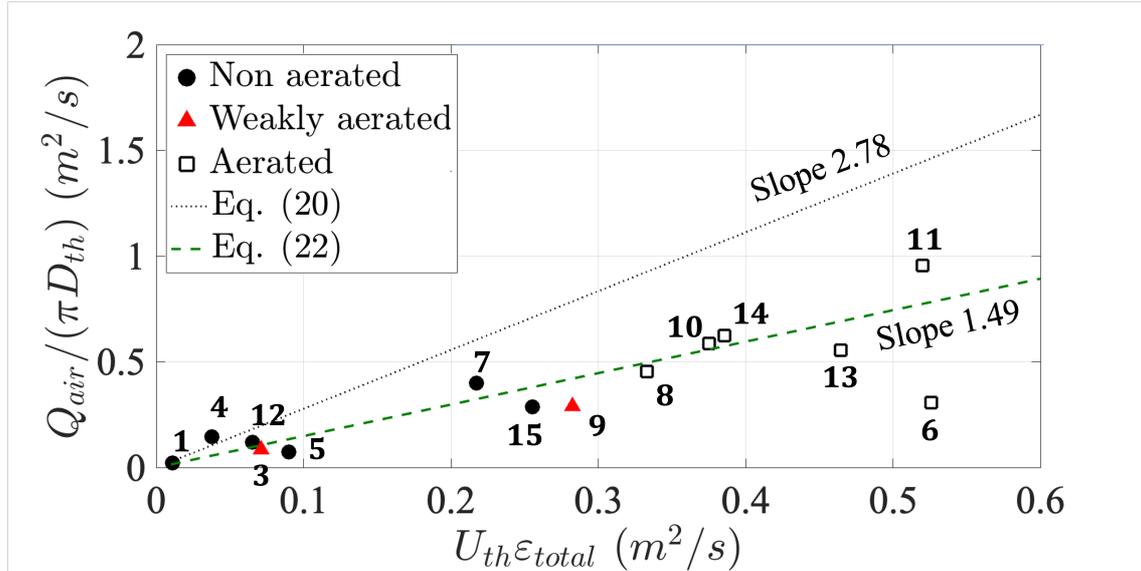

Fig. 29: Evolution of $Q_{air}\,/\,(\pi\,D_{th})$ versus $U_{th}$ times the total deformation for the large-scale experiments. The continuous line represents eq.(20) that is the fit of measurements in the small-scale experiment at large $H/D_0$ shown Fig.26. The dashed line represents eq.(22). The numbers refer to the experimental conditions detailed in Table 4 and illustrated in Fig.30. The break-up jet (run #2) is not included in this plot.

---

[2] When $Q_{air}/(\pi\,D_{th})$ is plotted versus $U_{th}\ \varepsilon_{varicose}$ (where $\varepsilon_{varicose}$ is relative to the two sides of the jet), a linear behavior is also observed (the correlation coefficient is 0.894) with a prefactor equals to 1.623. Hence, the effective roughness for the large-scale experiment at large $H/D_0$ amounts also to $1.623\ \varepsilon_{varicose}$.

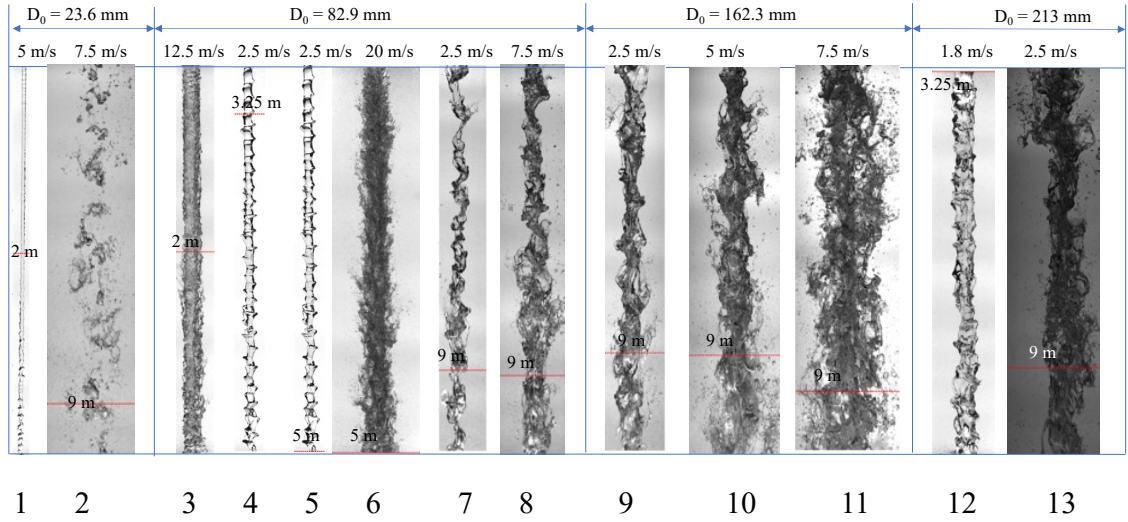

Fig. 30: Illustration of the jet allure for the flow conditions generated in the large-scale facility. The numbers refer to the runs mentioned in Table 4.

As before, the deformation $\varepsilon_{90}$ deduced from $D_{90}$ measurements has also been exploited in combination with eq.(10) and eq.(21). Table 3 summarizes the estimates of the effective deformation $\varepsilon_{eff}$ with respect to air entrainment deduced from experiments in the large-scale facility. In the last row of Table 3, the effective deformation has been expressed as a function of $\varepsilon_{total}$ using the relationship between $\varepsilon_{90}$ and $\varepsilon_{total}$ identified in Section 4.2. Considering either $\varepsilon_{total}$ or $\varepsilon_{90}$ leads to the same results. Besides, the difference between the linearized and the full formulae remains modest. Overall, the effective deformation in the large-scale facility is 1.5 times the total deformation measured on one side of the jet. Globally, the results of Table 3 validate both Sene and Henderson et al. proposals despite some dispersion. The average dispersion around the mean trend is about 30 to 33%, which is acceptable. However, the maximum dispersion recorded is rather large as it reaches about 110% for the linearized formula and about 80% for the full formula.

| Effective deformation with respect to air entrainment in the large-scale experiment | | | |
|---|---|---|---|
| Measured deformation exploited | from the linearized formula eq.(12) | from the full formula (see eq.(21)) | Range of deformation covered by experiments |
| $\varepsilon_{total}$ | 1.49 $\varepsilon_{total}$ 1 side | 1.075 $\varepsilon_{total}$ 1 side | 0.13 ≤ $\varepsilon_{total}$ 1side /Rth ≤ 0.9 |
| $\varepsilon_{90}$ | 0,993 $\varepsilon_{90}$ 1 side ≈ 1.46 $\varepsilon_{total}$ 1 side | 0.77 $\varepsilon_{90}$ 1 side ≈ 1.13 $\varepsilon_{total}$ 1 side | 0.11 ≤ $\varepsilon_{90}$ 1 side / Rth ≤ 1.1 |

Table 3. Effective jet deformation relative to air entrainment determined from experiments in the large-scale facility.

| Large-scale experiment | | | Water jet in air | | | |
|---|---|---|---|---|---|---|
| Run # | D0 (mm) | U0 (m/s) | H (m) | H/D0 | Jet aeration | Jet allure |
| 1 | 23.6 | 5 | 2 | 84.7 | Non aerated | Coherent jet ≈ Rayleigh mode |
| 2 | 23.6 | 7.5 | 9 | 381.4 | Non aerated | Broken-up jet |
| 3 | 82.9 | 12.5 | 2 | 24.1 | Weakly aerated | Coherent jet + Corrugations |
| 4 | 82.9 | 2.5 | 3.25 | 39.2 | Non aerated | Buckling |
| 5 | 82.9 | 2.5 | 5 | 60.3 | Non aerated | Buckling |
| 6 | 82.9 | 20 | 5 | 60.3 | Aerated | Distorted jet + Strong stripping |
| 7 | 82.9 | 2.5 | 9 | 108.6 | Non aerated | Buckling + Stripping |
| 8 | 82.9 | 7.5 | 9 | 108.6 | Aerated | Distorted jet + Stripping |
| 9 | 162.3 | 2.5 | 9 | 55.5 | Weakly aerated | Distorted Jet + Stripping |
| 10 | 162.3 | 5 | 9 | 55.5 | Aerated | Distorted Jet + Stripping |
| 11 | 162.3 | 7.5 | 9 | 55.5 | Aerated | Distorted Jet + Stripping |
| 12 | 213 | 1.8 | 3.25 | 15.3 | Non aerated | Coherent jet + Corrugations |
| 13 | 213 | 2.5 | 9 | 42.3 | Aerated | Distorted Jet + Stripping |
| 14 | 213 | 5 | 8 | 37.6 | Aerated | Distorted Jet + Stripping |
| 15 | 162.3 | 2.5 | 8 | 49.3 | Non aerated | Distorted Jet + Stripping |

Table 4: Flow conditions generated in the large-scale facility (see also Table C-2).

Overall, Sene's model (which is a linearized version of Henderson et al.'s proposal) happens to be also applicable to the large-scale experiments but with a slope that is nearly half the one identified from the small-scale experiments. Similarly, as shown in Tables 2 and 3, the complete Henderson et al.'s proposal also applies to our large-scale experiments but with a coefficient that is about 3/4 of the one found from the small-scale experiment when considering the full formulae. That coefficient drops to 54% when considering the linearized formulae.

We tentatively suggest that the difference in the slopes is connected with the apparition of jet aeration and/or of intense interface stripping. Indeed, the jets remain quite coherent in the small-scale facility while, in the large-scale facility, the air could deeply penetrate inside the jets and induce internal aeration of the jet as reported in Table 4. In addition, stripping eventually associated with significant atomization is absent from the small-scale facility, while it occurs in the large-scale facility over a significant range of flow conditions (see comments in Table 4). Overall, atomization and/or aeration are two mechanisms that create complex interface deformations, and one may tentatively foresee two consequences of these changes of jet topology.

A first possible consequence would be a change in the actual velocity at the jet periphery as liquid structures generated by stripping and/or atomization are more prone to experience air friction than when considering coherent and moderately rough jets. With such complex jet boundaries, one expects that the local liquid velocity along the jet boundary becomes less than the free fall estimate $U_{th}$. If so, keeping the reasoning of Henderson et al. or of Sene, the air flow rate entrained below the free surface per unit contact length should evolve as $U_{interface}$ times the deformation with $U_{interface}$ being less than $U_{th}$. That may explain why the data collected in the large-scale facility seem to be horizontally translated to the right in Fig.29, leading to a decrease of the slope. This qualitative explanation is supported by the fact that the lateral displacement to the right of the data is more marked at large fall heights, that is when the action of the surrounding air becomes more effective. That effect is especially marked for the run #6 which is the only run corresponding to a very strong stripping of the interface: in that case, the thin ligaments and the droplets emanating from them experience a strong deceleration, and are not expected to be very effective in terms of air entrainment. So far, we have no direct experimental evidence of such a deceleration of the jet boundary. The velocity measurements presented in Section 4.1 correspond to an average taken over the whole jet width, and this is probably the reason why no lateral variations in velocity could be detected. Localized velocity measurements near the jet boundary would possibly exhibit such a change: this is a question to address in future investigations using, for example, optical probes.

A second possible consequence could be related to qualitative changes in the interface deformation. Indeed, modifying the shape of the waves and notably their depth/wavelength aspect ratio is expected to modify their capability to trap air and to entrain it below the free surface. The results presented by Ramirez de la Torre et al., 2020 support that statement for coherent jets. That trend is expected to hold for the strong interface deformations that arise in aerated conditions. Refined measurements of the impact of the shape of waves would be useful to clarify this point.

The two above-mentioned consequences may also be competing depending on flow conditions. Indeed, strong interface deformations involve larger masses of water than stripping (notably with thicker liquid lumps, see Fig.30), and are more prone to embark air below the free surface. This seems to be case of run #11 in Fig.29, which entrains a significantly larger amount of air compared with run #6 although they both have nearly the same abscissa, that is similar $U_{th} \varepsilon_{total}$ products.

Before comparing with literature data, it is worth mentioning another point. Indeed, we have also carried out the same analysis using the varicose deformation instead of the total deformation: the same trends and conclusions are obtained with this alternative choice (of course, prefactors were not the same as when using the total deformation). Hence, at least over the data collected here, it was not possible to identify a type of deformation, namely varicose or sinuous, that was more relevant than the other when dealing with the entrained air flow rate. From now on, the discussion will exclusively rely on the total deformation.

### 4.3.4 Analysis of the air flow rate behavior

To put the results into a broader perspective, the ratio $Q_{air}/(\pi D_{th})$ measured in the small-scale and in the large-scale facilities are plotted together versus $U_{th} \varepsilon_{total}$ in Fig.31, where we have also reported data from the literature. To plot these data, we computed the quantities $U_{th}$ and $D_{th}$ for all series extracted from the literature. Concerning the deformation, both Van de Sande and Smith, 1973 and Cumming, 1975 provide data on the maximum extent D* of the jet at impact. From D* data, we deduced a maximum deformation $\varepsilon*$ for one side of the jet such that $D*= D_{th} + 2 \varepsilon*$. Note that Van de Sande and Smith, 1973 provide D* as a correlation that is not applicable to all the data they have collected (see the discussion in Section 1.3.2). The three data series from McKeogh et al., 1981 are also plotted in Fig.31. McKeogh et al. provide the deformation $\varepsilon_{max}/R$ where R + $\varepsilon_{max}$ represents the maximum deformation of one side jet and where R is the local jet radius. It is unclear whether their data on $\varepsilon_{max}/R$ were obtained using a measured local jet diameter D=2R or with the nozzle diameter $D_0$ (see the discussion in Section 1.3.2). To evaluate $\varepsilon_{max}$ from their data, we considered that the local jet diameter 2R equals $D_{th}$. When doing so, $\varepsilon_{max}$ has the same definition as $\varepsilon*$.

Setting aside the data corresponding to H/D ≤ 4 in the small-scale facility that are relevant for the air film scenario (see Section 4.3.2), Fig.31 shows that all data sets are consistent with each other. In particular, the McKeogh et al. series are close to eq.(22) at high $U_{th} \varepsilon_{max}$ abscissa. It is worth mentioning that McKeogh et al.'s data plotted in Fig.31 include flow conditions near to, and even slightly beyond break-up. Accordingly, the deformation $\varepsilon_{max}/R$ reported by McKeogh et al. ranges from 0.16 up to unity. Nearly all plotted data are globally comprised between eq.(20) and eq.(22). There are two exceptions.

- Some data from Van de Sande and Smith, 1973, whose abscissae in Fig.31 is below $\approx 5\ 10^{-3}$ deviate from the main trend. These data correspond to the smallest jet deformations recorded by the authors with $\varepsilon^*$ comprised between 1% and $\approx$20% of the jet radius, while for the rest of their data, one has $0.2 \leq \varepsilon^*/R_{th} \leq 0.76$.
- A few data from Cumming, 1975, whose abscissae in Fig.31 is below $10^{-3}$, neatly deviate from the main trend. Again, these data correspond to the smallest deformations with $\varepsilon^*$ comprised between 1% and 8% of the jet radius, while $0.15 \leq \varepsilon^*/R_{th} \leq 0.68$ for the remaining of their data.

Since Van de Sande and Smith, 1973 and Cumming, 1975 exploited rather small jet diameters ($D_0$ was within the range 1.9 mm to 10 mm), it is plausible that the smallest jet deformations in Van de Sande and Smith, 1973 and Cumming, 1975 were affected by some measurement uncertainty. We cannot definitely conclude on this aspect as the resolution of their measuring techniques is not known.

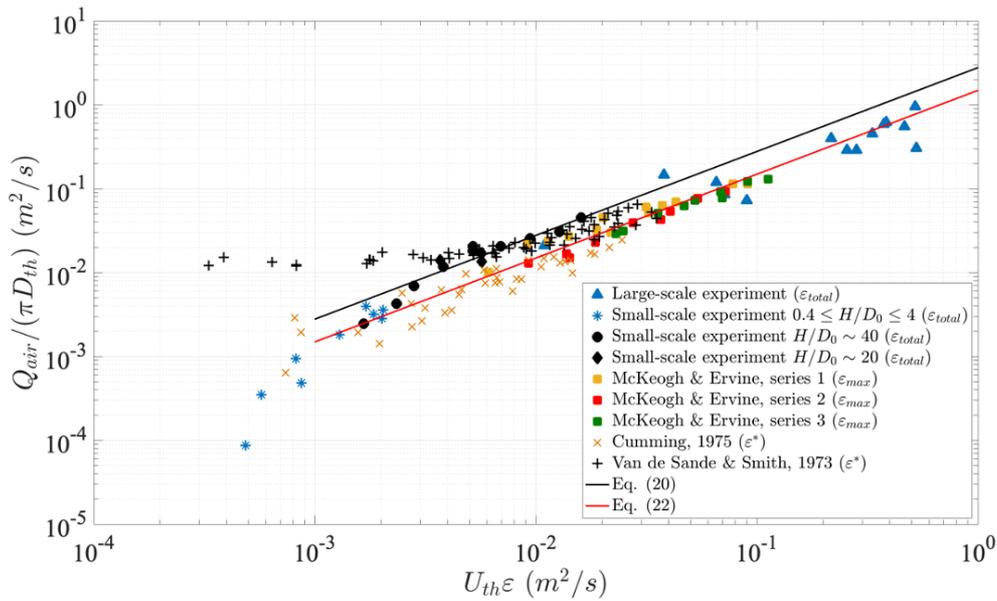

Fig. 31: Plot of $Q_{air}/(\pi\ D_{th})$ versus $U_{th}\ \varepsilon_{total}$ from the small-scale, the large-scale experiments and from literature. The latter include all the data from Van de Sande and Smith, 1973 and from Cumming, 1975 for which the deformation has been deduced from D* measurements, and all series from McKeogh et al., 1981, that are based on the deformation $\varepsilon_{max}$. One data from Van de Sande and Smith, 1973 and one data from Cumming, 1975 that have an abscissa below $10^{-4}$, do not appear in the above figure.

As done before, the effective deformation with respect to air entrainment has been determined using eq.(12) and eq.(21). The results are summarized in Table 5. For Van de Sande and Smith, 1973 and for Cumming, 1975, an analysis including all their data or considering only conditions with a significant deformation (the limit was set to a deformation larger than 20% of the jet radius) marginally changes the results. Similarly, for McKeogh et al., 1981 series, the results are similar when flow conditions near from or beyond break-up are excluded (data selected were such that $H/L_B \leq 0.7$) or when they are included. Globally, the effective deformation amounts to 1 to 1.9 $\varepsilon^*$ when using the linear formulae, and from 0.9 to 1.5 $\varepsilon^*$ from using the full formula. The upper values of these intervals come from the data of Van de Sande and Smith, 1973, as the coefficients deduced from their experiments are 40% to 80% larger than the coefficients identified from other authors. Finally, let us also mention that, for rectangular jets inclined at a 45° angle, Bagatur et al., 2003 found a prefactor of 0.95 when considering $\varepsilon^*$: their result is fully consistent with those of Table 5.

| Effective deformation with respect to air entrainment from littérature data | | | | |
|---|---|---|---|---|
| Authors | Measured deformation exploited | from the linearized formula eq.(12) | from the full formula (see eq.(21)) | Range of deformation covered by experiments |
| Van de Sande and Smith, 1972, 1973 and 1976 | $\varepsilon^*$ deduced from D* measurements | 1.92 $\varepsilon^*$ 1side | 1.52 $\varepsilon^*$ 1side | 0.04 ≤ $\varepsilon^*$ / Rth ≤ 0.76 |
| | | 1.90 $\varepsilon^*$ 1side | 1.46 $\varepsilon^*$ 1side | not too small deformations 0.2 ≤ $\varepsilon^*$ / Rth ≤ 0.76 |
| Cumming, 1975 | $\varepsilon^*$ deduced from D* measurements | 1.03 $\varepsilon^*$ 1side | 0.925 $\varepsilon^*$ 1 side | 0.008 < $\varepsilon^*$(1 side) /Rth ≤ 0.683 |
| | | 1.02 $\varepsilon^*$ 1side | 0.91 $\varepsilon^*$ 1 side | not too small deformations 0.2 < $\varepsilon^*$(1 side) /Rth ≤ 0.683 |
| McKeogh and Ervine 1981 | $\varepsilon_{max}$ | 1.33 $\varepsilon_{max}$ 1side | 0.96 $\varepsilon_{max}$ 1side | 0.16 ≤ $\varepsilon_{max}$ 1side / Rth ≤ 1.27 |
| | | 1.38 $\varepsilon_{max}$ 1side | 1.1 $\varepsilon_{max}$ 1side | far from break-up 0.16 ≤ $\varepsilon_{max}$ 1side / Rth ≤ 0.85 |

Table 5: Effective deformation deduced from eq.(12) or eq.(23) for the different data series.

To compare literature results with those presented in Table 2 and 3, let us rely on $\varepsilon_{90}$ that is in a way a measure of the maximum deformation experienced by the jet. Indeed, $\varepsilon_{90}$ is similar to $\varepsilon^*$ or to $\varepsilon_{max}$ measurements presented in the literature (although they are not identical) contrary to the standard deviation of the position of one jet boundary $\varepsilon_{total}$. The prefactors identified in the small-scale experiments (Table 2) agree with those deduced from published data (Table 5) when one considers $\varepsilon_{90}$ as the reference deformation. The effective deformations observed in large-scale experiments correspond to $\approx 0.7$ $\varepsilon_{90}$ (with the full formulae) up to $\approx 1.0$ $\varepsilon_{90}$ (with the linear formulae). These prefactors are less (or equal for the linearized approach) than those reported in the literature when using $\varepsilon^*$. This is probably because most, if not all, published data remain far from the aeration limit as shown in Fig.32 where the conditions examined by Van de Sande and Smith, 1973, by Cumming, 1975 and by McKeogh et al., 1981 are compared with the tentative aeration limit proposed in Fig.22. A possible exception comes from Van de Sande and Smith, 1973 who performed runs at a high initial velocity ($U_0$ up to 25 m/s). However, an analysis of their data at large initial jet velocity provides a prefactor K12 that does not significantly change (K12 increases up to 2.3 to be compared with 1.9). Hence, these jets do not behave as aerated jets as discussed in Section 4.3.3. A possible explanation for that may come from the fact that the high jet velocity data from Van de Sande and Smith, 1973 concern quite small jet diameters ($D_0$ from 3 to $\approx 5$ mm). High initial velocity favors interface stripping (similar to what is observed on the run #6). Such interface stripping develops from the injector exit (e.g. Descamps et al., 2008; Marmottant and Villermaux, 2004) and it continuously ejects liquid mass from the interface, leading thus to a quick decrease of the jet radius. Hence, it may not be possible to observe a significant aeration on thin high-speed jets, simply because they lost much of their liquid mass before air could penetrate deeply inside them.

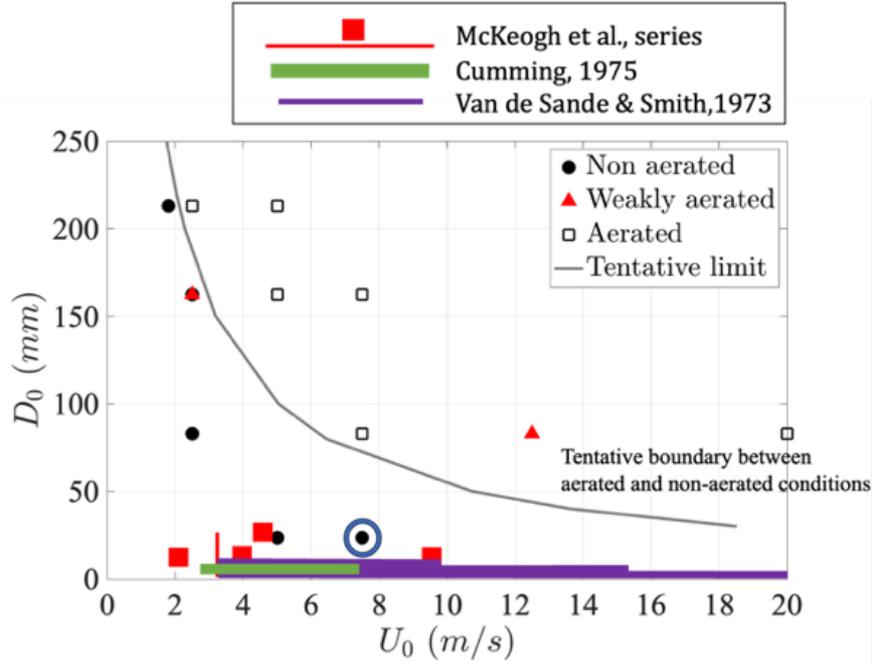

Fig. 32: Flow conditions of previous contributions exploited in Fig.31 mapped in a $D_0$, $U_0$ plane and compared with the tentative aeration boundary proposed in Fig.22.

As a way to synthesise the results for rough jets, we report in Fig.33 the measured ratio $Q_{air}$ / $Q_{water}$ versus the quantity $\{2\ K21\ \varepsilon/R_{th} + (K21\ \varepsilon/R_{th})^2\}$ where the coefficients K21 identified for each data series and for the relevant deformation $\varepsilon$, are provided in Tables 2, 3 and 5. Overall, and despite a significant dispersion, all data series gather around the bisector, meaning that Henderson et al.'s proposal applies to all the experiments considered with a coefficient K21 close to unity when the deformation considered is close to the maximum deformation (i.e. when it is $\varepsilon^*$, $\varepsilon_{max}$ or $\varepsilon_{90}$). The only exception concerns aerated flow conditions for which the coefficient K21 is about 0.77. Let us recall that aerated flow conditions have been defined as jets whose diameter exceeds the free-fall diameter by 20-30% (*weakly aerated* conditions) or by 50% or more (*aerated* conditions).

Finally, let us mention that run #6 (which corresponds to the largest ejection velocity, see Table C-2) provides an isolated data whose behavior is tentatively due to very strong stripping conditions. So far, only one measurement has been achieved for this flow condition: this measurement should be repeated, and extended to similar conditions of strong stripping.

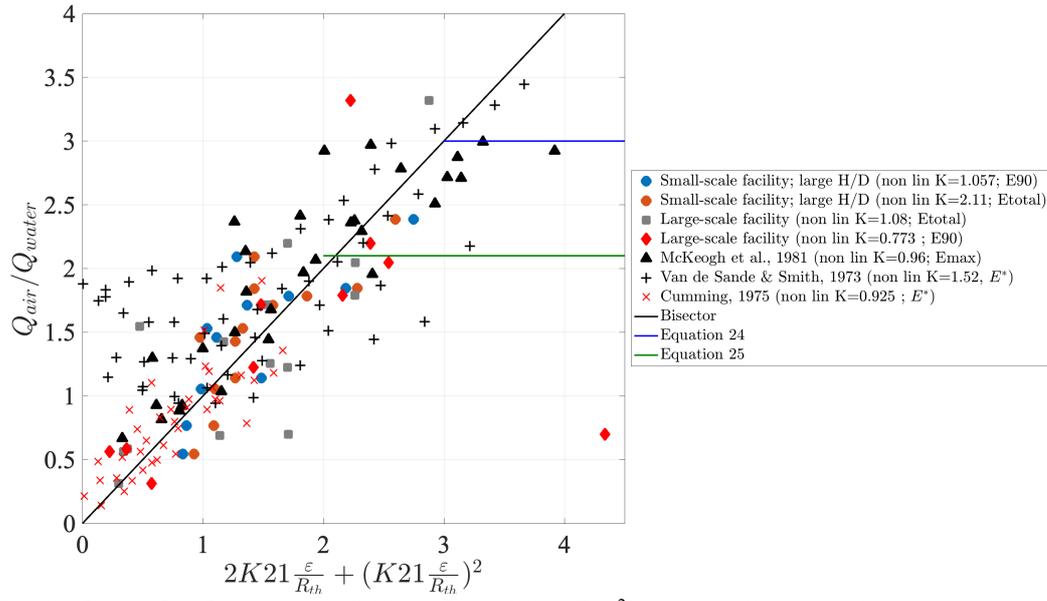

Fig. 33: Plot of $Q_{air}/Q_{water}$ versus $\{2\ K21\ \varepsilon/R_{th} + (K21\ \varepsilon/R_{th})^2\}$ from small-scale and large-scale experiments, from Van de Sande and Smith, 1973, from Cumming, 1975 and from McKeogh et al., 1981. The definition of the deformation considered and the corresponding factor K21 are indicated in the legend for each set. Note that all McKeogh et al. data including those near or at break-up have been reported in the figure.

## 5. Air entrainment: discussion

In the previous section, the scaling of the entrained air flow rate by a plunging jet has been clarified for smooth jets and for rough jets. A tentative frontier between these two regimes has also been proposed (see Section 4.3.2). For rough jets, the linear approach by Sene that leads to eq.(12) happens to be applicable at small to moderate deformations, while Henderson et al. formulation $Q_{air} / Q_{water} = \{2\ \varepsilon_{eff}/R_{th} + (\varepsilon_{eff}/R_{th})^2\}$ is to be preferred for large deformations. The value of the effective deformation $\varepsilon_{eff}$ involved in the expression of $Q_{air}$ has been identified in Section 4. In the following, we discuss some consequences of these findings.

### 5.1 Maximum entrained air flow rate

For turbulent, coherent, and non-aerated jets, $\varepsilon_{eff}$ was found equal to $K_{21}\ \varepsilon_{90}$. For that type of jet, van de Sande and Smith, 1972 report a very slow increase of the entrained air flow rate when the fall height exceeds the break-up length, while McKeogh and Ervine (1981) observed that $Q_{air}$ reaches a clear maximum for $H = L_B$. Hence, the maximum entrained air flow rate occurs at break-up. Besides, at break-up, the deformation of one side of the jet becomes of the order of the local jet radius, that is $\varepsilon_{90}/R_{th} \approx 1$. Therefore, the entrained air flow rate at break-up expresses as:

$$Q_{air}\ \text{at break-up}/Q_{water} = 2\ K_{21} + K_{21}^2 \qquad (23)$$

As $K_{21}$ is close to unity for turbulent, coherent and non-aerated jets, the maximum entrained air flow rate by a coherent plunging jet is expected to be:

$$\text{Max}(Q_{air}\ \text{non aerated})/Q_{water} = 3 \qquad (24)$$

When eq.(23) is applied to aerated jets, for which $\varepsilon_{eff} \approx K_{21}\,\varepsilon_{90}$ with $K_{21} \approx 0.77$, the maximum air flow rate predicted becomes:

$$\text{Max}(Q_{air}\text{ aerated})/Q_{water} = 2.1 \tag{25}$$

As shown in Fig.(34), eq.(25) is compatible with the data for aerated jets collected in the large scale facility. However, data for aerated jets are scarce and complementary experiments are needed to thoroughly test that limit.

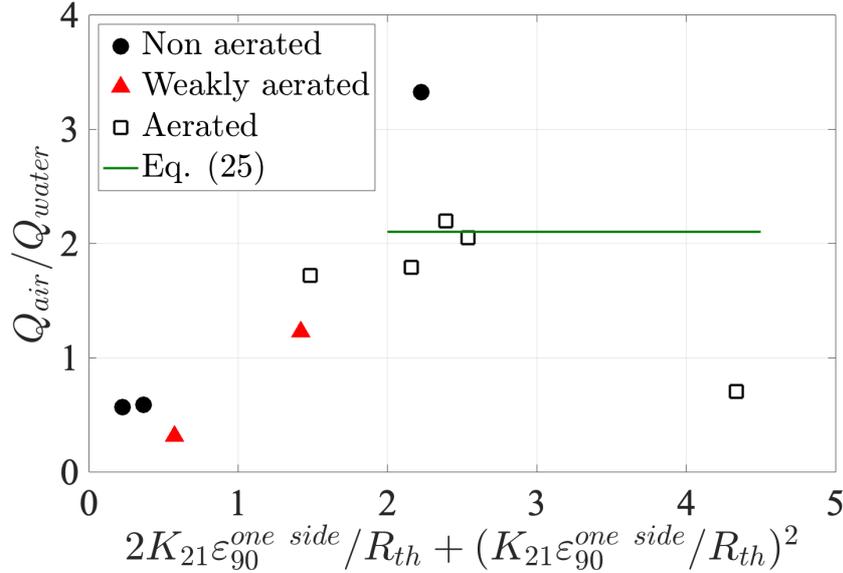

Fig. 34: Plot of $Q_{air}/Q_{water}$ versus $\{2\,K_{21}\,\varepsilon_{90}/R_{th} + (K_{21}\,\varepsilon_{90}/R_{th})^2\}$ for the large-scale experiments compared with the limit provided by eq.(25).

Concerning non-aerated jets, when one searches through the extensive review by Bin (1993), only a handful of experimental data collected among the $\approx$150 publications that Bin has scrutinized do reach or exceed the limit given by eq.(24). A first set comprising about 10 data comes from Van de Sande and Smith, 1973 (see Fig.17 in Bin, 1993). These data were obtained on small-diameter and high-speed jets, more precisely for $D_0 = 1.95$ mm diameter jets with velocities $U_0$ between 6 and 10 m/s, and for $D_0 = 3$ mm diameter jets for velocities $U_0$ above 20 m/s. The height of fall was H= 0.1 m, and the angle with the vertical was 30°. Over this set, the maximum value of $Q_{air}/Q_{water}$ measured is 3.5 which is close to the limit given by eq.(24). The second set corresponds to five to six data gathered by Henderson et al., 1970 (see Fig.20 in Bin, 1993) that were collected on $D_0 \approx$ 6 mm jets at high initial velocity ($U_0 \approx 10$ m/s). The maximum value of $Q_{air}/Q_{water}$ measured by Henderson et al., 1970 does not exceed 4, which is not so far from the expected limit. Note that we don't know the uncertainty related to these early contributions. When examining the literature since 1993, and to the best of our knowledge, only the contribution from Bagatur et al., 2003 mentions measurements such that $Q_{air}/Q_{water} = 3$. Finally, and referring to our measuring campaign, Fig.33 indicates that only one condition, namely run #7, for which $Q_{air}/Q_{water} = 3.32$, is above the expected limit. However, owing to the ±11 to 12% uncertainty on the gas flow rate estimate as discussed in Section 3.2, that value remains within the expected limit. Overall, Eq.(24) seems to provide a valid estimate of the upper bound of the air flow rate entrained by a turbulent, coherent, and non-aerated jet. Although complementary experiments would be useful to confirm that statement, this order of magnitude is relevant for engineering purposes.

## 5.2 Void fraction

The knowledge of the entrained air flow rate $Q_{air}$ can be exploited to provide an estimate of the void fraction just below the free surface. Indeed, let us assimilate the entrance region of the jet beneath the free surface to a steady, quasi-fully developed, one-dimensional two-phase flow. This is an idealized situation that neglects the growth of the two-phase mixing layers below the free surface (e.g., Cummings and Chanson, 1997) as well as air detrainment (Bertola, Wang and Chanson, 2018b). Applying a kinematic model (Zuber & Findlay, 1965) to a cylindrical region of radius $R_{th}$, and considering that the jet velocities are much larger than the bubble terminal velocity (this is reasonable as the latter is at most 0.30 m/s for bubbles less than 20 mm in equivalent diameter), the global gas fraction should equal the gas flow rate fraction $\beta = Q_G/(Q_G+Q_L)$. Here, $Q_G$ is the entrained gas flow rate, and $Q_L$ is assimilated to the liquid flow rate of the jet because the entrainment of the liquid surrounding the immersed jet is negligible at very short depths below the free surface. As shown in Fig.35, the local void fraction on the axis and the maximum local void fraction along transverse profiles established at short depths below the free surface (measurements have been achieved at depths comprised between 3 and 8 $D_0$) are nearly linear with the gas flow rate fraction $\beta$. Combining data series for $H/D \leq 4$ and for $H/D \approx 40$, the void fraction on the axis amounts to 0.42 $\beta$. When accounting for all the flow conditions investigated (except run #6 those behavior is odd, see comments in Section 4.3.3), the maximum void fraction equals 0.86 $\beta$, with an average dispersion of 12%, and extreme dispersions of -43%/+49%. Note that the void fraction on the axis in the large-scale facility experiences a shift from value close to 0.42 $\beta$ to a value close to 0.86 $\beta$. That shift globally corresponds to the transition from saddle-shaped void fraction profiles to bell-shaped profiles. These data demonstrate that the gas flow rate fraction is the correct order of magnitude of the void fraction just below the free surface.

As the maximum entrained flow rate is about 3 times liquid flow rates for non-aerated jets, the gas flow rate fraction is at most 3/4=0.75, so that the maximum local void fraction reachable in these systems should be about 0.65. This figure is compatible with the data collected in the small-scale facility. Similarly, for aerated jets, the gas flow rate fraction is at most 0.68, the maximum void fraction in these systems should be about 0.59. If we set aside run #6 which as mentioned before corresponds to very distinct conditions, the extreme recorded void fractions are 0.63 on the axis and 0.69 over the whole profiles: these figures are slightly above but still comparable to the expected limit.

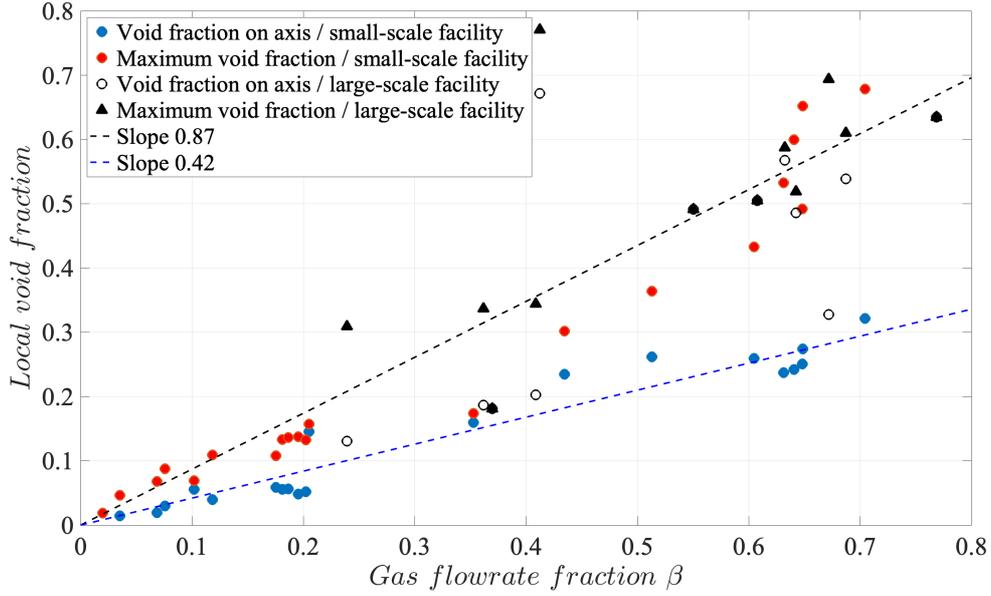

Fig. 35: Measured local void fractions in the small-scale and in the large-scale facilities versus the gas flow rate fraction β. Note the different behavior of run #6 (two data at an abscissa ≈0.4 with an ordinate between 0.65 and 0.8).

### 5.3 Scaling of the entrained air flow rate with flow parameters

In this work, we have chosen to bypass the question of the relationship between injection conditions (nozzle design, feeding circuitry, vibrations, etc…) and the state of the jet at impact. Instead, we have considered the jet characteristics at impact as an input parameter. However, many attempts to connect the entrained air flow rate with key flow parameters have been produced in the literature, among which those of Ma et al., 2010 summarized in Section 1.3. Our goal here is to discuss how the entrained air flow rate scales with the jet velocity.

- In the small-scale facility, and for very short heights of fall, we have shown in Section 4.3.1 that, when the jet is smooth enough, $Q_{air}$ grows as $U_{th}^{3/2}$ as predicted by the air film scenario proposed by Sene, 1988.

- At larger heights of fall, say above 20 $D_0$, the roughness scenario prevails, and eq.(20) is valid (for sake of simplicity, we stick to the linearized formula in this discussion): that equation indicates that $Q_{air}$ grows as $U_{th}$ times the jet deformation. According to Fig.23, the deformation increases with the jet velocity at impact at a rate that is at most linear. Such an increase does not follow the $U_{th}^2$ dependency expected by Sene in eq.(15). As a consequence, the growth of $Q_{air}$ as $U_{th}^3$ expected from eq.(16) and mentioned by Ma et al., 2010 is not recovered in the present experiments.

- In the large-scale experiment, eq.(22) tells that $Q_{air}$ grows as $U_{th}$ times the jet deformation. However, the connection between the deformation and the jet velocity illustrated in Fig.24 is quite complex, and it is difficult to identify a clear trend.

To put the results into a broader perspective, we regrouped in Fig.36 our results as well as some literature data in a plot of $Q_{air}/Q_{water}$ versus a Froude number. For the latter, we used the Froude at impact $Fr_{th} = U_{th}^2/(g\ D_{th})$. The choice of a Froude number is subjective and arbitrary as the problem is governed by many dimensionless parameters (e.g., Bertola, Wang and Chanson, 2018b). However, such a presentation will be sufficient for the conclusion we intend to draw. The literature data include Van de Sande and Smith, 1972 and 1973; Cumming, 1975; Van de Donk, 1981; McKeogh and Ervine, 1981; Sene, 1984 and 1988. Under the name "Chanson and coll." the following contributions have been regrouped: Cummings and Chanson, 1997; Brattberg and Chanson, 1998; Chanson, Aoki and Hoque, 2004; Wang, Slamet, Zhang and Chanson, 2018; Bertola, Wang and Chanson, 2018; Müller and Chanson, 2020. We first tested that none of these data are relevant for Sene's film scenario. Accordingly, the data collected at small H/$D_0$ in the small-scale facility have been excluded from the figures. All data except a few happen to be bounded between two limits: an upper bound of equation 0.4 $Fr^{0.35}$ and a lower bound of equation 0.013 $Fr^{0.55}$. Therefore, and even though a specific series could exhibit a different exponent, we globally observe that $Q_{air}$ increases as $U_{th}^n$ with an exponent n comprised between 1.7 and 2.1. This is far from the $U^3$ dependency predicted by Sene, 1984 in eq.(16), or reported by Ma et al., 2010.

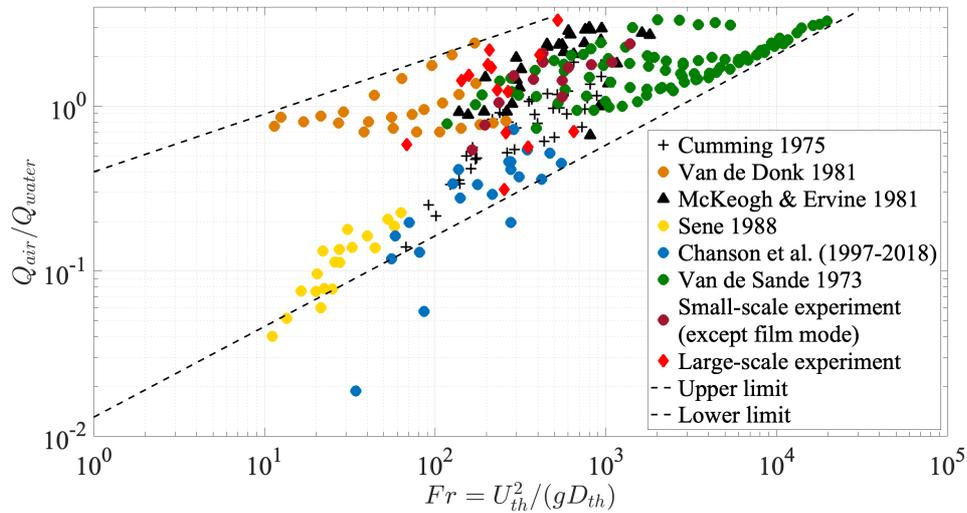

Fig. 36: Plot of $Q_{air}$/$Q_{water}$ versus the impact Froude number $Fr_{th}$. Data from the small-scale facility at large H/$D_0$, data from the large-scale facility, data from literature as indicated in the legend. The upper and lower bounds of the data are indicative.

Bertola, Wang and Chanson (2018b) attempted to scale their data with a Froude number based on the difference between the jet velocity and the critical velocity $U_c$ corresponding to the onset of air entrainment. Introducing the modified Froude number at impact defined as $Fr_{th}^* = (U_{th} - U_c)^2/(g\ D_{th})$, the same data as those presented in Fig.36 are plotted versus $Fr_{th}^*$ in Fig.37. The critical velocity $U_c$ is directly provided by some authors. When it is not, $U_c$ was extracted from the raw data when available (by plotting $Q_{air}$ as a function of the jet velocity), or deduced from the correlation proposed by Cummings and Chanson, 1990. Again, all data but a few are bounded between two limits: a lower bound of equation 0.022 $Fr^{*0.5}$ and an upper bound of equation 0.42 $Fr^{*0.35}$. These exponents are slightly lower than those proposed by Bertola, Wang and Chanson, 2018b as the latter were comprised between a 0.5 exponent observed at jet velocities above 4 m/s, and a 0.9 exponent for jet velocities below 4 m/s. According to the above results, the growth of the entrained air flow rate with $U_{th} - U_c$ remains moderate. Again, the $U^3$ dependency expected from eq.(16) is not recovered. The above discussion underlines that the question of the scaling of the entrained air flow rate with flow parameters remains open.

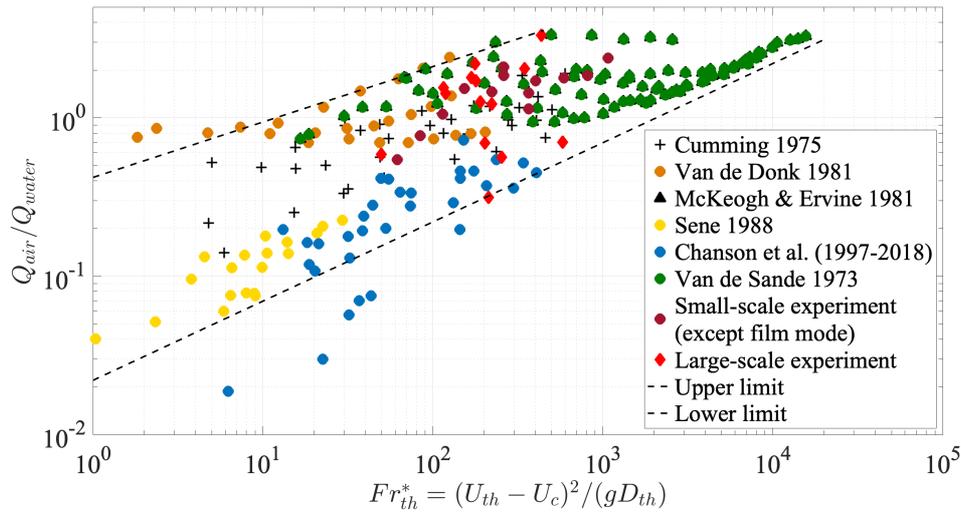

Fig. 37: Plot of $Q_{air}$ /$Q_{water}$ versus the modified Froude number at impact $Fr_{th}^*$= $(U_{th}$ - $U_c)^2$/(g $D_{th}$). Data from the small-scale facility at large $H/D_0$, data from the large-scale facility, data from literature as indicated in the legend. The upper and lower bounds of the data are indicative.

## Conclusion

Two main phenomenological models provide the flow rate of air entrained by a vertical plunging jet. A first scenario, originated by Henderson, McCarthy and Molly, 1970 for cylindrical jets, and later adapted by Sene, 1984 to planar jets, states that the air trapped within the jet corrugations is entrained below the free surface. A second scenario, proposed by Sene, 1984 for smooth jets, argues that a continuous air film is maintained open below the free surface by viscous stresses that equilibrate the assumed hydrostatic pressure gradient in the receiving pool. As very limited (if any) experimental evidence supports these scenarios, we revisited these proposals. In particular, our goal was to analyze the connection between the jet roughness and the entrained air flow rate.

To vary the jet characteristics at impact, experiments on vertical plunging jets impacting a pool have been achieved in two facilities. The small-scale facility had a fixed nozzle diameter of 7.6 mm with fall heights up to ≈ 0.3 m and jet velocities at nozzle up to 10 m/s. In the large-scale facility, fall heights from 2 to 9 m were accessible while nozzle diameters ranged from 23 to 213 mm: the maximum jet velocities at nozzle vary with the diameter. It was 20 m/s for the smallest diameter. All experiments concerned water jets with a radius larger than the capillary length scale. Also, to investigate coherent jets, fall heights less than the break-up length were considered. The topology of jets and the interface deformation were quantified using high-resolution and high-speed imaging. The entrained air flow rate was measured using conical and/or Doppler optical probes.

Quite different jet topologies were generated including, smooth jets, corrugated turbulent jets, jets experiencing buckling, jets subject to interface stripping and droplet ejection. In addition, aerated jets were identified in the large-scale facility as jets whose diameter increased significantly with the fall height. A tentative frontier between non-aerated jets and aerated jets has been established from experiments. It happens that nearly all contributions in the open literature concerned non-aerated jets.

For coherent, non-aerated jets, we show that Henderson, McCarthy and Molly's proposal is valid provided that the effective deformation with respect to air entrainment is set equal to the maximum jet deformation or the 90% limit. Alternately, the effective deformation also amounted to about twice the deformation $\varepsilon_{total}$ defined as the standard deviation of one jet edge position. These findings are supported by turbulent coherent jets with deformations ranging from $\approx 0.2$ to $\approx 1$ jet radius.

For these jets when they are smooth enough, the film scenario proposed by Sene, 1984 happens to approach the measurements at jet velocities below about 2 m/s but the discrepancy neatly increases for velocities above 4-5m/s. We tentatively proposed a frontier between the air film scenario and the roughness scenario that compares the jet roughness with the air film thickness expected from Sene's model: this criterion happens to be supported by the few experiments available. More experiments on that question are required to further test the proposed criterion.

For aerated jets, Henderson, McCarthy and Molly's proposal happens to be also valid, at least over the limited set of experiments presented here, but the effective deformation is weaker than for non-aerated jets: it is about 80% of the maximum deformation, or equivalently, about one time the deformation $\varepsilon_{total}$ defined as the standard deviation of one jet edge position. These findings are supported by turbulent coherent aerated jets with deformations ranging from $\approx 0.1$ to $\approx 1$ jet radius. The behavior of aerated jets compared with non-aerated situations is tentatively attributed to the velocity field at the jet boundary and/or to qualitative changes in the shape of jet deformations. Complementary and refined experiments would be welcome to test these options. Another related open question concerns the limit between aerated and non-aerated plunging jets.

Globally, if one sets aside smooth enough jets, corrugated jets do act as volumetric pumps but with an effective deformation that depends on their aeration. Let us finally mention a situation that escapes all the above scenarios: the latter concerns jet experimenting strong stripping. Only one flow condition of that category has been analyzed here, and that condition provides very distinctive results. Such high initial velocity jets subject to strong stripping just downstream of the injection nozzle deserve to be investigated further.

### CRediT authorship contribution statement

**Ivan Redor:** Investigation, Software, Visualization **Gregory Guyot:** Investigation, Software, Visualization, Software, Data curation **Martin Obligado:** Investigation, Methodology, Software, Validation **Jean-Philippe Matas:** Investigation, Methodology, Software, Validation **Alain Cartellier:** Conceptualization, Methodology, Investigation, Data curation, Validation, Formal analysis, Writing - original draft.

### Declaration of Competing Interest

The authors declare that they have no conflict of interests.

### Acknowledgments

We thank François Bonnel for his assistance on imaging techniques, and the CERG for implementing the large-scale experiment. This work has been supported by ANR-21-

CE05-0029, and by EdF under grant H-30575706-2019-000183 5500-5920075569. The LEGI is part of LabEx Tec21 Investissement d'Avenir, ANR-11-LABX-0030.

**Annex A**: Onset of air entrainment for a plunging water jet.
The figure below gathers available experimental data on the critical Capillary number (i.e. corresponding to the onset of air entrainment) for water jets in air versus the jet diameter scaled by the capillary length scale. The figure includes the data for tap water from Clanet (private communication 1989), those for demineralized water from Cartellier and Lasheras (2003), and those measured in the small-scale experiment described in Section 2 (red crosses in the figure below). The upper and lower boundaries of the empirical correlation proposed by Cummings and Chanson (1999) (see also Chanson (2009)) are indicated over the range of $a_c/(D_0/2)$ covered by the experiments these authors have exploited.

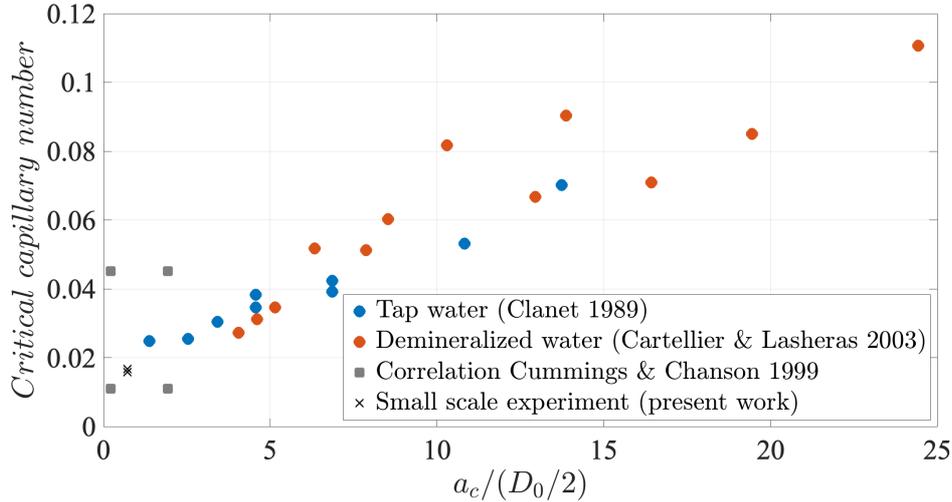

Fig. A-1: Critical Capillary number versus the ratio $a_c/(D_0/2)$ for a water jet in air.

**Annex B:** Break-up length dependence on flow parameters

In the small-scale experiment, the break-up length $L_B$ is monotonously increasing with the jet velocity (Fig.5), possibly with some saturation above 9-10 m/s. In the large-scale experiment, and as shown in the insert of Fig.8-b, $L_B$ is slightly increasing with the jet velocity for $U_0$ between 2.5 and 7.5 m/s, while for jet velocities between 7.5 and 20 m/s, $L_B$ remains almost constant. The difference in the behaviors of $L_B$ in the small-scale and in the large-scale experiments deserves to be discussed.

Let us recall that for a capillary instability and in absence of viscous effects, the comparison of the transit time of the liquid along the liquid lump connected to injector exit with the growth rate of the most amplified instable mode leads to:

$$L_B/D_0 \propto We_L^{1/2} \qquad\qquad (B-1)$$

where the involved Weber number $We_L = \rho_{liquid} U_0^2 D_0 / \sigma$. According to the above equation, $L_B$ is expected to linearly increase with the jet velocity. Trettel (2020) thoroughly revisited available experiments and found that $L_B$ grows as $U_0^{0.66}$ in the so-called turbulent surface breakup regime. These trends are compatible with the behavior observed in the small-scale experiment at moderate velocities (say up to $\approx 6$ m/s, see Fig.5), and also in the large-scale experiment in the limit of small jet velocities (see insert in Fig.8-b).

As the jet velocity increases, so does the air Weber number $We_{air} = \rho_{air} U_0^2 D_0 / \sigma$, up to a point such that the shear instability becomes dominant compared with a capillary destabilization. Introducing the growth rate of a shear instability, the resulting break-up length obeys (Eggers & Villermaux, 2008):

$$L_B/D_0 \propto [\rho_{liquid}/\rho_{air}]^{1/2} \qquad\qquad (B-2)$$

According to Eggers & Villermaux (2008), this result is valid whatever the origin and the thickness of the shear layer, as the latter could be thin (as when the wavelength is controlled by capillarity in a pure Kelvin-Helmholtz mode) or thick (which could be the case for a vorticity layer imposed by a co-current gas stream). The specificity of that regime is that $L_B$ becomes independent of the jet velocity: this is the behavior observed on the large-scale facility as shown in the insert of Fig.8-b. Interestingly, our measurements for the large-scale experiment lie between the correlation proposed by Chehrouki et al. (1985), namely $L_B/D_0 = 7$ $[\rho_{liquid}/\rho_{air}]^{1/2}$, and that proposed by Sallam et al. (2002) that writes $L_B/D_0 = 11$ $[\rho_{liquid}/\rho_{air}]^{1/2}$.

Let us now examine the transition between a situation where the break-up length increases with the jet velocity and a situation where $L_B$ does not depend on $U_0$. According to Sallam et al. (2002), the transition arises for $We_{air}$ about 10-30 in air-water systems under ambient conditions. Trettel (2020) argued that the transition depends on the turbulent intensity in the jet (see his equation 38). For turbulent intensities between 1% and 10%, Trettel found a critical air Weber number $We_{air}$ in the range 10-40: this is similar to Sallam et al.'s findings. As shown in Fig.B-1, the above criterion based on $We_{air}$ happens to be consistent with the data gathered in the two facilities exploited here since the break-up length remains constant for $U_0$ above about 7.5 m/s, that is for $We_{air} \geq 21$.

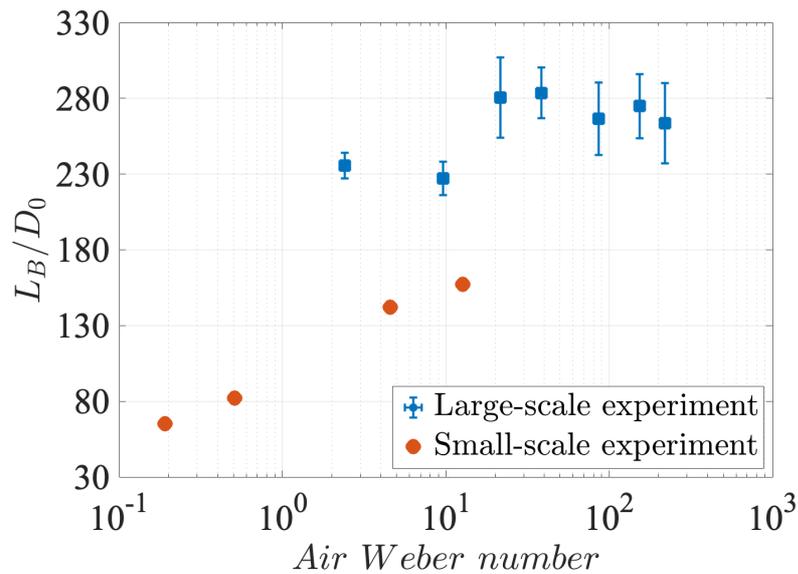

Fig. B-1: Break-up length versus the air Weber number as measured in the large-scale experiment with $D_0$=23 mm, and in the small-scale experiment ($D_0$=7.6 mm). For the latter series, the error bar is less than the symbol height.

**Annex C:** Raw data from small-scale and large-scale experiments.

| U0 (m/s) | H/D0 | Uth (m/s) | Dth (mm) | Qair (m3/s) | Symetry (%) | Varicose deformation (mm) | Total deformation relative to one jet side (mm) | D impact measured (mm) | D 90 measured (mm) |
|---|---|---|---|---|---|---|---|---|---|
| 9.8 | 0.4 | 9.80 | 7.60 | 9.45E-05 | 7.2 | 0.23 | 0.17 | 7.68 | 10.09 |
| 8.57 | 0.7 | 8.58 | 7.60 | 8.61E-05 | 2.2 | 0.32 | 0.24 | 7.78 | 9.84 |
| 7.35 | 1.3 | 7.36 | 7.59 | 7.64E-05 | 19.9 | 0.34 | 0.25 | 7.765 | 9.34 |
| 6.12 | 1.6 | 6.14 | 7.59 | 6.75E-05 | 9.31 | 0.44 | 0.33 | 7.956 | 9.72 |
| 4.9 | 2.2 | 4.93 | 7.57 | 4.31E-05 | 46.78 | 0.35 | 0.26 | 8.082 | 9.68 |
| 3.67 | 2.5 | 3.72 | 7.55 | 2.24E-05 | 2.3 | 0.29 | 0.22 | 8.022 | 9.22 |
| 3.06 | 2.5 | 3.12 | 7.53 | 1.14E-05 | 2.7 | 0.37 | 0.28 | 7.827 | 9.15 |
| 2.45 | 3.2 | 2.55 | 7.46 | 8.20E-06 | 9.44 | 0.30 | 0.23 | 7.656 | 8.69 |
| 1.84 | 3.2 | 1.97 | 7.35 | 2.01E-06 | 67.59 | 0.33 | 0.25 | 7.473 | 8.41 |
| 1.53 | 3.7 | 1.70 | 7.21 | 1.41E-06 | 3.28 | - | - | - | - |
|  |  |  |  |  |  |  |  |  |  |
| 9.8 | 39.1 | 10.09 | 7.49 | 1.06E-03 | half profile | 2.07 | 1.59 | 9.16 | 14.12 |
| 8.57 | 39.9 | 8.91 | 7.45 | 7.14E-04 | 22.4 | 1.84 | 1.43 | 8.47 | 12.99 |
| 7.35 | 39.7 | 7.74 | 7.40 | 5.95E-04 | 7.36 | 1.52 | 1.21 | 8.08 | 11.94 |
| 6.12 | 39.7 | 6.59 | 7.33 | 4.76E-04 | 13.67 | 1.26 | 1.05 | 7.77 | 11.06 |
| 4.9 | 40.9 | 5.49 | 7.18 | 4.10E-04 | 3.39 | - | - | - | - |
| 4.9 | 40.9 | 5.49 | 7.18 | 4.65E-04 | 3.37 | 1.11 | 0.95 | 7.40 | 10.65 |
| 4.29 | 40.4 | 4.94 | 7.08 | 2.67E-04 | 14.52 | - | - | - | - |
| 3.67 | 41.1 | 4.43 | 6.92 | 2.55E-04 | 5.45 | 1.05 | 0.86 | 7.23 | 9.71 |
| 3.06 | 41.2 | 3.94 | 6.70 | 1.46E-04 | 18.73 | 0.88 | 0.71 | 7.00 | 9.29 |
| 2.45 | 41.7 | 3.50 | 6.36 | 8.54E-05 | 21.34 | 0.85 | 0.67 | 6.55 | 8.56 |
| 1.84 | 41.1 | 3.08 | 5.87 | 4.54E-05 | half profile | 0.72 | 0.54 | 6.18 | 7.84 |
| 1.53 | 41.8 | 2.93 | 5.49 | 1.79E-05 | 31.2 | - | - | - | - |
|  |  |  |  |  |  |  |  |  |  |
| 6.12 | 21.1 | 6.37 | 7.45 | 3.97E-04 | 6.6 | - | - | - | - |
| 6.12 | 21.1 | 6.37 | 7.45 | 3.17E-04 | half profile | 1.18 | 0.89 | 8.54 | 11.51 |
| 4.9 | 21.1 | 5.21 | 7.37 | 3.24E-04 | 13.1 | 0.9 | 0.71 | 8.15 | 10.53 |

Table C-1: Air entrainment rate, jet size and deformation at impact. Small-scale facility.

| Run # | D0 (mm) | U0 (m/s) | H (m) | H/D0 | Uth (m/s) | Dth (mm) | Qair (m3/s) | Symetry (%) | D impact measured (mm) | Varicose deformation (mm) | Total deformation (mm) | D90 (mm) | Jet aeration | Jet allure |
|---|---|---|---|---|---|---|---|---|---|---|---|---|---|---|
| 1 | 23.6 | 5.0 | 2 | 84.7 | 8,0 | 18.6 | 0.00124 | 13.8 | 17.5 | 2.7 | 1.35 | 20.68 | Non aerated | Coherent jet = Rayleigh mode |
| 2 | 23.6 | 7.5 | 9 | 381.4 | 15.3 | 16.5 | - | - | - | - | - | - | Non aerated | Broken-up jet |
| 3 | 82.9 | 12.5 | 2 | 24.1 | 14,0 | 78.4 | 0.0212 | 16.5 | 91.5 | 7.8 | 5.1 | 104.38 | Weakly aerated | Coherent jet + Corrugations |
| 4 | 82.9 | 2.5 | 3.25 | 39.2 | 8.4 | 45.3 | 0.0209 | 18.7 | - | 9.1 | 4.55 | - | Non aerated | Buckling |
| 5 | 82.9 | 2.5 | 5 | 60.3 | 10.2 | 41,0 | 0.00933 | 25.2 | - | 12.3 | 8.85 | - | Non aerated | Buckling |
| 6 | 82.9 | 20 | 5 | 60.3 | 22.3 | 78.5 | 0.0757 | 3.6 | 154.5 | 15.7 | 23.55 | 211.76 | Aerated | Distorted jet + Strong stripping |
| 7 | 82.9 | 2.5 | 9 | 108.6 | 13.5 | 35.6 | 0.0448 | 25.9 | 44.3 | 12.8 | 16.05 | 72.43 | Aerated | Buckling + Stripping |
| 8 | 82.9 | 7.5 | 9 | 108.6 | 15.3 | 58.1 | 0.0829 | 41.2 | 82.4 | 11.6 | 21.8 | 124.54 | Aerated | Distorted Jet + Stripping |
| 9 | 162.3 | 2.5 | 9 | 55.5 | 13.5 | 69.8 | 0.0633 | 0.9 | 81.2 | 24.4 | 20.9 | 119.71 | Weakly aerated | Distorted Jet + Stripping |
| 10 | 162.3 | 5.0 | 9 | 55.5 | 14.2 | 96.3 | 0.178 | 17.5 | 100.1 | 24,0 | 26.45 | 167.64 | Aerated | Distorted Jet + Stripping |
| 11 | 162.3 | 7.5 | 9 | 55.5 | 15.3 | 113.8 | 0.341 | 28.2 | 157.3 | 26.1 | 34.05 | 237.13 | Aerated | Distorted Jet + Stripping |
| 12 | 213 | 1.8 | 3.25 | 15.3 | 8.2 | 99.9 | 0.0376 | 17.0 | 101.6 | 10.0 | 8,00 | 121.74 | Non aerated | Coherent jet + Corrugations |
| 13 | 213 | 2.5 | 9 | 42.3 | 13.5 | 91.6 | 0.16 | 10.7 | 106.6 | 27.5 | 34.35 | 183.68 | Aerated | Distorted Jet + Stripping |
| 14 | 213 | 5.0 | 8 | 37.6 | 13.5 | 129.7 | 0.254 | - | - | - | - | - | Aerated | Distorted Jet + Stripping |
| 15 | 162.3 | 2.5 | 8 | 49.3 | 12.8 | 71.8 | 0.0649 | - | - | - | - | - | Non aerated | Distorted Jet + Stripping |

Table C-2: Air entrainment rate, jet size and deformation at impact, aeration and jet allure. Large-scale facility. The total deformation in the table represents $\varepsilon_{max}$: it is thus relative to one side of the jet while the varicose deformation is relative to two sides of the jets (see Fig.11 and section 3.1).